\documentclass[a4paper,11pt]{article}
\usepackage[dvipdfmx,hiresbb]{graphicx}

\usepackage{jheppub} 
                   
\usepackage[T1]{fontenc} 

\title{\boldmath 
Gauge fields in the presence of the electroweak bubble wall
}

\author[a]{Takahiro Kubota  
}

\affiliation[a]{CELAS, The University of Osaka, Toyonaka, Osaka, 560-0043, Japan}

\emailAdd{takahirokubota859@hotmail.com}

\abstract{
The gauge field theory  of  the standard  electroweak model in the presence of the electroweak bubble wall   is investigated with a view to   its applications to microscopic phenomena,  which are believed to have occurred during the phase transition in the early universe.    The asymptotic fields are defined anew so that   the effects of the  position-dependent Higgs condensate are taken into account through   the position-dependent $W$ and $Z$ boson masses.  A novel method of  massive gauge field quantization  in the $R_{\xi}$-gauge with $\xi =1$   is proposed for the case of the position-dependent masses. Our procedure  is based on  the eigenfunction expansion  method associated with second-order differential operators,  i.e., a sort of generalized Fourier expansion. The commutation relations of creation and annihilation operators of various   wave  propagation modes are given in terms of   what is known as   the  spectral function. The decoupling of unphysical states  from the physical S-matrix is also investigated    along the line of Kugo-Ojima's quartet mechanism on the basis of the BRST symmetry. It is pointed out that one of the quartet fields  is not merely the unphysical scalar field but should be  a linear combination of the unphysical scalar and the gauge fields.   The physical and unphysical polarizations of the gauge field waves  are unambiguously  distinguished and this  will help us evaluate the friction caused by the physical polarization states of $W$ and $Z$ boson waves  on the bubble wall during the phase transition  in the early universe.  
}

\begin{document} 
\maketitle
\flushbottom

\section{Introduction}

The present  paper is a sequel to the  author's pervious one \cite{kubota1}, in which two-point 
Green's functions of scalar, spinor and vector fields in the presence of the electroweak bubble wall 
were constructed.   The purpose  of Ref.\cite{kubota1} and of the present paper is to prepare field 
theoretical tools for   analyzing  phenomena  that must have occurred near the electroweak  bubble wall during the expansion of the Higgs condensate in the early universe.  Detailed analyses of those  phenomena are indispensable  for  precise microscopic understanding  of  the magnitude of   the friction caused on the bubble wall by the plasma consisting of standard model particles.  The eventual goal of these considerations is  to evaluate hopefully the bubble wall velocity as accurately as possible and to get  information  on the gravitational wave emitted during the  bubble expansion.

The standard model phase transition is smooth cross-over, but could be of first-order if the electroweak model is extended somehow in the future \cite{kajantie} - \cite{grojean}.   The bubble wall velocity during the phase transition in the early universe is one of the key parameters  to determine the strength and spectral shape of the  gravitational wave that was  presumably emitted during the first-order phase transition \cite{witten} - \cite{kosowsky2}.    The  gravitational wave projects planned in the near future such as LISA \cite{caprini1} - \cite{auclair},   DECIGO \cite{kawamura0} - \cite{kawamura},  Taiji \cite{ruan} and TianQin \cite{luo}  - \cite{luo2} are expected to have suitable sensitivity  to probe the electroweak phase transition and  therefore detailed analysis on the bubble wall velocity is  pressingly important.   It is also of great significance for the electroweak baryogenesis \cite{kuzmin} - \cite{garbrecht}.

The Higgs vacuum expectation value varies through the bubble wall interface  and therefore the standard model particles have position dependent masses. The Green's functions  in such a case are expressed   in \cite{kubota1}   in the form of integral representations,    where  the so-called eigenfunction expansion method initiated by Weyl \cite{weyl}, developed further by  Stone \cite{stone} and completed in its final form   by Titchmarsh \cite{titchmarsh} and Kodaira \cite{kodaira1} -   \cite{kodaira2} was shown  to be  an indispensable and at the same time  the most suitable technique.  (See also \cite{yoshida} -   \cite{yosida2} for a pedagogical exposition.)  It has been shown in \cite{kubota1} that the spectral function of the eigenvalue problem of 
a self-adjoint differential operator plays a crucial role in constructing the Green's functions. 

The Green's functions can be obtained  simply by solving partial differential equations under  given  boundary conditions without touching upon the procedure of field quantization or the details  of the Hilbert space. From the standpoint of quantum field theory, however, the method of field quantization must be reexamined  before applying the Green's functions to quantum phenomena.  Since masses are position dependent, the notions of asymptotic fields and their wave equations have to be reconsidered. By studying the various solutions of the wave equations, we are able to define a set of  operators which  are  analogous to creation and annihilation operators, but  are no more  associated with   conventional  particle pictures. They can create or annihilate only particular wave propagation modes. The electroweak bubble wall breaks both translational and Lorentz invariance,  and therefore we are not able to use the conventional approach  of classifying particles according to their spin and mass. Since the particle picture is veiled, the Hilbert space constructed by applying creation operators on the vacuum will become necessarily different from the usual  Fock  space

In  the electroweak gauge theory,  the Hilbert space contains several kinds of unphysical negative norm states as well as physical but zero-norm states.  In the early days of non-abelian gauge theories, 
Kugo and Ojima \cite{kugoojima1} - \cite{kugoojima5} developed a powerful  method   of classifying physical and unphysical  states by employing the   BRST (Becchi, Rouet, Stora \cite{brst3} -\cite{brst1}  and  Tyutin \cite{tyutin}) invariance. They advocated the so-called ``BRST quartet mechanism", by which unwanted  states such as  ghost, anti-ghost, scalar polarization and  unphysical scalar field states   conspire  together to  decouple  from the physical S-matrix. In the presence of the electroweak bubble wall, although we  have still the BRST invariance of the Lagrangian, the Lorentz invariance is no more respected and it is far from obvious how the BRST quartet mechanism works.

The purpose of the present paper is to scrutinize how and to what extent  the presence of the electroweak bubble modifies  the framework of and notions in gauge theories, such as  asymptotic fields, field quantization method,   \footnote{See Refs.  \cite{boedeker2}  - \cite{ai3} for other  approaches to  field quantization under bubble wall  backgrounds.}  creation and annihilation operators,  classification of vector field polarizations,  the decoupling of unphysical states from the physical S-matrix, and  so forth.  It  should be emphasized  that these are  not totally  academic problems   but    are strongly motivated by the recent intense studies on the bubble wall expansion dynamics in the early universe.

The electroweak bubble wall feels  differential  vacuum energy  across the bubble interface and is driven  forward so that the bubble expands.  In addition,  as argued in \cite{boedeker1},    there occur     friction effects caused by plasma particles passing through the bubble wall. Particles get massive when they pass through the wall from symmetry-restored  to symmetry-broken regions.  The energy and the transverse momentum of the particles are conserved  and  some amount of  momentum in the  direction perpendicular to the wall   necessarily decreases in correspondence with the increase of the mass. The  momentum  lost by  plasma particles is transferred to the wall, as   a result of which the wall feels friction impeding the bubble expansion. 

 In the meanwhile the effects due to the transition radiation   \cite{ginzburgfrank}  -   \cite{jackson} have also   been  studied in \cite{boedeker2} -  \cite{GouttennoireJinnoSala}. Namely, the process of a particle impinging on the wall and emitting another particle (or several particles) was investigated.  It was argued that soft $W$- and $Z$- vector boson emission dominates the friction on the wall. The  pressure exerted on the wall is enhanced   and is featured by  logarithmic terms    that come from the soft region  in  the phase space. The fixed order calculation of the thermal pressure breaks down quite possibly and  the method of summation  of multiple soft gauge boson emission effects are  proposed.    According to the common lore of  infrared singularities in field theories,   a great care must be exerted to handle  the masses, in particular, of  gauge bosons in infrared problems.  The present paper is intended to offer theoretical bases for such subtleties   connected with   the gauge fields whose masses are position-dependent. 

The structure of the present article is as follows.  In Section \ref{sec:bubblewall}, the electroweak bubble wall is prescribed by a differential equation ensuring the absence of tadpoles.  In Section \ref{sec:wavepropagation}, the asymptotic fields for the case of position-dependent mass are defined and are expanded in terms of mode functions by taking the scalar field case as an example.  The procedure of the scalar field quantization is exemplified by using the spectral function.  The gauge fixing procedure, the ghost  and anti-ghost fields and the BRST symmetry are summarized in Section \ref{sec:electroweakbrst} for the sake of   preparation to quantize   the $Z$-boson field  in Section \ref{sec:gaugefieldquantization}.  The properties of auxiliary fields that are introduced through the gauge fixing procedure  are given in Section \ref{sec:moreabout} and it is argued that  the Kugo-Ojima's quartet mechanism works  well  by modifying one of the quartet fields appropriately. The physical polarization states are explicitly given in Section \ref{sec:physicalpolarization}.   Section \ref{sec:summary} is devoted to the summary of the present paper. 
\section{The electroweak bubble wall}
\label{sec:bubblewall}

First of all we would like to  specify   the electroweak bubble wall to be treated in the  present paper,  lest our scope  should become   too much stretched.  Let us begin with the scalar part of the  Lagrangian in the standard electroweak theory  
\begin{eqnarray}
{\cal L}_{\rm scalar}= ( \nabla^{\mu} \Phi )^{\dag} \nabla_{\mu}\Phi - V(\Phi^{\dag}\Phi )\:.
\label{eq:scalarlagrangian}
\end{eqnarray}
We parameterize  the standard model Higgs doublet field $\Phi$ in the following way, 
\begin{eqnarray}
\Phi =
\frac{1}{\sqrt{2}}
\bigg \{
v+H -i  \sum_{a=1}^{3}\tau ^{a}\chi^{a}
\bigg \}
 \left (
\begin{tabular}{c}
$0$
\\
\\
$1$
\end{tabular}
\right )
=
\frac{1}{\sqrt{2}} \left (
\begin{tabular}{c}
$-\chi^{2} -i \chi^{1}$
\\
\\
$ v+H + i \chi^{3} $
\end{tabular}
\right )\:,
 \label{eq:higgsparameterization}
\end{eqnarray}
where $H$   and $\chi^{a} $  $(a=1,2,3)$ are the Higgs boson and unphysical scalar fields, respectively.   The Pauli matrices are denoted by $\tau^{a}$  $(a=1,2,3)$.  The vacuum expectation value  $v$  is not necessarily a constant in the present context,   but may depend on the space-time  coordinates. We will go as far as possible   without specifying the explicit  form of the Higgs potential $V(\Phi^{\dag} \Phi )$ in  (\ref{eq:scalarlagrangian}).    This is because the standard   electroweak theory will be possibly  extended in the future  so that the phase transition is of the first order  and therefore the Higgs potential should be kept  general enough    in the form amenable to future modification.

\begin{figure}[h]
\begin{center}
   \input{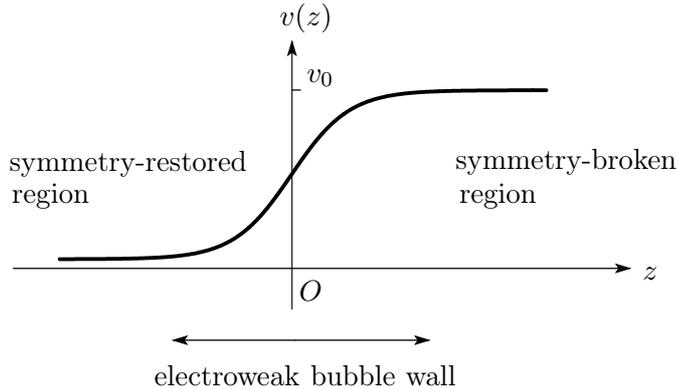}
   \vskip0.5cm
\caption{A typical profile of the Higgs condensate $v(z)$,  that connects the symmetry-restored region ($z \to -\infty$)  and the symmetry-broken region ($z \to + \infty$).}
\label{fig:illustration}
\end{center}
\end{figure}

Putting  (\ref{eq:higgsparameterization}) into    (\ref{eq:scalarlagrangian}),   we find a   so-called 
tadpole term  which is linear in the Higgs field  $H$, namely,  
\begin{eqnarray}
{\cal L}_{\rm tadpole}
=- H \left \{   
\square \: v 
+  v V^{\:\prime} \left ( \frac{v^{2}}{2} \right ) \right \}    \:,
\label{eq:tadpole}
\end{eqnarray}
where the prime $(')$ in $V^{\:\prime}(\Phi^{\dag}\Phi )$ means the derivative of the potential with respect to $\Phi^{\dag}\Phi$.  The d'Alembertian operator in (\ref{eq:tadpole})  is given as usual by
\begin{eqnarray}
\square =g^{\mu \nu} \frac{\partial}{\partial x^{\mu}}
\frac{\partial }{\partial x^{\nu}}=\frac{\partial^{2}}{\partial t^{2}} - \frac{\partial^{2}}{\partial x^{2}}-\frac{\partial^{2}}{\partial y^{2}}-\frac{\partial^{2}}{\partial z^{2}}\:,
\end{eqnarray}
where our convention of the metric is $g^{\mu \nu} = {\rm diag}(1, -1, -1, -1)$.  Here  and hereafter we  use  the hybrid notations   for the space-time coordinates, namely, $x^{\mu}=(t, x, y, z)$  together with       $x^{\mu}= (x^{0},\:  \cdots ,\:  x^{3})$.   To ensure  the   stability, this tadpole term should vanish and we impose a condition that must  be satisfied   by    $v$, 
\begin{eqnarray}
\square \: v +   v V^{\:\prime} \left ( \frac{v^{2}}{2} \right )  =  0\:.
\label{eq:stability}
\end{eqnarray}
{\color{black}{
By discarding the terms linear in the Higgs field,  the separation between background and quantum fluctuations is well defined.
}}
The trivial constant solutions to (\ref{eq:stability})  are  $v=0$ and another   one  $v=v_{0} \neq 0$  satisfying   $V^{\prime}(v_{0}^{\: 2}/2)=0$,   which correspond  to symmetry-restored and symmetry-broken vacuum solutions, respectively.   If, however, $v$ is not a constant but depends on one of the three space coordinates, say $z$, then the equation (\ref{eq:stability}) for $v=v(z)$ becomes a non-trivial ordinary differential equation, 
\begin{eqnarray}
-\frac{d^{2} v(z)}{dz^{2}}  +  v(z) V^{\:\prime} \left ( \frac{v(z)^{2}}{2} \right ) 
=
0\:.
\label{eq:ordinarydiffeq}
\end{eqnarray}
The $z \to \pm \infty$ behavior of the solution to (\ref{eq:ordinarydiffeq}) should be a constant that is either $v=0$ or $v=v_{0}$ with $V^{\:\prime} (v_{0}^{\:2}/2)=0$.  An illustrative shape of such a solution is given   in Figure \ref{fig:illustration} as an example, which connects the symmetry-restored region ($z \to - \infty$) with the symmetry-broken region   ($z \to + \infty$).    The electroweak bubble wall is the region in-between      around $z \sim 0$ where   $v(z)$ is varying.   See Ref.\cite{polyakov1} for an early attempt of making use of the   solutions to (\ref{eq:ordinarydiffeq}) and also see   Refs. \cite{ayala} and \cite{farrar}   for wave propagation analyses through domain walls  described by   eq. (\ref{eq:ordinarydiffeq}).

In the present work, we simply assume the existence of such a  bubble wall solution $v(z)$ as  illustrated  in  Figure \ref{fig:illustration} without going into the detailed form of $V(\Phi^{\dag}\Phi )$  or $v(z)$, either. All we require is  the asymptotic behavior,  $v(-\infty )=0$ and $v(+\infty )=v_{0}$ \:.  We would like to   focus  our attention to scrutinize whether the conventional field theory technique works   just as well in the presence of the   $z$-dependent Higgs condensate $v(z)$.   The procedure of field quantization itself becomes non-trivial, because all the masses of the standard   model particles are   now $z$-dependent.  The $W$- and $Z$-boson masses for example are given, respectively, by
\begin{eqnarray}
M_{W}^{2} =M_{W}(z)^{2}  = \frac{1}{4} g^{2} v(z)^{2}, 
\hskip1cm
M_{Z}^{2} =  M_{Z} (z)^{2} = \frac{1}{4} \left ( g^{2} + g^{\:\prime \:2}  \right ) v(z)^{2}, 
\end{eqnarray}
where $g$ and $g^{\:\prime}$ are the $SU(2)_{L}$ and $U(1)_{Y}$ gauge couplings.  We have to go into the details of the wave propagation modes along the $z$-direction, which is no more   expressed as  plane waves.

In the electroweak phase transition in the early universe, the bubble wall must have expanded very rapidly  in the rest frame of electroweak plasma particles.   The solution $v=v(z)$ to (\ref{eq:ordinarydiffeq}) on  the other hand is time-independent, and the use of this solution indicates that we are not working   in such a Lorentz frame but in the bubble wall rest frame.  By setting $v=v(z)$, the bubble wall is supposed to be planar and this can  be justified if we note that the length scale of the microscopic phenomena which we would like to analyze should be much smaller than the size of the Higgs condensate.

\section{Wave propagation modes}
\label{sec:wavepropagation}

\subsection{Asymptotic fields}
\label{subsec:asymptoticfield}

In the presence of the electroweak bubble wall, a generic class of standard model   Heisenberg fields           $\varphi (t, \vec{x}_{\perp}, z) $   in the $R_{\: \xi}$ gauge with $\xi =1$  satisfies the following type of equations of motion  (except for the electromagnetic and matter sector fields)
\begin{eqnarray}
\left \{  \square +M_{W}(z)^{2} \right \} \varphi (t, \vec{x}_{\perp}, z) &= & j_{\small W}(t, \vec{x}_{\perp}, z)\:,
\label{eq:heisenbergeq1}
\\
\left \{  \square +M_{Z}(z)^{2} \right \} \varphi (t, \vec{x}_{\perp}, z) &=&  j_{\small Z}(t, \vec{x}_{\perp}, z)\;,
\label{eq:heisenbergeq2}
\end{eqnarray}
where $j_{\small W}(t, \vec{x}_{\perp}, z)$   and $j_{\small Z}(t, \vec{x}_{\perp}, z)$   represent interactions of $\varphi (t, \vec{x}_{\perp}, z) $.  These equations differ from the usual one in that the gauge boson masses,  $M_{W}(z)$ and $M_{Z}(z)$,    are position $z$-dependent.   Due to these masses,  there is no translational invariance in the $z$-direction, and   we have to handle the $z$-coordinate  in a different way  from the transverse coordinates,  $\vec{x}_{\perp} = (x, y, 0)$.  The analyses to be   given below on  (\ref{eq:heisenbergeq1}) and  those on (\ref{eq:heisenbergeq2}) go in perfect parallel   and we take only the latter equation   (\ref{eq:heisenbergeq2}) hereafter.

The formal solution to (\ref{eq:heisenbergeq2}) can be expressed as \cite{yangfeldman} -  \cite{lsz2} 
\begin{eqnarray}
\varphi (t, \vec{x}_{\perp}, z) 
&=& 
 \varphi  ^{\rm in} (t, \vec{x}_{\perp}, z) 
\nonumber \\
& & \hskip-0.5cm
+ \int dt^{\:\prime} \: d^{2}\vec{x}_{\perp}^{\:\prime} \: dz^{\:\prime} 
\: \Delta_{R}(t-t^{\:\prime}, \vec{x}_{\perp}-\vec{x}_{\perp}^{\:\prime}, z, z^{\:\prime}; M_{Z}(*))
 j_{\small Z}(t^{\:\prime}, \vec{x}_{\perp}^{\:\prime}, z^{\:\prime})\:,
\\
\varphi (t, \vec{x}_{\perp}, z) 
&=&
\varphi  ^{\rm out} (t, \vec{x}_{\perp}, z) 
\nonumber \\
& & \hskip-0.5cm
+ \int dt^{\:\prime} \: d^{2}\vec{x}_{\perp}^{\:\prime} \: dz^{\:\prime} 
\: \Delta_{A}(t-t^{\:\prime}, \vec{x}_{\perp}-\vec{x}_{\perp}^{\:\prime}, z, z^{\:\prime}; M_{Z}(*))
 j_{\small Z}(t^{\:\prime}, \vec{x}_{\perp}^{\:\prime}, z^{\:\prime})\:.
\end{eqnarray}
We introduced  retarded ($\Delta_{R}$) and advanced  ($\Delta_{A}$)    Green's functions that satisfy, respectively,  the following differential equations,  
\begin{eqnarray}
\left \{  \square +M_{Z}(z)^{2} \right \} \Delta_{R} (t - t^{\:\prime}, \vec{x}_{\perp} 
- \vec{x}_{\perp}^{\:\prime}, z, z^{\:\prime}; M_{Z}(*)) 
  &=& \delta(t - t^{\:\prime})\delta^{2} (\vec{x}_{\perp} - \vec{x}_{\perp}^{\;\prime}) \delta(z- z^{\:\prime})\:,
 \nonumber \\
 \left \{  \square^{\;\prime} +M_{Z}(z^{\:\prime})^{2} \right \} \Delta_{R} (t - t^{\:\prime}, \vec{x}_{\perp} 
- \vec{x}_{\perp}^{\:\prime}, z, z^{\:\prime}; M_{Z}(*)) 
&=&  \delta(t - t^{\:\prime})\delta^{2} (\vec{x}_{\perp} - \vec{x}_{\perp}^{\;\prime}) \delta(z- z^{\:\prime})\:,
\nonumber \\
\label{eq:retardedgreensfunction}
\end{eqnarray}
and 
\begin{eqnarray}
\left \{  \square +M_{Z}(z)^{2} \right \} \Delta_{A} (t - t^{\:\prime}, \vec{x}_{\perp} 
- \vec{x}_{\perp}^{\:\prime}, z, z^{\:\prime}; M_{Z}(*)) 
&=&  \delta(t - t^{\:\prime})\delta^{2}
(\vec{x}_{\perp} - \vec{x}_{\perp}^{\;\prime})
\delta(z- z^{\:\prime})\:,
\nonumber \\
\left \{  \square^{\:\prime} +M_{Z}(z^{\:\prime})^{2} \right \} \Delta_{A} (t - t^{\:\prime}, \vec{x}_{\perp} 
- \vec{x}_{\perp}^{\:\prime}, z, z^{\:\prime}; M_{Z}(*)) 
&=&  \delta(t - t^{\:\prime})\delta^{2}
(\vec{x}_{\perp} - \vec{x}_{\perp}^{\;\prime})
\delta(z- z^{\:\prime})\:.
\nonumber \\
\label{eq:advancedgreensfunction}
\end{eqnarray}
We used the notation of the d'Alembertian $\square ^{\: \prime}$  defined by
\begin{eqnarray}
\square ^{\:\prime}=
\frac{\partial^{2}}{\partial t^{\:\prime \:2}} - \frac{\partial^{2}}{\partial x^{\:\prime \: 2}}-\frac{\partial^{2}}{\partial y^{\:\prime \: 2}}-\frac{\partial^{2}}{\partial z^{\:\prime \: 2}}\:.
\end{eqnarray}
We also made use of the notation $M_{Z}(*)$ as opposed to  $M_{Z}(z)$ or $M_{Z}(z^{\:\prime})$,   so that we keep in our mind that   the dependence of the Green's functions on   $M_{Z}(z)$ or  $M_{Z}(z^{\:\prime})$   is only indirect as we will confirm later.   Since we do not discuss  renormalization procedures in the present paper,  we omit the renormalization factor in front of  the incoming  $ \varphi  ^{\rm in} (t, \vec{x}_{\perp}, z) $  and outgoing $ \varphi  ^{\rm out} (t, \vec{x}_{\perp}, z) $ fields.   The retarded and advanced Green's functions are non-vanishing for $t > t^{\:\prime}$ and   $t^{\:\prime} > t$, respectively, and therefore $\varphi ^{\rm in}(t, \vec{x}_{\perp}, z)$ and   $\varphi ^{\rm out}(t, \vec{x}_{\perp}, z ) $   describe  the asymptotic behavior of the field in  the infinite   past and infinite  future, respectively  (in the sense of  so-called weak asymptotic condition),  i.e.,
\begin{eqnarray}
\varphi (t, \vec{x}_{\perp}, z) 
& \longrightarrow &
\varphi ^{\rm in}(t,  \vec{x}_{\perp}, z) \:, \hskip1.2cm (t \longrightarrow -\infty )\:,
\label{eq:asymptoticfields1}
\\
\varphi (t, \vec{x}_{\perp}, z) 
& \longrightarrow &
\varphi  ^{\rm out}(t, \vec{x}_{\perp}, z)  \:, \hskip1cm (t \longrightarrow +\infty )\:.
\label{eq:asymptoticfields2}
\end{eqnarray}
These are  what we call asymptotic fields  which  satisfy the ``free" field equations of motion, 
\begin{eqnarray}
\left \{  \square +M_{Z}(z)^{2} \right \} \varphi ^{\rm in}(t, \vec{x}_{\perp}, z) = 0, 
\hskip1.0cm 
\left \{  \square +M_{Z}(z)^{2} \right \} \varphi ^{\rm out}(t, \vec{x}_{\perp}, z) =0\:.
\label{eq:inoutwaveequation}
\end{eqnarray}

The  asymptotic field has been  one of  the basic concepts in the formulation of field theories  {\` a} la Lehmann, Symanzik and Zimmermann   (LSZ)  \cite{lsz} -  \cite{lsz2},  but note    that, in our case,  the mass term in (\ref{eq:inoutwaveequation}) is   not a constant but is $z$-dependent.    Very roughly speaking, the concept of asymptotic fields is considered in the situation in which interaction regions are restricted  in space-time and forces between particles are switched off in the remote past and remote future. Nevertheless in our case,   interactions with the background of   Higgs condensate   are   still taken into account.   Such a modification of the notion of  asymptotic fields must be reexamined in detail carefully to see whether the LSZ techniques remain intact.   In particular we should be concerned with the expansion formulas of interacting Heisenberg field operators  with respect  to our asymptotic fields \cite{haag1} - \cite{bogoliubov}. 

For now,  we have to put our definition of the asymptotic fields on a sound basis by  specifying, first of all,  a mathematical method of constructing  the retarded and  advanced Green's functions that appear in (\ref{eq:retardedgreensfunction}) and (\ref{eq:advancedgreensfunction}). Technical tools for elaborating  the Green's functions are prepared in the   course of  solving (\ref{eq:inoutwaveequation}) and of  getting all  the wave propagation modes for   given mass function $M_{Z}(z)$.   Therefore we    investigate  the solutions to  (\ref{eq:inoutwaveequation}) first  in Section  \ref{sec:kleingordonpropagation} and then will come back later in   Section \ref{sec:retardedadvancedgreensfunctions}    to the retarded and advanced Green's functions.

{\color{black}{
A   remark to be inserted here for our better understanding is that,  in  (\ref{eq:asymptoticfields1}) and  (\ref{eq:asymptoticfields2}), we take only the temporal infinity ($t \to \pm \infty$) and the spatial coordinates (in particular the $z$-coordinate)  are kept fixed.     This is of course the usual procedure in the LSZ formalism to define the asymptotic fields.  The overall profile of the Higgs condensate $v(z)$ is unaltered as  depicted in Figure \ref{fig:illustration} and the symmetry-broken and symmetry-restored regions  coexist when  we solve the wave equation (\ref{eq:inoutwaveequation}).   For those who  would like to compute the friction on the bubble wall  in  the plasma rest frame, the bubble wall  is expanding very rapidly and the Higgs condensate is filling almost all corners of the space    in the infinite future. Therefore   it might  be puzzling for them to think  of wave propagation modes   of  (\ref{eq:inoutwaveequation}) which are streching not only in  symmetry-broken but also  symmetry-restored regions  in the  $t \to + \infty$ limit.  In the present paper we are just aiming at the LSZ-like formulation in the presence of the Higgs condensate and the bubble wall rest frame  turns out  to be    the most convenient amenable to the LSZ-like formulation. At present, however, we do not have  much to say about other Lorentz frames for the use of  the LSZ-like formulation.
}}

\subsection{The Klein-Gordon type real scalar field}
\label{sec:kleingordonpropagation}

Let us investigate wave-propagation modes of the real scalar field satisfying 
(\ref{eq:inoutwaveequation}).   Here and hereafter we drop the superscript ``in" and ``out"  for simplicity, 
hoping that there would not arise any confusion,   
and we rewrite   (\ref{eq:inoutwaveequation}) again  
\begin{eqnarray}
\left \{
\square  +M_{Z}(z)^{2}
\right \} \varphi (t, \vec{x}_{\perp}, z)=0\:.
\label{eq:scalarwaveequation}
\end{eqnarray}
Since the translational invariance is preserved in the time  and   ${\vec x}_{\perp}$-directions,   
we consider a superposition of the plane waves in the ${\vec x}_{\perp }$-direction as the solution  to (\ref{eq:scalarwaveequation}), i.e., 
\begin{eqnarray}
e^{-iEt}\:  e^{i\vec{p}_{\perp} \cdot \vec{x}_{\perp}} \:  \phi (z)\:,
\label{eq:modes}
\end{eqnarray}
where $\vec{p}_{\perp}=(p_{x}, p_{y}, 0 )$ is the wave vector in the transverse direction.
 By putting (\ref{eq:modes})   into  (\ref{eq:scalarwaveequation}),  we get an equation  for $\phi (z) $, 
 which turns out to be of the Schr{\" o}dinger type 
\begin{eqnarray} 
\left \{ -\frac{d^{2}}{dz^{2}}  + M_{Z}(z)^{2}  \right \} \phi (z; \lambda ) =\lambda \phi (z; \lambda ) , 
\hskip0.3cm
\lambda = E^{2} -  \vert \vec{p}_{\perp} \vert ^{2}\:,
\hskip0.3cm
( - \infty  <  z <  +\infty  )\:.
\label{eq:schroedingertypeeq}
\end{eqnarray}
Since this is a second order differential equation, there are two independent solutions which are denoted 
 by $\phi_{i}(z; \lambda )\:\: (i=1,2)$  and are characterized  by the initial conditions at $z=0$, 
\begin{eqnarray}
\phi_{1}(0;\lambda)&=&1, \hskip1cm \phi_{1}^{\: \prime}(0;\lambda )=0 \:,
\label{eq:initialconditions1}
\\
\phi_{2}(0;\lambda)&=&0, \hskip1cm \phi_{2}^{\: \prime}(0;\lambda )=1\:. 
\label{eq:initialconditions}
\end{eqnarray}
Here the prime ($'$) means the derivative with respect to $z$. Since there appear only real numbers  on the right hand side of    (\ref{eq:initialconditions1}) and (\ref{eq:initialconditions}),   the solutions $\phi_{i}(z; \lambda ), \:  (i=1,2)$ are both real functions.

According to Refs. \cite{weyl} and  \cite{stone},  there exists a $2 \times 2$  matrix function $\rho_{ij}  (\lambda ) \:\:(i,j=1,2)$ that describes the spectra of the self-adjoint differential operator on the left hand side of  (\ref{eq:schroedingertypeeq}) and satisfies the completeness relation
\begin{eqnarray}
\int 
\sum_{i, j =1, 2}
\phi_{i}  (z; \lambda ) d \rho _{ij} (\lambda) _{} \phi_{j}  (z^{\prime}; \lambda)^{}
=
\delta (z -z^{\prime})\:.
\label{eq:completenessrelation0}
\end{eqnarray}
Here the integration $d\rho _{ij}(\lambda )$ in (\ref{eq:completenessrelation0}) is  to be understood as 
the Stieltjes integral, but it turns out  to be 
the ordinary Riemann integral if  the $\lambda$-spectra in (\ref{eq:schroedingertypeeq}) do
 not allow discrete ones.   Hereafter,  for the sake of simplicity, we assume  that the  $\lambda$-spectra   in 
 (\ref{eq:schroedingertypeeq})  are continuous.  It should be kept in our mind that the algorithm of deriving  the spectral function $\rho_{ij}(\lambda )$  has been well established. See Refs. \cite{titchmarsh} - \cite{yosida2}, \cite{kubota1}   for more details.

\subsection{Scalar field quantization in the presence of the electroweak bubble wall}
\label{sebsec:fieldquantization}

Now the solution to the wave equation (\ref{eq:scalarwaveequation})  may be expanded in terms of the 
complete set of the solutions as 
 \begin{eqnarray}
 \varphi (t, \vec{x}_{\perp}, z) 
 &=&
  \int d\lambda \sum_{i=1, 2} \int \frac{d^{2}\vec{p}_{\perp}}  {\sqrt{(2\pi)^{2} \: 2E }} 
 \nonumber \\
 & & 
 \times   \bigg \{
 \alpha_{i}(\vec{p}_{\perp}, \lambda ) e^{i \vec{p}_{\perp} \cdot \vec{x}_{\perp} -iEt} 
+   \alpha_{i}^{\dag }(\vec{p}_{\perp}, \lambda ) e^{ - i \vec{p}_{\perp} \cdot \vec{x}_{\perp} + iEt}
  \bigg \}   \phi_{i}(z; \lambda)   \:,
 \label{eq:scalaralphacretionannihilationX}
 \end{eqnarray}
 where $E$ is subject to the ``on-shell" condition and is positive
 \begin{eqnarray}
E=\sqrt{\vec{p}^{\: 2}_{\perp}  + \lambda } \: > 0 \:.
\end{eqnarray}
We should attach a superscript ``in" or ``out" to  the coefficients $\alpha_{i}$   and $\alpha_{i}^{\dag}$, but we  omit it   here and hereafter only for    simplicity.   In quantum theories the  coefficients $\alpha_{i}$   and $\alpha_{i}^{\dag}$ are regraded as operators satisfying certain commutation relations   in such a way that the equal-time canonical commutation relations
\begin{eqnarray}
& & 
\left [ \dot{ \varphi } (t, \vec{x}_{\perp}, z) ,  \varphi (t, \vec{x}_{\perp}^{\: \prime}, z^{\prime} ) \right ]
=
-i \: \delta ^{2}(\vec{x}_{\perp} - \vec{x}_{\perp}^{\:\prime} ) \delta (z - z^{\prime})\:,
\label{eq:canonicalcommutaionrelation1}
\\
& & 
\left [ { \varphi } (t, \vec{x}_{\perp}, z) ,  \varphi (t, \vec{x}_{\perp}^{\: \prime}, z^{\prime} ) \right ]
=0\:,
\label{eq:canonicalcommutaionrelation2}
\\
& & 
\left [ \dot{ \varphi } (t, \vec{x}_{\perp}, z) ,  \dot \varphi (t, \vec{x}_{\perp}^{\: \prime}, z^{\prime} ) \right ]
=0\:,
\label{eq:canonicalcommutaionrelation3}
 \end{eqnarray}
 are realized.    Here 
$ \dot{ \varphi } (t, \vec{x}_{\perp}, z)$ is supposed to be canonically conjugate to  
 $ \varphi (t, \vec{x}_{\perp}, z)$ 
 and    the dot ($\cdot$) means the derivative with respect to $t$.   By looking at the completeness relation  (\ref{eq:completenessrelation0}), we immediately notice that the  relations  (\ref{eq:canonicalcommutaionrelation1}),   (\ref{eq:canonicalcommutaionrelation2})  and  (\ref{eq:canonicalcommutaionrelation3})  can be derived just by postulating  the following   commutation relations
 \begin{eqnarray}
 & & 
\left [  \alpha_{i}(\vec{p}_{\perp}, \lambda ) ,\: 
  \alpha_{j}^{\dag }(\vec{p}_{\perp}^{\: \prime}, \lambda^{\prime} )  \right ]
  =\delta^{2}(\vec{p}_{\perp} - \vec{p}_{\perp}^{\:\prime})\delta(\lambda - \lambda ^{\prime})
  \frac{d\rho_{ij} ( \lambda )}{d\lambda}\:, 
  \label{eq:scalarcommutationrelation1}
 \\
 & &  \left [  \alpha_{i}(\vec{p}_{\perp}, \lambda ) ,\: 
  \alpha_{j}^{ }(\vec{p}_{\perp}^{\: \prime}, \lambda^{\prime} )  \right ]
  =
  \left [  \alpha_{i}^{\dag}(\vec{p}_{\perp}, \lambda ) ,\: 
  \alpha_{j}^{\dag }(\vec{p}_{\perp}^{\: \prime}, \lambda^{\prime} )  \right ]=0\:.
  \hskip1.0cm (i, j=1,2)
   \label{eq:scalarcommutationrelation2}
 \end{eqnarray}
 These relations  differ from those of  creation and/or  annihilation of particles by the presence of $d\rho_{ij}(\lambda)/d\lambda$ on the right hand side of  (\ref{eq:scalarcommutationrelation1}).     Note, however, that  $d\rho_{ij}(\lambda)/d\lambda$ is a symmetric positive definite matrix and therefore  we are able to interpret  $\alpha _{i}^{\dag}  (\vec{p}_{\perp}, \lambda )$ and $\alpha _{i} (\vec{p}_{\perp}, \lambda )$ as the creation and annihilation operators, respectively, of the  wave represented  by  (\ref{eq:modes})  and labelled    by $\vec{p}_{\perp}$,   $\lambda = E^{2} - \vert \vec{p}_{\perp} \vert ^{2}$ and $ i \:  (=1$ or $2)$.   More specifically speaking, the states $\alpha _{i}^{\dag}  (\vec{p}_{\perp}, \lambda )\vert 0 \rangle $ are all positive norm states and there is no fear of introducing negative norm states  via    (\ref{eq:scalarcommutationrelation1})  into our framework. 

By employing   (\ref{eq:scalarcommutationrelation1}) and (\ref{eq:scalarcommutationrelation2}), we are  able to derive the four-dimensional commutation relations  in the same form as in the ordinary  constant mass case, i.e.,
 \begin{eqnarray}
 \left [ \varphi(t, \vec{x}_{\perp}, z ),   \: \varphi(t^{\: \prime}, \vec{x}_{\perp}^{\:\prime}, z^{\:\prime}  )   \right ]
 =
 i \Delta \left ( t-t^{\:\prime}, \vec{x}_{\perp} - \vec{x}_{\perp}^{\:\prime}, z, z^{\:\prime} ; M_{Z}(*) \right ) \:.
 \label{eq:4dimcommutationrelation}
 \end{eqnarray}
Here we introduced a novel  function  
 \begin{eqnarray}
& & \hskip-0.8cm 
\Delta ( t - t^{\:\prime}, \vec{x}_{\perp} -\vec{x}_{\perp}^{\:\prime},  z, z^{\:\prime} ; M_{Z}(*))
\nonumber \\
&=& -i 
\int \frac{d^{2}\vec{p}_{\perp}}{(2\pi)^{2}} \: e^{i \vec{p}_{\perp} \cdot ( \vec{x}_{\perp} - 
\vec{x}^{\;\prime}_{\perp} ) }
\sum_{i,j=1,2}
\int \phi_{i}(z; \lambda )\frac{d\rho_{ij}(\lambda)}{2E}\phi_{j}(z^{\:\prime}; \lambda )
\left \{
e^{-iE(t-t^{\:\prime} )} -  e^{iE(t-t^{\:\prime} )}
\right \}\:,
\nonumber \\
\label{eq:deltafunction}
\end{eqnarray}
which is analogous to the familiar invariant $\Delta$-function.  (See \cite{bjorkendrell} or    \cite{itzyksonzuber} for example.)     As will be confirmed later,   when the mass-function $M_{Z}(z)$ is independent of $z$, then  (\ref{eq:deltafunction}) reduces   to the conventional invariant $\Delta$-function in the form of Fourier transformation.   It should also be noted that   the function (\ref{eq:deltafunction}) satisfies
\begin{eqnarray}
\left \{  \square + M_{Z}(z)^{2}  \right \} \Delta ( t - t^{\:\prime}, \vec{x}_{\perp} - \vec{x}_{\perp}^{\:\prime}, 
 z, z^{\:\prime} ; M_{Z}(*))   =  0\:,
 \label{eq:formula1}
\\
\left \{  \square^{\:\prime} + M_{Z}(z^{\:\prime})^{2}  \right \} \Delta ( t - t^{\:\prime}, \vec{x}_{\perp} - {\vec x}_{\perp}^{\:\prime}, z, z^{\:\prime} ; M_{Z}(*))  = 0\:.
\label{eq:formula2}
\end{eqnarray}
 The properties at $t=t^{\:\prime}$
\begin{eqnarray}
& &
\Delta ( 0, \vec{x}_{\perp} - {\vec x}_{\perp}^{\:\prime},  z, z^{\:\prime} ; M_{Z}(*))  =0\:,
\label{eq:formula3}
\\
& &
\frac{\partial }{\partial t}\Delta ( t-t^{\:\prime}, \vec{x}_{\perp} - {\vec x}_{\perp}^{\:\prime}, z, z^{\:\prime} ; M_{Z}(*))  
\bigg \vert_{t=t^{\:\prime}} = - \delta^{2}( \vec{x}_{\perp} - \vec{x}^{\:\prime}_{\perp})\delta (z - z^{\:\prime})\:,
\label{eq:formula5}
\\
& &
\frac{\partial }{\partial t^{\;\prime}}  \frac{\partial }{\partial t}\Delta ( t-t^{\:\prime}, \vec{x}_{\perp} - {\vec x}_{\perp}^{\:\prime}, z, z^{\:\prime} ; M_{Z}(*))    \bigg \vert_{t=t^{\:\prime}} =0   \:.
\label{eq:formula4}
\end{eqnarray}
should also  be mentioned, and  these properties enable us to reach 
(\ref{eq:canonicalcommutaionrelation1}),  (\ref{eq:canonicalcommutaionrelation2})   and 
(\ref{eq:canonicalcommutaionrelation3})    consistently  from  (\ref{eq:4dimcommutationrelation}) .

\subsection{The retarded and advanced  Green's functions}
\label{sec:retardedadvancedgreensfunctions}

We are now well equipped to  present the retarded and advanced Green's functions, which are given by 
\begin{eqnarray}
& &  \hskip-0.5cm
\Delta_{R} ( t-t^{\:\prime}, \vec{x}_{\perp} - \vec{x}_{\perp}^{\:\prime}, z, z^{\:\prime}; M_{Z}(*))
\nonumber \\
& & = -
\int \frac{dE \: d^{2}\vec{p}_{\perp}}{(2\pi)^{3}} e^{-iE(t - t^{\:\prime} ) }
 e^{i\vec{p}_{\perp} \cdot  (  \vec{x}_{\perp} - \vec{x}_{\perp}^{\:\prime} ) }
 \sum_{i,j=1,2}  \int  \frac{\phi_{i}(z; \lambda ) d\rho_{ij}(\lambda ) \phi_{j}(z^{\:\prime}; \lambda )}
 { (E+ i \varepsilon )^{2}  - \vec{p}_{\perp}^{\: 2}  - \lambda }\:,
\label{eq:retardedgreensfunction2}
 \end{eqnarray}
 \begin{eqnarray}
& &  \hskip-0.5cm
\Delta_{A} ( t-t^{\:\prime}, \vec{x}_{\perp} - \vec{x}_{\perp}^{\:\prime}, z, z^{\:\prime}; M_{Z}(*))
\nonumber \\
& & = -
\int \frac{dE \: d^{2}\vec{p}_{\perp}}{(2\pi)^{3}} e^{-iE(t - t^{\:\prime} ) }
 e^{i\vec{p}_{\perp} \cdot  (  \vec{x}_{\perp} - \vec{x}_{\perp}^{\:\prime} ) }
 \sum_{i,j=1,2}   \int   \frac{\phi_{i}(z; \lambda ) d\rho_{ij}(\lambda ) \phi_{j}(z^{\:\prime}; \lambda )}
 { (E -  i \varepsilon )^{2}  - \vec{p}_{\perp}^{\: 2}  - \lambda }\:.
 \label{eq:advancedgreensfunction2}
 \end{eqnarray}
 It is almost obvious that  (\ref{eq:retardedgreensfunction2}) and (\ref{eq:advancedgreensfunction2}) satisfy     (\ref{eq:retardedgreensfunction}) and (\ref{eq:advancedgreensfunction})    respectively  thanks to  the formula  (\ref{eq:completenessrelation0}).  Using the Cauchy theorem on the complex $E-$plane   by adding the integration path of an infinite semi-circle, we can confirm that $\Delta_{R}$ ($\Delta_{A}$)  is non-vanishing   only  for $t > t^{\:\prime}$   $(t < t^{\:\prime})$.  With these explicit formulas of the Green's functions,  the definition of the asymptotic fields given in Section   \ref{subsec:asymptoticfield} is well established.

\subsection{The special case of $z$-independent mass (no bubble wall) }

For   illustration, let us discuss the simplest familiar case  by employing  the above rather unfamiliar  method. Namely we consider the case of $z$-independent vacuum expectation value,  $v(z) = v_{0} \approx 246 \:{\rm GeV}$  and suppose that the $Z$-boson mass is just a number $M_{Z}(z) = m_{Z} \approx 91.2 \: {\rm GeV}$.   The solutions satisfying  (\ref{eq:initialconditions1}) and  (\ref{eq:initialconditions}) are 
\begin{eqnarray}
\phi_{1}( z; \lambda ) = {\rm cos} \left (z \sqrt{\lambda - m_{Z}^{\:2}} \right ), 
\hskip0.5cm
\phi_{2}( z; \lambda ) = \frac{1}{\sqrt{\lambda - m_{Z}^{\: 2}}} {\rm sin} \left (z \sqrt{\lambda - m_{Z}^{\:2}} \right )
\end{eqnarray}
for $\lambda > m_{Z}^{\:2}$ and 
\begin{eqnarray}
\phi_{1}( z; \lambda ) = {\rm cosh} \left (z \sqrt{ m_{Z}^{\:2} - \lambda }  \right ), 
\hskip0.5cm
\phi_{2}( z; \lambda ) = \frac{1}{\sqrt{ m_{Z}^{\: 2} - \lambda }} {\rm sinh} \left (z \sqrt{ m_{Z}^{\:2} - \lambda }  \right )
\end{eqnarray}
for $ m_{Z}^{\:2} > \lambda $. 
Commanding the method developed in Refs. \cite{titchmarsh} and \cite{kodaira1}, we are able to get 
\begin{eqnarray}
\frac{d\rho_{ij} (\lambda ) }{d\lambda}= \frac{1}{2\pi}\left (
\begin{tabular}{cc}
$\displaystyle{\frac{1}{\sqrt{\lambda - m_{Z}^{\:2}}}}$ & $0$ 
\\
$0$ &$\displaystyle{  \sqrt{\lambda - m_{Z}^{\;2}}}$
\end{tabular}
\right )
\end{eqnarray}
for $\lambda > m_{Z}^{\:2}$ and  $d\rho_{ij}(\lambda )/d\lambda =0$  for $ m_{Z}^{\:2} > \lambda $.
Thus the integration region over $\lambda $ in  the completeness relation   (\ref{eq:completenessrelation0}) is automatically restricted to $\lambda > m_{Z}^{\:2}$ and we are able to get a familiar  formula
\begin{eqnarray}
\int 
\sum_{i, j =1, 2}
\phi_{i}  (z; \lambda ) d \rho _{ij} (\lambda) _{} \phi_{j}  (z^{\prime}; \lambda)^{}
&=&
 \frac{1}{2\pi} \int _{m_{Z}^{\:2}}^{+\infty} \frac{d\lambda}{\sqrt{\lambda - m_{Z}^{\:2}}} 
{\rm cos} \left ( (z-z^{\;\prime} ) \sqrt{\lambda - m_{Z}^{\:2} }\right )
\nonumber \\
&=&
 \frac{1}{\pi} \int _{0}^{+\infty} dq  \: 
{\rm cos} \left ( (z-z^{\;\prime} ) \: q \right )
\nonumber \\
&=& \delta (z - z^{\:\prime})\:.
   \label{eq:completenessrelationX}
\end{eqnarray}
Note that $q \equiv \sqrt{\lambda - m_{Z}^{\:2}} \geq  0$ in (\ref{eq:completenessrelationX}) plays the role of  the $z$-component of the wave vector.   This example shows clearly that our expansion by using $\phi_{i} ( z; \lambda )\; (i=1,2)$  for the general $z$-dependent $M_{Z}(z)$ case  is a natural generalization of the conventional Fourier expansion method. 

The mode expansion  (\ref{eq:scalaralphacretionannihilationX}) of the scalar field $\varphi (t, \vec{x}_{\perp}, z)$ can also be cast into the familiar form
\begin{eqnarray}
\varphi (t, \vec{x}_{\perp}, z) 
&=&
 \int _{0}^{+ \infty} dq \: \int \frac{d^{2}\vec{p}_{\perp} }{ \sqrt{ (2\pi)^{3} 2E} }
  \bigg \{ 
  \nonumber \\
  & & \hskip0.5cm
  a_{-} (\vec{p}_{\perp}, q) e^{i \vec{p}_{\perp} \cdot \vec{x}_{\perp} + i qz - iEt}
 +
 a_{+} (\vec{p}_{\perp}, q) e^{i \vec{p}_{\perp} \cdot \vec{x}_{\perp} - i qz -iEt}
 \nonumber \\
 & & \hskip0.5cm  
 + a_{-}^{\dag}  (\vec{p}_{\perp}, q) e^{ - i \vec{p}_{\perp} \cdot \vec{x}_{\perp} - i qz + iEt}
 +
 a_{+}^{\dag}  (\vec{p}_{\perp}, q) e^{- i \vec{p}_{\perp} \cdot \vec{x}_{\perp} + i qz + iEt}  \bigg \}\:,
 \label{eq:familiarformmodeexpansion}
\end{eqnarray}
if we define linear combinations of the operators  
\begin{eqnarray}
a_{\pm} (\vec{p}_{\perp}, q) = \sqrt{2\pi} 
\left \{  \sqrt{\lambda - m_{Z}^{\:2} } \:   \alpha_{1}(\vec{p}, \lambda ) \pm i\alpha_{2}(\vec{p}_{\perp}, \lambda)   \right \}, 
\hskip0.5cm
\left ( q=\sqrt{\lambda - m_{Z}^{2}} \right )\:.
\label{eq:newoperators}
\end{eqnarray}
Note that $E$ in (\ref{eq:familiarformmodeexpansion}) is given by 
$  E=\sqrt{ \vert \vec{p}_{\perp}\vert ^{2} + \lambda } = \sqrt{ \vert \vec{p}_{\perp} \vert ^{2} + q^{2} + m_{Z}^{\:2}} $\: .
The non-vanishing commutation relations of the operators (\ref{eq:newoperators}) are confirmed to be 
\begin{eqnarray}
\left [ a_{+} (\vec{p}_{\perp}, q ), a_{+}^{\dag} (\vec{p}_{\perp}^{\:\prime},  q ^{\:\prime } ) \right ]
=
\delta^{2}(\vec{p}_{\perp} - \vec{p}_{\perp}^{\:\prime} ) \delta (q-q^{\:\prime} )\:,
\label{eq:conventional1}
\\
\left [ a_{-} (\vec{p}_{\perp}, q ), a_{-}^{\dag} (\vec{p}_{\perp}^{\:\prime},  q ^{\:\prime } ) \right ]
=
\delta^{2}(\vec{p}_{\perp} - \vec{p}_{\perp}^{\:\prime} ) \delta (q-q^{\:\prime} )\:.
\label{eq:conventional2}
\end{eqnarray}
These are the ordinary relations between creation and annihilation operators,  and we are led to interpret $a_{+} (\vec{p}_{\perp}, q) $ and $a_{-} (\vec{p}_{\perp}, q) $ as annihilation operators of a particle with  momentum $( E, \vec{p}_{\perp}, -q) $ and $( E, \vec{p}_{\perp}, +q) $,  respectively.   Since we use the real functions $\phi_{i}(z; \lambda ), (i=1,2)$ instead of complex ones as the basis of the mode expansion, we have necessarily to introduce two kinds of creation and annihilation operators as in   (\ref{eq:conventional1}) and (\ref{eq:conventional2}). This situation continues to be the case in the vector field mode expansion that will be  discussed later.

Incidentally the function (\ref{eq:deltafunction}) can be   shown,  for $M_{Z}(z)=m_{Z}$,  to reduce to the familiar invariant $\Delta$-function in the following manner:
 \begin{eqnarray}
& & \hskip-0.8cm 
\Delta ( t - t^{\:\prime}, \vec{x}_{\perp} -\vec{x}_{\perp}^{\:\prime},  z, z^{\:\prime} ; m_{Z})
\nonumber \\
&=& -i 
\int \frac{d^{2}\vec{p}_{\perp}}{(2\pi)^{2}} \: e^{i \vec{p}_{\perp} \cdot ( \vec{x}_{\perp} - 
\vec{x}^{\;\prime}_{\perp} ) }
\frac{1}{\pi}\int _{0}^{+\infty} dq \: {\rm cos} \left ( (z-z^{\:\prime} ) q \right ) \frac{1}{2E}
\left \{
e^{-iE(t-t^{\:\prime} )} -  e^{iE(t-t^{\:\prime} )}
\right \}
\nonumber \\
&=& -i 
\int _{-\infty} ^{+\infty} dq\: 
\int \frac{d^{2}\vec{p}_{\perp}}{(2\pi)^{3} 2E } \: 
e^{i \vec{p}_{\perp} \cdot ( \vec{x}_{\perp} - 
\vec{x}^{\;\prime}_{\perp} ) + iq(z-z^{\:\prime}) }
\left \{
e^{-iE(t-t^{\:\prime} )} -  e^{iE(t-t^{\:\prime} )}
\right \}\:.
\end{eqnarray}
In the last step above,  the integration region  $0 < q < + \infty$ is extended to $-\infty < q < +\infty$, owing  to the symmetry property of the integrand.

\section{The  BRST analyses in the electroweak theory}
\label{sec:electroweakbrst}

\subsection{The gauge fixing and auxiliary fields}
\label{sec:gaugefixingauxiliaryfields}

In order to quantize the gauge fields,  we have to fix the gauge, thereby introducing the Faddeev-Popov ghost fields.  We now summarize  the gauge-fixing procedure below, and it should be warned in advance that there is nothing new in the content of Section \ref{sec:electroweakbrst}.  This section is only for preparing various quantities to be used later. 
{\color{black}{
Lest the equations below should become too much  cluttered, we define the following notations   of gauge-fixing   functions, 
\begin{eqnarray}
F^{a}&\equiv & \partial^{\mu} A_{\mu}^{a} +\frac{1}{2} gv \chi^{a}, \hskip0.5cm ( a=1,2,3 ) \:,
\label{eq:fa}
\\
F^{0}&\equiv & \partial ^{\mu}B_{\mu} -\frac{1}{2} g^{\:\prime} v \chi^{3}\:.
\label{eq:fzero}
\end{eqnarray}
The gauge fields associated with  $SU(2)_{L}$ and $U(1)_{Y}$ gauge groups are denoted by  $A_{\mu}^{a}$ $(a=1,2,3)$ and $B_{\mu}$, respectively. The gauge coupling constants are  $g$ and $g^{\:\prime} $ as  introduced before.     

With the use  of  (\ref{eq:fa}) and (\ref{eq:fzero}),   the $R_{\:\xi}$ gauge with $\xi =1$  
  is  chosen  by adding the  following gauge fixing terms to the original symmetric Lagrangian 
 }} 
\begin{eqnarray}
{\cal L}_{\rm GF}
&=&
-\sum_{a=1}^{3} {\cal B}^{a}F^{a} - {\cal B}^{0} F^{0} 
+ \frac{1}{2} \sum _{a=1}^{3}  \left ( {\cal B}^{a}  \right )^{2}
+\frac{1}{2} \left ( {\cal B}^{0} \right )^{2} 
{\color{black}{
+ ({\rm surface \:\: terms})\:.}}
 \label{eq:gaugefixing0}
\end{eqnarray}
Note that  ${\cal B}^{a}$   $(a=1,2,3)$ and ${\cal B}^{0}$  are auxiliary fields of the same type as those   made use of extensively in quantum electrodynamics by Nakanishi \cite{nakanishinoboru3} and   by Lautrup \cite{lautrup}.    
  {\color{black}{The surface terms  in (\ref{eq:gaugefixing0}) are given by 
 \begin{eqnarray}
  ({\rm surface \:\: terms})
  \equiv 
 \partial ^{\mu} \left ( \sum_{a = 1}^{3}  {\cal B}^{a} A_{\mu}^{a} + {\cal B}^{0} B_{\mu} \right )\:,
 \label{eq:sufaceterms}
 \end{eqnarray}
 and of course these  surface terms do not alter the equations of motion.    The derivatives acting on gauge fields contained in $F^{a}$ and $F^{0}$  in (\ref{eq:gaugefixing0}) are  turned into those acting on the    auxiliary ${\cal B}$-fields, i.e., 
\begin{eqnarray}
- \sum_{a=1}^{3} {\cal B}^{a} ( \partial ^{\mu}A_{\mu}^{a}) -   {\cal B}^{0} ( \partial^{\mu} B_{\mu}) 
+
  ({\rm surface \:\: terms})
  =
   \sum_{a=1}^{3} ( \partial ^{\mu} {\cal B}^{a} )  A_{\mu}^{a} +  ( \partial^{\mu} {\cal B}^{0} )  B_{\mu} \:.
   \label{eq:derivativeonbfieldonly}
\end{eqnarray}
Note  that     the momentum variables canonically conjugate to ${\cal B}^{a}$ and ${\cal B}^{0}$ turn out    to be $A_{0}^{a}$ and $B_{0}$, respectively.    The reason for adding these surface terms is that this makes  the Lagrangian density    invariant under the BRST transformation as argued in \cite{kugoojima2}.    See Eq. (\ref{eq:lagrangiandensityinvariant})  to be given later for more details.
 }}

As an  additional   remark,  it should be mentioned   that the gauge fixing Lagrangian   (\ref{eq:gaugefixing0}) can be  rewritten  identically  as 
\begin{eqnarray}
{\cal L}_{\rm GF}
& \equiv &
{\color{black}{-}} \left ( F^{+} {\cal B}^{-} + F^{-}{\cal B}^{+}\right ) 
{\color{black}{-}}   F^{Z}{\cal B}^{Z}  {\color{black}{-}}   F^{A}{\cal B}^{A} 
+ {\cal B}^{+}{\cal B}^{-}  + \frac{1}{2}  ({\cal B}^{Z})^{2}   + \frac{1}{2}  ({\cal B}^{A})^{2}
\nonumber \\
& & 
{\color{black}{
+ ({\rm surface \:\: terms})\:,
}}
\label{eq:gaugefixing1}
\end{eqnarray}
where we have introduced the following linear combinations of 
(\ref{eq:fa}) and   (\ref{eq:fzero}), 
\begin{eqnarray}
F^{\pm}
&\equiv&
\frac{F^{1} \mp i F^{2}}{\sqrt{2}}
=
\partial^{\mu}   W_{\mu}^{\pm} + M_{W}  \: \chi^{\pm}\:,
\label{eq:fplusminus}
\\
F^{Z}&\equiv&
\frac{gF^{3}- g^{\:\prime}F^{0}}{\sqrt{g^{2} + g^{\:\prime \: 2}}}
=\partial^{\mu}  Z_{\mu} +  M_{Z}  \:  \chi^{{\color{black}{3}}}\:,
\label{eq:fz}
\\
F^{A}&\equiv&
\frac{g^{\:\prime} F^{3} + g F^{0}}{\sqrt{g^{2} + g^{\:\prime \:2}}}
= \partial^{\mu}  A_{\mu} \:.
\label{eq:fcapitala}
\end{eqnarray}
together with the redefined auxiliary fields, 
{\color{black}{
\begin{eqnarray}
{\cal B}^{\pm} \equiv \frac{{\cal B}^{1} \mp i {\cal B}^{2}}{\sqrt{2}}\:, 
\hskip0.5cm  
{\cal B}^{Z} \equiv  \:\:\: \frac{g {\cal B}^{3} - g^{\:\prime} {\cal B}^{0}}{\sqrt{g^{2}+g^{\:\prime \: 2}}}\:, 
\hskip0.5cm
{\cal B}^{A}\equiv  \frac{g^{\:\prime} {\cal B}^{3} + g {\cal B}^{0}}{\sqrt{g^{2}+g^{\:\prime \: 2}}}\:.
  \label{eq:bpmbzba}
\end{eqnarray}
}}
We also defined $\chi^{\pm}$ in (\ref{eq:fplusminus})   by
\begin{eqnarray}
\chi^{\pm}=\frac{\chi^{1}\mp i \chi^{2}}{\sqrt{2}}\:.
\label{eq:chiplusminusdef}
\end{eqnarray} 
The point of rewriting ${\cal L}_{\rm GF}$ as in (\ref{eq:gaugefixing1}) is that the gauge fields in 
(\ref{eq:fplusminus}), (\ref{eq:fz}) and (\ref{eq:fcapitala}) are arranged so that they turn out to be  
$W^{\pm}$-,    $Z$-    and the electromagnetic ($A_{\mu}$)-fields
which are related to the original gauge fields $A_{\mu}^{a}$ $(a=1,2,3)$   and $B_{\mu}$ via
\begin{eqnarray}
W_{\mu}^{\pm}\equiv  \frac{A_{\mu}^{1} \mp i A_{\mu}^{2}}{\sqrt{2}}\:, \:\:\:
Z_{\mu} \equiv \frac{gA_{\mu}^{3}  - g^{\:\prime}B_{\mu}}{\sqrt{g^{2}+g^{\:\prime \:2}}}\:, \:\:\:
A_{\mu} \equiv \frac{g^{\:\prime} A_{\mu}^{3}  + g B_{\mu}}{\sqrt{g^{2}+g^{\:\prime \:2}}}\:.
\label{eq:wpmza}
\end{eqnarray}
The path integration over the auxiliary fields ${\cal B}^{\pm}$, ${\cal B}^{Z}$ and ${\cal B}^{A}$  gives us  the  familiar  gauge fixing Lagrangian that we often use in  diagrammatic computations.  We can  go back and forth freely between   (\ref{eq:gaugefixing0})   and (\ref{eq:gaugefixing1}) simply by redefining the auxiliary fields as in      (\ref{eq:bpmbzba}).    For practical purposes  we often use  (\ref{eq:gaugefixing1}) instead of (\ref{eq:gaugefixing0}), while from the view point of BRST analyses we find (\ref{eq:gaugefixing0}) more convenient to deal with.

{\color{black}{
The surface terms in (\ref{eq:gaugefixing1}) are given by  
\begin{eqnarray}
({\rm surface \:\: terms}) = \partial^{\mu} \left (
W_{\mu}^{+} {\cal B}^{-} + W_{\mu}^{-} {\cal B}^{+} + Z_{\mu}{\cal B}^{Z} + A_{\mu}{\cal B}^{A}
\right )\:,
\end{eqnarray}
as confirmed easily by rewriting (\ref{eq:sufaceterms}) with the use  of 
(\ref{eq:bpmbzba}) and (\ref{eq:wpmza}). 
The momentum variables which are canonically conjugate to ${\cal B}^{\pm}$, ${\cal B}^{Z}$ and 
${\cal B}^{A}$ are thus $W_{0}^{\mp}$, $Z_{0}$ and $A_{0}$, respectively.
}}

\subsection{The Faddeev-Popov ghost Lagrangian}

We just follow  the standard procedure to write down the Faddeev-Popov Lagrangian  
${\cal L}_{\rm FP}$ associated with the gauge fixing (\ref{eq:gaugefixing0}), that is, 
{\color{black}{
\begin{eqnarray}
{\cal L}_{\rm FP}
&=&
\sum_{a=1}^{3} ( \partial^{\mu} {\overline c}^{a})  \: ({\cal D}_{\mu} c)^{a}
+ ( \partial^{\mu}{\overline c}^{0}) ( \partial_{\mu}c^{0} )
+  \frac{1}{4} g^{2}v \sum_{a,b,c=1}^{3} \varepsilon ^{abc} \: {\overline c}^{a} \chi^{b} c^{c}
\nonumber \\
& & -\frac{1}{4} g^{2}v \left ( {\overline c}^{1} c^{1} + {\overline c}^{2}c^{2} \right )  \left ( v+H \right )
- \frac{1}{4} v \left ( g {\overline c}^{3} - g^{\:\prime} {\overline c}^{0} \right )
\left (  gc^{3} - g^{\:\prime} c^{0}  \right ) (v+H)
\nonumber \\
& & + \frac{1}{4} gg^{\:\prime} v \left \{
\left(   {\overline c}^{1}c^{0} +{\overline c}^{0} c^{1}   \right ) \chi^{2}
-
\left(   {\overline c}^{2}c^{0} +{\overline c}^{0} c^{2}   \right ) \chi^{1}
\right \}\:.
    \label{eq:symmetryrespecedFPghostLagrangian}
\end{eqnarray}
}}
Here the gauge covariant derivative ${\cal D}_{\mu}$ is given as usual by
\begin{eqnarray}
({\cal D}_{\mu} c)^{a} \equiv  \partial_{\mu} c^{a} + g \varepsilon^{abc} c^{b} A_{\mu}^{c} \:.
\end{eqnarray}
On the ground of argument in  Refs.\cite{kugoojima1} -   \cite{kugoojima5}  in regard to the (anti-)hermiticity properties of ghost and anti-ghost fields, we take
\begin{eqnarray}
& &  c^{a\: \dagger}=c^{a}, \hskip0.5cm  \overline{c}^{a\: \dagger}=- \overline{c}^{a},  \hskip0.5cm  (a=1,2,3) 
\nonumber \\
& &  c^{0\: \dagger} = c^{0}, \hskip0.5cm \overline{c}^{0\: \dagger} = - \overline{c}^{0}\:, 
\label{eq:hermiticity}
\end{eqnarray} 
for granted.  (In some of literatures, $i \: \overline{c}^{a}$ and $i \: \overline{c}^{0}$ are sometimes called anti-ghost fields,  and with such naming    the anti-ghost fields would become   hermitian.)  Thanks to the  properties (\ref{eq:hermiticity}),  the  BRST charge $Q_{B}$ to be introduced  later turns  out to be a  hermitian operator.

If we use the expression (\ref{eq:gaugefixing1}) as the  gauge fixing Lagrangian,  we have to rewrite the gauge fields $A_{\mu}^{\:a}$  in (\ref{eq:symmetryrespecedFPghostLagrangian}) in terms of  (\ref{eq:wpmza}). We also had better make use of the redefined ghost and anti-ghost fields
\begin{eqnarray}
& & 
c^{(\pm)}=\frac{c^{1} \mp i c^{2}}{\sqrt{2}}\:, \hskip1cm
c^{A} = \frac{g^{\:\prime} c^{3} + g c^{0}}{\sqrt{g^{2}+ g^{\:\prime \:2}}}\:, \hskip1cm
c^{Z} = \frac{g c^{3} - g^{\:\prime} c^{0}}{\sqrt{g^{2} + g^{\prime \: 2}}}\:,
\label{eq:anotherghost}
\\
& & 
{\overline c}^{(\pm)}=\frac{{\overline c}^{1} \mp i {\overline c}^{2}}{\sqrt{2}}\:, 
 \hskip1cm
{\overline c}^{A} = \frac{g^{\:\prime} {\overline c}^{3} + g {\overline c}^{0}}{\sqrt{g^{2}+g^{\:\prime \:2}}}\:,
\hskip1cm
{\overline c}^{Z} = \frac{g {\overline c}^{3} - g^{\:\prime}  {\overline c}^{0}}{\sqrt{g^{2}+g^{\:\prime \: 2}}}\:.
\label{eq:anotherantighost}
\end{eqnarray}
We note here an identity 
\begin{eqnarray}
{\overline{c}}^{(+)}F^{-} + {\overline{c}}^{(-)}F^{+} 
+ {\overline{c}}^{Z}F^{Z} + {\overline{c}}^{A}F^{A} 
=
\sum_{a=1}^{3}
{\overline{c}}^{a}F^{a} + {\overline{c}}^{0}F^{0} \:,
\label{eq:antighostrelations}
\end{eqnarray}
which determines the relations (\ref{eq:anotherantighost})  among the anti-ghost fields. 
It is almost straightforward to rewrite  (\ref{eq:symmetryrespecedFPghostLagrangian})  in terms of 
(\ref{eq:anotherghost}) and (\ref{eq:anotherantighost}) and we relegate the rewritten  form of   (\ref{eq:symmetryrespecedFPghostLagrangian})   in Appendix  \ref{sect:FPghostlagrangian}.  See also Refs.  \cite{aoki1} -  \cite{aoki3}.

\subsection{The BRST transformation}
\label{sec:43brsttransformation}

The BRST transformation (to be denoted by $\delta _{B}$) of the gauge and Higgs doublet fields is obtained simply by replacing the 
$SU(2)_{L}$ and   $U(1)_{Y}$ gauge transformation parameters by the ghost fields 
$c^{a}, \:\: (a=1,2,3)$ and $c^{0}$, respectively,  i.e., 
\begin{eqnarray}
& & 
\delta_{B} A_{\mu}^{a} = \delta \lambda \: ({\cal D}_{\mu} c )^{a}, 
\hskip1.0cm
\delta_{B} B_{\mu} = \delta \lambda \: \partial _{\mu} c^{0}, 
\\
& & 
\delta_{B}\Phi =\delta \lambda \left (
-\frac{i}{2} g  \sum_{a=1}^{3}\tau^{a} c^{a} -\frac{i}{2} g^{\:\prime} c^{0}
\right ) \Phi \:,
\end{eqnarray}
where $\delta \lambda$ is a Grassmann parameter. The transformation rules for the ghost fields 
are fixed by requiring the nilpotency, i.e.,  $\delta_{B}^{2} A_{\mu}^{a}=0$, $\delta_{B}^{2}B_{\mu}=0$ 
and $\delta_{B}^{2}\Phi =0$   as follows, 
\begin{eqnarray}
& & 
\delta_{B}c^{a} = \frac{1}{2} g \: \delta \lambda \: \varepsilon^{abc} c^{b}c^{c} ,  \:\:\:(a=1,2,3), 
\hskip0.8cm
\delta_{B}c^{0}=0\:.
\end{eqnarray}
The anti-ghost fields, on the other hand,  are transformed into the auxiliary fields
\begin{eqnarray}
& & 
\delta_{B}  \overline{c}^{a} = - \delta \lambda  \: {\cal B}^{a}, \:\:\:(a=1,2,3)\:,
\hskip2.0cm
\delta_{B}\overline{c}^{0} = - \delta \lambda \: {\cal B}^{0}\:,
\end{eqnarray}
and the  properties of nilpotency, 
$\delta_{B}^{2}\overline{c}^{a}=0$ and $\delta_{B}^{2}\overline{c}^{0}=0$, are guaranteed by setting 
the following rules, 
\begin{eqnarray}
& & 
\delta_{B} {\cal B}^{a}=0, \:\:\:(a=1,2,3)\:,
\hskip3.cm
\delta_{B} {\cal B}^{0}=0.
\end{eqnarray}
By construction,   the sum of gauge fixing Lagrangian (\ref{eq:gaugefixing0}) and  the Faddeev-Popov Lagrangian  (\ref{eq:symmetryrespecedFPghostLagrangian}) is expressed as 
{\color{black}{
    \begin{eqnarray}
    & &  \delta \lambda \left ( {\cal L}_{\rm GF} + {\cal L}_{\rm FP} \right ) 
    =\delta_{B} \Xi\:,
\label{eq:lagrangiandensityinvariant}
    \\
       & & \Xi   \equiv  \sum_{a=1}^{3} \overline{c}^{a}  \bigg ( F^{a} -\frac{1}{2} {\cal B}^{a} \bigg ) 
     + \overline{c}^{0} \bigg ( F^{0}  - \frac{1}{2} {\cal B}^{0} \bigg ) 
     - \partial^{\mu} \bigg ( \sum_{a=1}^{3} \overline{c}^{a} A^{a}_{\mu} + \overline{c}^{0} B_{\mu} \bigg ) \:,
     \nonumber 
    \end{eqnarray}
    }}
and the invariance of the total Lagrangian density (including the surface terms) under the BRST transformation is obvious   by virtue of the nilpotency .

\subsection{The conserved current and the conserved charge}

Now that we have the BRST invariant Lagrangian, it is straightforward to derive the conserved Noether current following the standard method, namely, 
\begin{eqnarray}
J_{B}^{\mu} 
&=& \sum_{a=1}^{3} \left \{ 
(\partial^{\mu} {\cal B}^{a} ) c^{a} - ({\cal D}^{\mu} c)^{a} {\cal B}^{a} \right \} 
+ (\partial^{\mu} {\cal B}^{0} ) c^{0}
-(\partial ^{\mu} c^{0}) {\cal B}^{0}
+\frac{1}{2} g \: \sum_{a,b,c=1}^{3} \varepsilon^{abc} ( \partial ^{\mu} \overline{c}^{a}) c^{b} c^{c}
\nonumber \\
& &
- \sum_{a=1}^{3} \partial_{\nu}\left (
F^{a\:\mu \nu} c^{a} \right )
-   \partial _{\nu}   \left \{ \left ( \partial^{\mu} B^{\nu} - \partial^{\nu}B^{\mu}  \right ) c^{0} \right \}   \:.
\label{eq:brstcurrent}
\end{eqnarray}
Here use has been made of the equations of motion of gauge fields to eliminate some part of  contributions of the scalar fields.   The conserved charge is the space integral of $J_{B}^{0}$ and is  given by
\begin{eqnarray}
Q_{B}= \int d^{2}\vec{x}_{\perp} \int dz \left [
\sum_{a=1}^{3} \left \{
\dot{{\cal B}}^{a} c^{a}- {\cal B}^{a}({\cal D}^{0} c)^{a} \right \} 
+\dot{{\cal B}}^{0} c^{0} 
 - {\cal B}^{0}\dot{c}^{0}  + \frac{1}{2}g \: \sum_{a,b,c=1}^{3}\varepsilon^{abc} \dot{\overline{c}}^{a}c^{b}c^{c}
\right ]\:,
\nonumber \\
\label{eq:brstchargebcterms}
\end{eqnarray}
where we discarded   the last surface terms in (\ref{eq:brstcurrent}).  As the property of the conserved charge we have the commutation relations
\begin{eqnarray}
i\left [ \delta \lambda Q_{B}, {\cal O} (t, \vec{x}_{\perp}, z)   \right ]= \delta_{B} {\cal O} (t, \vec{x}_{\perp}, z)\:,
\label{eq:brstcummmutationrelations}
\end{eqnarray}
where ${\cal O}  (t, \vec{x}_{\perp}, z) $ denotes  generic Heisenberg fields  in the electroweak  theory such as gauge, scalar and auxiliary fields, together with ghost and anti-ghost fields.  In terms of the physical  field variables, i.e., (\ref{eq:wpmza}),   (\ref{eq:anotherghost}) and (\ref{eq:anotherantighost}),  $\delta_{B} {\cal O} (t, \vec{x}_{\perp}, z)$'s are all given in Appendix  \ref{sec:brsttransformation}.

We also have another  conservation law of the  ghost number, assigning $+1$ for the ghost fields  and   $-1$ for the anti-ghost fields. The conserved Noether current turns out to be 
\begin{eqnarray}
J_{c}^{\mu}= \sum_{a=1}^{3} \left \{
\overline{c}^{a} ( {\cal D}^{\mu} c )^{a} -\partial^{\mu} \overline{c}^{a} \:c^{a} \right \} 
+
\overline{c}^{0} ( \partial ^{\mu} c^{0})  - (\partial^{\mu} \overline{c}^{0})  c^{0}
\end{eqnarray}
and the conserved charge is 
\begin{eqnarray}
Q_{c}=\int d\vec{x}_{\perp} dz \: J_{c}^{0}\:.
\end{eqnarray}
The Kugo-Ojima's subsidiary condition of the physical states
($\vert {\rm phys} \rangle $)  is expressed,  in terms of $Q_{B}$ and $Q_{c}$,   by 
\begin{eqnarray}
Q_{B} \vert {\rm phys} \rangle  =0, \hskip1cm 
Q_{c} \vert {\rm phys} \rangle  =0.
\label{eq:subsidiarycondition}
\end{eqnarray}
Note that the fields appearing  in $Q_{B}$ and $Q_{c}$ are all interacting Heisenberg fields and the commutation relations of $Q_{B}$    with asymptotic fields have to be discussed  later.

{\color{black}{
In the previous subsection,   we have seen that ${\cal L}_{\rm GF}+ {\cal L}_{\rm FP}$ can be expressed as in  (\ref{eq:lagrangiandensityinvariant}), which is equivalently rewriten as 
\begin{eqnarray}
\delta \lambda  \left  ({\cal L}_{\rm GF} + {\cal L}_{\rm FP}  \right )= i \left [ \delta \lambda Q_{ B}, \Xi  \right ]\:.
\label{eq:addedequation1}
\end{eqnarray} 
The subsidiary condition (\ref{eq:subsidiarycondition}) indicates 
\begin{eqnarray}
\langle {\rm phys} \vert  \left ({\cal L}_{\rm GF} + {\cal L}_{\rm FP}  \right ) \vert {\rm phys}^{\prime} \rangle  =0\:,
\label{eq:addedequation2}
\end{eqnarray}
which implies that the effects of adding $\left ({\cal L}_{\rm GF} + {\cal L}_{\rm FP}  \right ) $  to the original symmetric Lagrangian by hand become null in the physical subspace and   physical quantities are independent of the choice of the gauge.  In the present paper, we set $\xi =1$ in the $R_{\xi}$ gauge, but the relations such as  (\ref{eq:addedequation1}) and (\ref{eq:addedequation2}) are both valid independently of the way of gauge fixing.  In fact the proof of the gauge independence of the S-matrix usually starts with (\ref{eq:addedequation1}), but we have to be careful about the  gauge dependence of renormalized quantities. In the present paper we do not enter the details of the renormalization procedure in the non-trivial $v(z)$ case,  and we will discuss the gauge independence of physical quantities after scrutinizing the renormalization in our separate publications. 
}} 

\section{Field quantization in the presence of the electroweak bubble wall  }
\label{sec:gaugefieldquantization}

\subsection{Asymptotic gauge fields}
\label{subsec:asymptoticfields}

Let us now write down  the equations of motion of asymptotic fields which are linear with respect to 
fields. The vector fields,  $W^{\pm}_{\mu}$, $Z_{\mu}$ and $A_{\mu}$,  satisfy
\begin{eqnarray}
\left (  \square + M_{Z}^{\:2} \right )  Z_{\mu} + 2M_{Z} \frac{\partial_{\mu} v}{v} \chi^{3}&=&0\:,
\label{eq:zbosoneqofmotion}
\\
\left (  \square + M_{W}^{\:2} \right )  W^{\pm}_{\mu} +2M_{W} \frac{\partial_{\mu} v}{v} \chi^{\pm}&=&0\:,
\label{eq:wbosoneqofmotion}
\\
\square A_{\mu}&=&0\:,
\label{eq:elemageqofmotion}
\end{eqnarray}
and we notice that  unphysical scalar fields appear in (\ref{eq:zbosoneqofmotion}) and 
(\ref{eq:wbosoneqofmotion}) due to $\partial_{\mu} v \neq 0$ .  Later we will assume that the 
vacuum expectation value $v$ depends on only one space coordinate, $z$ (i.e., $v=v(z)$), 
but for now  we set $\partial_{\mu} v \neq 0$ $(\mu=0,1,2,3)$, keeping generality as much as possible.
The masses of the gauge bosons are given as before, 
\begin{eqnarray}
M_{Z}^{2}&=&\frac{1}{4} \left ( g^{2} + g^{\:\prime \:2} \right ) v^{2}\:,
\hskip1cm
M_{W}^{2}= \frac{1}{4}  g^{2}  v^{2}\:.
\end{eqnarray}
and are supposed to depend on the space-time coordinates through $v$.

The linearized equations of motion of the unphysical scalar fields  also contain coupling 
terms with the gauge field through $\partial_{\mu}v\neq   0$, 
\begin{eqnarray}
\left ( \square + M_{1}^{\:2} \right ) \chi^{3} - 2M_{Z}\frac{\partial ^{\mu} v}{v}  Z_{\mu} &=& 0\:,
\label{eq:spacetimedependentmass2}
\\
\left ( \square + M_{2}^{\:2} \right ) \chi^{\pm} - 2M_{W}\frac{\partial ^{\mu} v}{v}  W_{\mu}^{\pm} &=&0\:.
\label{eq:spacetimedependentmass}
\end{eqnarray}
In contrast with the usual case, the mass terms  $M_{1}^{2}$ and $M_{2}^{2}$   in (\ref{eq:spacetimedependentmass2}) and (\ref{eq:spacetimedependentmass})   differ from the gauge bosom masses in the following way, 
\begin{eqnarray}
M_{1}^{\:2}=M_{Z}^{\:2}+
V^{\:\prime}\left (  \frac{v^{2}}{2} \right )\:,
\hskip1cm
M_{2}^{\:2}=M_{W}^{\:2}+  V^{\:\prime}\left (  \frac{v^{2}}{2} \right )                    \:.
\label{eq:m1m2}
\end{eqnarray}
We must not fail to include the second term $V^{\:\prime}\left (  v^{2}/2 \right )$  which will play   non-trivial  roles when  the absence of tadpoles is considered.

Let us turn to the auxiliary fields, which are expressed in terms of  the gauge and unphysical scalar fields via equations of motion, 
\begin{eqnarray}
{\cal B}^{Z}=\partial_{\mu}  Z^{\mu} + M_{Z}  \chi^{3}\:,
\hskip0.5cm
{\cal B}^{\pm}=\partial _{\mu} W^{\pm \: \mu } + M_{W}  \chi^{\pm}\:,
\hskip0.5cm
{\cal B}^{A}=\partial_{\mu}A^{\mu}\:.
\label{eq:auxiliaryfieldbz}
\end{eqnarray}
Since the gauge  and unphysical scalar fields satisfy the second-order differential equations, 
we can likewise derive the second-order equations for ${\cal B}^{Z}$,  ${\cal B}^{\pm}$ and 
${\cal B}^{A}$, which turn out to be 
\begin{eqnarray}
\left ( \square + M_{Z}^{\:2} \right ) {\cal B}^{Z} +M_{Z} \left \{
\frac{1}{v} \square \: v +
V^{\:\prime}\left (   \frac{v^{2}}{2} \right )
\right \} \chi^{3}&=&0\:,
\label{eq:beforekleingordon1}
\\
\left ( \square + M_{W}^{\:2} \right ) {\cal B}^{\pm} +M_{W} \left \{
\frac{1}{v} \square \: v +
V^{\:\prime}\left (   \frac{v^{2}}{2} \right )
\right \} \chi^{\pm}&=&0\:,
\label{eq:beforekleingordon2}
\\
\square \: {\cal B}^{A}&=&0\:.
\end{eqnarray}
It is extremely curious to remark that once we impose the vanishing tadpole condition (\ref{eq:stability}), 
Eqs.  (\ref{eq:beforekleingordon1}) and  (\ref{eq:beforekleingordon2})
become simple Klein-Gordon type equations,  
\begin{eqnarray}
\left ( \square + M_{Z}^{\:2} \right )   {\cal B}^{Z} 
 &=&0\:,
\hskip1cm 
\left ( \square   + M_{W}^{\:2} \right ) {\cal B}^{\pm}   = 0\:.
\label{eq:afterkleingordon}
\end{eqnarray}
This implies that, under the condition (\ref{eq:stability}), the mode expansion of the auxiliary fields 
should go along the line in Section  \ref{sec:wavepropagation}. 

\subsection{The mode-expansion and field quantization for $v=v(z)$}

Let us scrutinize the $Z$-boson field's  propagation  in the similar manner to  the    scalar field case given  in Section  \ref{sec:kleingordonpropagation}. The $W^{\pm}$-boson field's propagation will be treated exactly in the same way as the $Z$-boson field and its exposition need not be repeated.  The equation of the electromagnetic field is unchanged as we see in  (\ref{eq:elemageqofmotion})  and will not be discussed  in the rest of the present paper.   Here and hereafter we assume that the vacuum expectation value depends only on the $z$-coordinate, namely,
\begin{eqnarray}
v=v(z)\:.
\end{eqnarray} 
With this setting, the wave equations of $Z^{\mu}$-field are handled  in two distinct ways,     
one for  $\mu=0,1, 2$ case  and the other for $\mu=3$   case.

\subsubsection{The case of $Z^{\mu}$ $(\mu=0,1, 2)$ fields }

The equations of $Z^{\mu}$ $(\mu =0,1,2)$ fields, as given in (\ref{eq:zbosoneqofmotion}), 
\begin{eqnarray}
\left \{  \square + M_{Z}(z)^{2} \right \}  Z^{\: \mu} (t, \vec{x}_{\perp}, z )&=&0\:, \hskip0.5cm (\mu =0,1,2)
 \label{eq:zbosoneqofmotion2}
\end{eqnarray}
are separated from the unphysical scalar fields and look much simpler than the 
$Z^{3} (t, \vec{x}_{\perp}, z )$ case. 
The mode expansion goes exactly in the same way as in Section \ref{sebsec:fieldquantization}. Before moving to the mode expansion, however, we have to prepare a set of  vectors describing the polarization of the propagating waves.  Since we are dealing with  three components in (\ref{eq:zbosoneqofmotion2}), we introduce only three basis vectors, which we choose as follows, 
\begin{eqnarray}
& &  
\varepsilon^{(0)\:\mu} (E, \vec{p}_{\perp})  
=  \frac{1}{\sqrt{E^{2} - \vert \vec{p}_{\perp} \vert^{2}}} \left (
\begin{tabular}{c}
$E$
\\
$p_{x}$
\\
$p_{y}$
\\
$0$
\end{tabular}
\right )
=  \frac{1}{\sqrt{\lambda } }\left (
\begin{tabular}{c}
$E$
\\
$p_{x}$
\\
$p_{y}$
\\
$0$
\end{tabular}
\right )\:,
\label{eq:polarizationvectors2}
   \\
   & & 
\varepsilon^{(1)\:\mu} (E, \vec{p}_{\perp}) 
= \frac{1}{ \vert \vec{p}_{\perp} \vert } \left (
\begin{tabular}{c}
$0$
\\
$-p_{y}$
\\
$p_{x}$
 \\
$0$
\end{tabular}
\right )\:,
\label{eq:polarizationvectors3}
    \\
    & &
\varepsilon^{(2)\:\mu} (E, \vec{p}_{\perp})  
=  
\frac{1}{ \vert \vec{p}_{\perp} \vert \sqrt{E^{2} - \vert \vec{p}_{\perp} \vert^{2}}} \left (
\begin{tabular}{c}
$\vert \vec{p}_{\perp} \vert^{2}$
\\
$E \: p_{x}$
\\
$E \: p_{y}$
\\
$0$
\end{tabular}
\right )
=  
\frac{1}{\vert \vec{p}_{\perp} \vert \sqrt{\lambda} } \left (
\begin{tabular}{c}
$\vert \vec{p}_{\perp} \vert^{2}$
\\
$E \: p_{x}$
\\
$E \: p_{y}$
\\
$0$
\end{tabular}
\right )\:.
\label{eq:polarizationvectors}
\end{eqnarray}
Note that $\varepsilon^{(0)\:\mu}$ is time-like, while $\varepsilon^{(1)\:\mu}$ and $\varepsilon^{(2)\:\mu}$ are space-like. They are mutually orthogonal and are normalized appropriately, i.e.,
\begin{eqnarray}
& & 
\varepsilon^{(0)} \cdot \varepsilon^{(0)} =1, \hskip0.5cm
\varepsilon^{(1)} \cdot \varepsilon^{(1)} =\varepsilon^{(2)} \cdot \varepsilon^{(2)} =-1\:,
\\
& & 
\varepsilon^{(0)} \cdot \varepsilon^{(1)} =\varepsilon^{(1)} \cdot \varepsilon^{(2)} =\varepsilon^{(2)} \cdot \varepsilon^{(0)}=0\:.
\end{eqnarray}
It should also be useful to note that these three vectors satisfy the summation formula
\begin{eqnarray}
\sum_{a=0}^{2} \frac{\varepsilon^{(a)\:\mu}\varepsilon^{(a)\:\nu}}{
\varepsilon^{(a)} \cdot \varepsilon^{(a)}
}
=
\sum_{a=0}^{2}
\sum_ {b=0}^{2}  \: \varepsilon^{(a)\:\mu}  \:     \eta^{ab}    \:   \varepsilon^{(b)\:\nu}
=
\left (
\begin{tabular}{cccc}
$1$ & $0$ & $0$ & $0$
\\
$0$ & $-1$ & $0$ & $0$
\\
$0$ & $0$ & $-1$ & $0$
\\
$0$ & $0$ & $0$ & $0$
\end{tabular}
\right )^{\mu \nu} \:,
\label{eq:summationformula}
\end{eqnarray}
where 
\begin{eqnarray}
& & 
 \eta^{ab}=
\left (
\begin{tabular}{ccc}
$1$ & $0$ & $0$
\\
$0$ & $-1$ & $0$
\\
$0$ & $0$ & $-1$
\end{tabular}
\right )^{ab}\:, \hskip1cm (a,b=0,1,2)\:.
\label{eq:etaab}
\end{eqnarray}
Both of $\varepsilon^{(1)\:\mu}$ and $\varepsilon^{(2)\:\mu}$    are  orthogonal in the four-dimensional sense   to the vector $p^{\mu} \equiv (E, p_{x}, p_{y}, p_{z})$, i.e., $p\cdot \varepsilon^{(1)}=0$,  $p\cdot \varepsilon^{(2)}=0$, whatsoever the $z$-component $p_{z}$ would be.   In the three-dimensional sense, on the other hand,   only $\varepsilon ^{(1)\: i} (i=1,2,3)$ is orthogonal to  $\vec{p}=(p_{x}, p_{y}, p_{z})$.
In the present study of the $z$-dependent mass, however,   the $z$-component $p_{z}$ of the gauge field wave vector is not   well-defined and we have  to refrain from  using the terminology such as longitudinal or transversal   polarizations.   The  vectors  in  (\ref{eq:polarizationvectors})  are simply  mutually-orthogonal   basis vectors to be used for the expansion   of $Z^{\mu}\:\:\:(\mu=0,1,2) $ according to the propagation  modes.  The expansion is thus given by 
 \begin{eqnarray}
 Z^{\: \mu} (t, \vec{x}_{\perp}, z) 
 &=&
  \int d\lambda \sum_{i=1, 2} \int \frac{d^{2}\vec{p}_{\perp}}
 {\sqrt{(2\pi)^{2} \: 2E }} \sum_{a=0}^{2} \bigg \{
 \alpha_{i}^{\: (a)} (\vec{p}_{\perp}, \lambda ) 
\:  e^{i \vec{p}_{\perp} \cdot \vec{x}_{\perp} -iEt} 
 \nonumber \\  
 & &
\hskip-1.2cm   + 
  \alpha_{i}^{\: (a) \: \dag }(\vec{p}_{\perp}, \lambda ) 
\:   e^{ - i \vec{p}_{\perp} \cdot \vec{x}_{\perp} + iEt}  \bigg \}
    \phi_{i}(z; \lambda)\:  \varepsilon^{(a)\:\mu} (E, \vec{p}_{\perp} )
  \:,\hskip0.3cm (\mu =0,1,2)\:.
 \label{eq:zmu012alphacretionannihilation2}
 \end{eqnarray}
The coefficients $ \alpha_{i}^{\: (a)} (\vec{p}_{\perp}, \lambda ) $ and 
$  \alpha_{i}^{\: (a) \: \dag }(\vec{p}_{\perp}, \lambda ) $ are operators, whose commutation relations 
we are about to discuss.

  \subsubsection{Quantization of the $Z^{\mu}$ ($\mu =0,1,2$)-fields}

In analogy with the scalar case, we postulate the commutation relation
\begin{eqnarray}
\left [ \alpha^{(a)}_{i} (\vec{p}_{\perp} , \lambda ) ,  \; \alpha^{(b) \: \dag }_{j} (\vec{p}_{\perp}^{\:\prime}, 
\lambda^{\:\prime} )  \right ]
&=&
- \eta^{ab} \: \delta^{2}(  \vec{p}_{\perp} - \vec{p}_{\perp}^{\:\prime} )
\: \delta ( \lambda - \lambda^{\:\prime} ) \: \frac{d \rho _{ij}(\lambda )}{d\lambda }\:,
\label{eq:alphakalphalcommutator}
\\
& & (a, b=0, 1,2;   \:\:\: i,j=1,2)\:,
\nonumber 
\end{eqnarray}
where $\eta^{ab}$ was defined by (\ref{eq:etaab}).    Although the four-dimensional Lorentz covariance is not preserved due to the bubble wall, we still have the $O(1,2)$ symmetry in the $(t, x, y)$ subspace, and we have to put $-\eta^{ab}$ on the right-hand side of (\ref{eq:alphakalphalcommutator}). In connection with this,   let us   recall that $d\rho_{ij}(\lambda )/ d\lambda $   is a positive semi-definite   matrix and note the sign difference on the right hand side,   $-\eta^{00}=-1$   in comparison with $-\eta^{11}=-\eta^{22}=+1$. The states in the Hilbert space   constructed by applying   the creation operator $\alpha_{i}^{(0)\:\dag } (\vec{p}_{\perp}, \lambda )$   on the vacuum become necessarily negative norm states. We will discuss later the decoupling  of negative   norm states from the physical space.  Here we just confirm the following formula, 
\begin{eqnarray}
& & \hskip-2.0cm
\sum_{a=0}^{2} \sum_{b=0}^{2}
\left [ \alpha^{(a)}_{i} (\vec{p}_{\perp} , \lambda ) \varepsilon ^{(a)\:\mu} (E, \vec{p}_{\perp})  \;,
\alpha^{(b) \: \dag }_{j} (\vec{p}_{\perp}^{\:\prime}, \lambda^{\:\prime} )  
 \varepsilon ^{(b)\:\nu} (E^{\:\prime}, \vec{p}_{\perp}^{\:\prime} ) \right ]
 \nonumber \\
&=&
- \eta^{\mu \nu} \: \delta^{2}(  \vec{p}_{\perp} - \vec{p}_{\perp}^{\:\prime} )
\: \delta ( \lambda - \lambda^{\:\prime} ) \: \frac{d \rho _{ij}(\lambda )}{d\lambda }\:,
 \hskip1cm   (\mu, \nu = 0, 1,2)\:,
 \label{eq:alphaepsiloncr}
\end{eqnarray}
where $E^{\:\prime}=\sqrt{\lambda^{\:\prime} +  \vert \vec{p}_{\perp}^{\:\prime} \vert^{2}} $ and use has been made of the summation formula (\ref{eq:summationformula}).

Now that we are equipped with the formula (\ref{eq:alphaepsiloncr}),  we are able to confirm  the four-dimensional commutation relation of the $Z^{\mu}( t, \vec{x}_{\perp}, z ) $ field of (\ref{eq:zmu012alphacretionannihilation2}) in the following form 
\begin{eqnarray}
\left [  Z^{\: \mu} ( t, \vec{x}_{\perp}, z ) , Z^{\: \nu} ( t^{\:\prime}, \vec{x}_{\perp}^{ \: \prime} , z^{\:\prime} )   \right ]
=
-i \: \Delta ( t-t^{\:\prime} , \vec{x}_{\perp} - \vec{x}_{\perp}^{\:\prime},  z, z^{\:\prime}; M_{Z}(*) )\:
\eta^{\mu \nu }\:, 
\\
 (\mu, \nu = 0, 1,2)\:.
 \nonumber 
\end{eqnarray}
The equal time commutation relations are also easily deduced from above with the help of the formulas 
(\ref{eq:formula3}) - (\ref{eq:formula4}), i.e., 
\begin{eqnarray}
\left [  Z^{\: \mu} ( t, \vec{x}_{\perp}, z ) , Z^{\:\nu} ( t, \vec{x}_{\perp}^{ \: \prime} , z^{\:\prime} )   \right ]
&=&0\;,
\\
\left [  \dot{Z}^{\: \mu} ( t, \vec{x}_{\perp}, z ) , Z^{\:\nu} ( t, \vec{x}_{\perp}^{ \: \prime} , z^{\:\prime} )   \right ]
&=&i\:\delta^{2}( \vec{x}_{\perp} - \vec{x}_{\perp}^{\:\prime} ) \delta(z-z^{\:\prime}) \: \eta^{\mu \nu }\:,
\\
\left [  \dot{Z}^{\: \mu} ( t, \vec{x}_{\perp}, z ) ,  \dot{Z}^{\:\nu} ( t, \vec{x}_{\perp}^{ \: \prime} , z^{\:\prime} )   \right ]
&=&0\;,\hskip1.8cm   (\mu, \nu = 0, 1,2)\:.
\nonumber 
\end{eqnarray}

\subsubsection{Coupled equations of   $Z^{3}  ( t, \vec{x}_{\perp}, z ) $ and  $\chi^{3} ( t, \vec{x}_{\perp}, z ) $}

Let us now turn to the equations of the $\mu=3$ component  
$Z^{3}  ( t, \vec{x}_{\perp}, z ) = - Z_{3}  ( t, \vec{x}_{\perp}, z ) $
and  $\chi^{3} ( t, \vec{x}_{\perp}, z ) $
which are,  according to  (\ref{eq:zbosoneqofmotion}) and (\ref{eq:spacetimedependentmass2}),  coupled equations 
\begin{eqnarray}
 \left \{  \square + M_{Z}(z)^{2}  \right \}  Z^{3}  ( t, \vec{x}_{\perp}, z ) - 2M_{Z}(z) \frac{ v^{\prime}(z)}{v(z)} \chi^{3} ( t, \vec{x}_{\perp}, z ) &=&0\:,
 \label{eq:z3chi3eqofmotion1}
\\
\left \{ \square + M_{1}(z)^{2} \right \} \chi^{3} ( t, \vec{x}_{\perp}, z )  - 2M_{Z}(z) \frac{ v^{\prime}(z)}{v(z)}  Z^{3}  ( t, \vec{x}_{\perp}, z ) &=& 0\:, 
 \label{eq:z3chi3eqofmotion2}
\end{eqnarray}
and therefore we have to take a different approach from the previous section.   Here again we look for  solutions of the plane wave type   
 $e^{i  ( \vec{p}_{\perp} \cdot \vec{x}_{\perp} -E t ) }$   
  in the $t$- and $\vec{x}_{\perp}$-directions 
 and the coupled equations can be put in the form of the one-dimensional Schr{\" o}dinger type  equation with two-component wave functions, 
\begin{eqnarray}
\left \{
-\frac{\partial^{2} }{\partial z^{2}} + U(z)
\right \}
\left (
\begin{tabular}{c}
$Z^{3} ( t, \vec{x}_{\perp}, z ) $
\\
\\
$\chi^{3} ( t, \vec{x}_{\perp}, z ) $
\end{tabular}
\right )
=
\lambda \:
\left (
\begin{tabular}{c}
$Z^{3} ( t, \vec{x}_{\perp}, z ) $
\\
\\
$\chi^{3} ( t, \vec{x}_{\perp}, z ) $
\end{tabular}
\right ), 
\hskip0.1cm
\left (  \lambda =E^{2} - \vert \vec{p}_{\perp} \vert^{2}  \right ).
\label{eq:2by2differentialequation}
\end{eqnarray}
The "potential term"  $U(z)$ is a 2 $\times $ 2 matrix whose components are given as follows,
\begin{eqnarray}
U(z)=
\left (
\begin{tabular}{cc}
$M_{Z}(z)^{2}$ & $\displaystyle{- 2M_{Z}(z)\frac{v^{\prime}(z)}{v(z)}}$
\\
\\
$\displaystyle{- 2M_{Z}(z)\frac{v^{\prime}(z)}{v(z)}}$ & $M_{1}(z)^{2}$
\end{tabular}
\right )\:.
\end{eqnarray}
Note that $U(z)$ is a symmetric matrix and that the differential operator on the left hand side of (\ref{eq:2by2differentialequation})  is a self-adjoint operator. 

With regard to   (\ref{eq:2by2differentialequation}),   let us recall that Kodaira  investigated  in \cite{kodaira2}  the  eigenfunction expansion method associated with any even order self-adjoint differential operators as  a straightforward generalization of his previous work  \cite{kodaira1}.  As he argued in \cite{kodaira2}, his   result can be readily  extended to the case of simultaneous differential equations. More specifically speaking,   second-order self-adjoint differential operators for two component wave functions as given in  (\ref{eq:2by2differentialequation})  can be   studied in the same way as  in the case of fourth-order differential equations  for a single wave function. In fact we can go in the following way. 

Let us consider the following ordinary differential equation in the infinite region $-\infty < z < +\infty$, 
\begin{eqnarray}
\left \{
-\frac{d^{2} }{d z^{2}} + U(z)
\right \}
\Phi ^{(k)}(z; \lambda )
=
\lambda \:\Phi ^{(k)}(z; \lambda )\:, 
\hskip1.0cm
 (k=1, \cdots , 4)\;,
 \label{eq:capitalphiequation}
\end{eqnarray}
where $\Phi^{(k)} (z; \lambda )$ has two components of  wave functions,  
\begin{eqnarray}
\Phi^{(k)} (z; \lambda )=\left (
\begin{tabular}{c}
$\Phi_{1}^{\:(k)}(z; \lambda )$
\\
\\
$\Phi_{2}^{\:(k)}(z;\lambda )$
\end{tabular}
\right ), \hskip0.5cm (k=1, \cdots , 4)\:.
\end{eqnarray}
The superscript ``$(k)$"     attached to $\Phi^{(k)} (z; \lambda )$ discriminates the initial conditions imposed at $z=0$ which are listed below,
\begin{eqnarray}
\Phi^{(1)} (0; \lambda )=\left (
\begin{tabular}{c}
$1$
\\
\\
$0$
\end{tabular}
\right ), 
\hskip1cm
\Phi^{(1)\; \prime } (0; \lambda )=\left (
\begin{tabular}{c}
$0$
\\
\\
$0$
\end{tabular}
\right ), 
\label{eq:intislconditions1}
\\
\Phi^{(2)} (0; \lambda )=\left (
\begin{tabular}{c}
$0$
\\
\\
$0$
\end{tabular}
\right ), 
\hskip1cm
\Phi^{(2)\; \prime } (0; \lambda )=\left (
\begin{tabular}{c}
$1$
\\
\\
$0$
\end{tabular}
\right ), 
\\
\Phi^{(3)} (0; \lambda )=\left (
\begin{tabular}{c}
$0$
\\
\\
$1$
\end{tabular}
\right ), 
\hskip1cm
\Phi^{(3)\; \prime } (0; \lambda )=\left (
\begin{tabular}{c}
$0$
\\
\\
$0$
\end{tabular}
\right ), 
\\
\Phi^{(4)} (0; \lambda )=\left (
\begin{tabular}{c}
$0$
\\
\\
$0$
\end{tabular}
\right ), 
\hskip1cm
\Phi^{(4)\; \prime } (0; \lambda )=\left (
\begin{tabular}{c}
$0$
\\
\\
$1$
\end{tabular}
\right ).
\label{eq:intislconditions4}
\end{eqnarray}
Here differentiation with respect to $z$ is denoted by prime ($\prime$).

Every  solution to  (\ref{eq:2by2differentialequation}) can be expressed as a linear combination of these 
four fundamental solutions, and the completeness  is guaranteed  by the formula 
\begin{eqnarray}
\int  \sum_{k, l=1}^{4} \Phi^{(k)}_{\sigma} (z; \lambda ) 
 d\widehat{\rho}_{kl}(\lambda )       \Phi^{(l)}_{\tau} 
(z^{\:\prime} ; \lambda )=\delta_{\sigma \tau} \delta (z-z^{\:\prime}) \:,
\hskip0.5cm (\sigma , \tau =1,2)\:.
\label{eq:bigformula}
\end{eqnarray}
Here $ \widehat{\rho}_{kl}(\lambda )$ is a 4 $\times $ 4 matrix and describes the spectrum of $\lambda$. The integration $d\widehat{\rho}_{kl}(\lambda )$ is, precisely speaking,   the Stieltjes integral.   We, however,  assume simply that the $\lambda$-spectrum is  continuous 
{\color{black}{or equivalently we assume the absence of bound state solutions of the potential $U(z)$
in (\ref{eq:capitalphiequation}). If there would be bound state solutions, they could affect the calculation of the friction on the bubble wall, but for now we leave it for future investigations. Under this assumption  
}}
we regard (\ref{eq:bigformula}) as the usual Riemann integration.  We are thus led to the expansion of the pair of the fields in (\ref{eq:2by2differentialequation}) in the following way, 
 \begin{eqnarray}
 \left (
 \begin{tabular}{c}
$ Z^{3} (t, \vec{x}_{\perp}, z) $
 \\
 \\
 $ \chi^{3} (t, \vec{x}_{\perp}, z) $
 \end{tabular}
 \right )
 &=&
  \int d\lambda \sum_{k=1}^{4} \int \frac{d^{2}\vec{p}_{\perp}}
 {\sqrt{(2\pi)^{2} \: 2E }} \bigg \{
 \beta_{k} (\vec{p}_{\perp}, \lambda ) e^{i \vec{p}_{\perp} \cdot \vec{x}_{\perp} -iEt} 
 \nonumber \\  
 & &
\hskip1.0cm   + 
  \beta_{k}^{\: \dag }(\vec{p}_{\perp}, \lambda ) e^{ - i \vec{p}_{\perp} \cdot \vec{x}_{\perp} + iEt}
  \bigg \}
   \Phi^{(k)}(z; \lambda) \:.
 \label{eq:scalaralphacreationannihilation3}
 \end{eqnarray}
 The coefficients $ \beta_{k} (\vec{p}_{\perp}, \lambda )$ and $  \beta_{k}^{\: \dag }(\vec{p}_{\perp}, \lambda ) $ are operators, which are going to describe, respectively,  the annihilation and creation  of wave modes of  $Z^{3} (t, \vec{x}_{\perp}, z) $ and  at the same time those of $ \chi^{3} (t, \vec{x}_{\perp}, z) $. 

\subsubsection{Quantization  of   $Z^{3}  ( t, \vec{x}_{\perp}, z ) $ and  $\chi^{3} ( t, \vec{x}_{\perp}, z )$ fields  
}

We now propose to set up the following commutation relations
\begin{eqnarray}
\left [ \beta_{k} (\vec{p}_{\perp} , \lambda ) ,  \beta^{ \: \dag }_{l} (\vec{p}_{\perp}^{\:\prime}, 
\lambda^{\:\prime} )  \right ]
&=&
 \delta^{2}(  \vec{p}_{\perp} - \vec{p}_{\perp}^{\:\prime} )
\delta ( \lambda - \lambda^{\:\prime} ) \frac{d\widehat{\rho}_{kl}(\lambda )}{d\lambda }
\label{eq:betakbetalcommutator}
\end{eqnarray}
analogously to (\ref{eq:alphakalphalcommutator}). Since $d\widehat{\rho}_{kl}(\lambda )/d\lambda $ is a symmetric positive definite matrix, this commutation relation does not generate negative norm states when $\beta_{k}^{\dag} ( \vec{p}_{\perp}, \lambda )$ is  regarded as a creation operator of the wave mode labeled  by $ (\vec{p}_{\perp}, \lambda , k)$.The four-dimensional commutation relations among $Z^{3}(t, \vec{x}_{\perp}, z)$ and $\chi^{3}(t, \vec{x}_{\perp}, z)$ are easily worked out as follows,  
\begin{eqnarray}
\left [ 
Z^{3}(t, \vec{x}_{\perp}, z), \: Z^{3}(t^{\:\prime}, \vec{x}_{\perp}^{\:\prime}, z^{\: \prime})
\right ]= i \widehat{\Delta}_{11}(t-t^{\:\prime}, \vec{x}_{\perp} - \vec{x}_{\perp}^{\:\prime}, z, z^{\: \prime} ;  
U(*))\:,
\\
\left [ 
Z^{3}(t, \vec{x}_{\perp}, z), \: \chi^{3}(t^{\:\prime}, \vec{x}_{\perp}^{\:\prime}, z^{\: \prime})
\right ]= i \widehat{\Delta}_{12}(t-t^{\:\prime}, \vec{x}_{\perp} - \vec{x}_{\perp}^{\:\prime}, z, z^{\: \prime} ;  
U(*))\:,
\\
\left [ 
\chi^{3}(t, \vec{x}_{\perp}, z), \: Z^{3}(t^{\:\prime}, \vec{x}_{\perp}^{\:\prime}, z^{\: \prime})
\right ]= i \widehat{\Delta}_{21}(t-t^{\:\prime}, \vec{x}_{\perp} - \vec{x}_{\perp}^{\:\prime}, z, z^{\: \prime} ;  
U(*))\;,
\\
\left [ 
\chi^{3}(t, \vec{x}_{\perp}, z), \: \chi^{3}(t^{\:\prime}, \vec{x}_{\perp}^{\:\prime}, z^{\: \prime})
\right ]= i \widehat{\Delta}_{22}(t-t^{\:\prime}, \vec{x}_{\perp} - \vec{x}_{\perp}^{\:\prime}, z, z^{\: \prime} ;  
U(*))\;.
\label{eq:chi3chi3fourdimcr}
\end{eqnarray}
Here we have newly defined a  $2\times 2$ matrix function,  
\begin{eqnarray}
& & \hskip-1.0cm 
\widehat{\Delta}_{\sigma \tau}  ( t - t^{\:\prime}, \vec{x}_{\perp} -\vec{x}_{\perp}^{\:\prime},  z, z^{\:\prime} ; U(*))
\nonumber \\
&=& -i 
\int \frac{d^{2}\vec{p}_{\perp}}{(2\pi)^{2}} \: e^{i \vec{p}_{\perp} \cdot ( \vec{x}_{\perp} - 
\vec{x}^{\;\prime}_{\perp} ) }
\sum_{k,l}
\int \Phi_{\sigma}^{(k)} (z; \lambda )\frac{d\widehat{\rho}_{kl}(\lambda)}{2E} 
\Phi_{\tau}^{(l)}(z^{\:\prime}; \lambda )
\nonumber \\
& & \hskip0.3cm  
\times  
\left \{
e^{-iE(t-t^{\:\prime} )} -  e^{iE(t-t^{\:\prime} )}
\right \}\:, \hskip1cm (\tau , \sigma =1,2)\:.
   \label{eq:deltafunction2}
\end{eqnarray}
Apparently (\ref{eq:deltafunction2}) satisfies the  differential equations
\begin{eqnarray}
\left \{   \square +U(z)   \right \} \left (
\begin{tabular}{cc}
$\widehat{\Delta}_{11}$ &  $\widehat{\Delta}_{12}$
\\
$\widehat{\Delta}_{21}$  & $\widehat{\Delta}_{22}$
\end{tabular}
\right )=0\:,
\hskip0.5cm
\left \{   \square^{\:\prime} +U(z^{\prime})   \right \} \left (
\begin{tabular}{cc}
$\widehat{\Delta}_{11}$ &  $\widehat{\Delta}_{21}$
\\
$\widehat{\Delta}_{12}$  & $\widehat{\Delta}_{22}$
\end{tabular}
\right )=0\;,
\end{eqnarray}
together with the characteristic properties at $t=t^{\:\prime}$,  
\begin{eqnarray}
& & 
\widehat{\Delta}_{\sigma \tau} \left ( 0, \vec{x}_{\perp} - \vec{x}^{\:\prime}_{\perp}, z, z^{\:\prime} ; U(*) \right  )=0\:,
\label{eq:544onecederivativeformula}
\\
& & 
\frac{\partial}{\partial t} \widehat{\Delta}_{\sigma \tau} \left ( t - t^{\:\prime}, \vec{x}_{\perp} - \vec{x}^{\:\prime}_{\perp}, z, z^{\:\prime} ; U(*) \right  )\Big \vert_{t=t^{\:\prime}}=-\delta_{\sigma \tau} \delta^{2}(\vec{x}_{\perp} -
\vec{x}^{\:\prime} )\delta(z - z^{\:\prime}) \:,
\label{eq:546onecederivativeformula}
\\
& & 
\frac{\partial }{\partial t^{\:\prime}}
\frac{\partial}{\partial t} \widehat{\Delta}_{\sigma \tau} \left ( t - t^{\:\prime}, \vec{x}_{\perp} - \vec{x}^{\:\prime}_{\perp}, z, z^{\:\prime} ; U(*) \right  )\Big \vert_{t=t^{\:\prime}}=0\:.
\end{eqnarray}
The formula  (\ref{eq:546onecederivativeformula}) leads us to derive the equal-time commutation relations 
\begin{eqnarray}
\left [ 
\dot{Z}^{3}(t, \vec{x}_{\perp}, z), \: Z^{3}(t^{}, \vec{x}_{\perp}^{\:\prime}, z^{\: \prime})
\right ]&=& -i
\delta^{2}(\vec{x}_{\perp} - \vec{x}_{\perp}^{\:\prime})\delta(z-z^{\:\prime})\;,
\\
\left [ 
\dot{Z}^{3}(t, \vec{x}_{\perp}, z), \: \chi^{3}(t^{}, \vec{x}_{\perp}^{\:\prime}, z^{\: \prime})
\right ]&=&0\;,
\\
\left [ 
\dot{\chi}^{3}(t, \vec{x}_{\perp}, z), \: Z^{3}(t^{}, \vec{x}_{\perp}^{\:\prime}, z^{\: \prime})
\right ]&=&0\;,
\\
\left [ 
\dot{\chi}^{3}(t, \vec{x}_{\perp}, z), \: \chi^{3}(t^{}, \vec{x}_{\perp}^{\:\prime}, z^{\: \prime})
\right ]&=&  - i 
\delta^{2}(\vec{x}_{\perp} - \vec{x}_{\perp}^{\:\prime})\delta(z-z^{\:\prime})\;.
\label{eq:chi3dotchi3equaltime}
\end{eqnarray}

\subsection{The auxiliary  field   ${\cal B}^{Z} ( t, \vec{x}_{\perp}, z )$}

Towards the end of Section \ref{sec:gaugefixingauxiliaryfields}, it was mentioned that $Z^{0}( t, \vec{x}_{\perp}, z)$ is the momentum variable canonically conjugate to the auxiliary field  ${\cal B}^{Z}(t, \vec{x}_{\perp}, z)$, and let us look at this point further. Expressing  ${\cal B}^{Z}(t, \vec{x}_{\perp}, z)$ by the formula (\ref{eq:auxiliaryfieldbz}) we are able to compute the commutator
\begin{eqnarray}
\left [  Z^{0} ( t, \vec{x}_{\perp}^{} , z ), \:\:
 {\cal B}^{Z} ( t, \vec{x}_{\perp}^{\:\prime}, z^{\:\prime} )   \right ]
&=&
\left [
Z^{\: 0} ( t, \vec{x}_{\perp}, z ) , \:\:
\dot{Z}^{\:0} ( t, \vec{x}_{\perp}^{ \: \prime} , z^{\:\prime} )  
\right ]
\nonumber \\
&=&-i \delta^{2}(\vec{x}_{\perp} - \vec{x}_{\perp}^{\:\prime})\delta(z- z^{\:\prime})\:,
\end{eqnarray}
which is a desirable relation. We also have to confirm  that $ {\cal B}^{Z} ( t, \vec{x}_{\perp}, z  )$ is a  variable independent of $\chi^{3}(t^{}, \vec{x}_{\perp}, z)$ and the commuting property
\begin{eqnarray}
\left [  \chi^{3} ( t, \vec{x}_{\perp}^{} , z ), \:\:
 {\cal B}^{Z} ( t, \vec{x}_{\perp}^{\:\prime}, z^{\:\prime} )   \right ] =0
\end{eqnarray}
follows from  (\ref{eq:chi3chi3fourdimcr}) and (\ref{eq:544onecederivativeformula}).  The formula (\ref{eq:chi3dotchi3equaltime}), on the other hand, tells us the relation
\begin{eqnarray}
\left [  \dot{\chi}^{3} ( t, \vec{x}_{\perp}^{} , z ), \:\:
 {\cal B}^{Z} ( t, \vec{x}_{\perp}^{\:\prime}, z^{\:\prime} )   \right ] 
 =- i   \delta^{2}(\vec{x}_{\perp} - \vec{x}_{\perp}^{\:\prime})\delta(z-z^{\:\prime})\:M_{Z}(z)\:.
\end{eqnarray}
Recall   that the momentum variable canonically conjugate to   $ \chi^{3} ( t, \vec{x}_{\perp},z)$
is not  $\dot{\chi}^{3} ( t, \vec{x}_{\perp} , z )$, but  is given by 
\begin{eqnarray}
\Pi_{\chi^{3}} (t, \vec{x}_{\perp}, z)  \equiv 
 \dot{\chi}^{3} ( t, \vec{x}_{\perp}^{} , z ) - M_{Z}(z)\: Z^{0}(t, \vec{x}_{\perp}, z)\:.
 \label{eq:canonicallyconjugatemomentum}
\end{eqnarray}
Thanks to the second term in (\ref{eq:canonicallyconjugatemomentum}),   the commuting property 
\begin{eqnarray}
\left [  \Pi_{\chi^{3}} ( t, \vec{x}_{\perp}^{} , z ), \:\:
 {\cal B}^{Z} ( t, \vec{x}_{\perp}^{\:\prime}, z^{\:\prime} )   \right ] 
 =0
\end{eqnarray}
is guaranteed.

Let us now move to the commutation relations of $ {\cal B}^{Z} ( t, \vec{x}_{\perp}, z  )$ with itself. 
We have already listed up all the non-vanishing equal time commutation relations involving $Z^{\mu}(t, \vec{x}_{\perp}, z)$ and $\chi^{3}(t, \vec{x}_{\perp}, z)$ and it is straightforward to confirm the vanishing  of the
following equal time commutators,  
\begin{eqnarray}
& & 
\left [  
 {\cal B}^{Z} ( t, \vec{x}_{\perp}, z ), \:\:
  {\cal B}^{Z} ( t^{}, \vec{x}_{\perp}^{\:\prime} , z^{\:\prime} )
    \right ]=0\:, 
\hskip0.5cm 
    \left [  
\dot{{\cal B}}^{Z} ( t, \vec{x}_{\perp}, z ), \:\:
  {\cal B}^{Z} ( t^{}, \vec{x}_{\perp}^{\:\prime} , z^{\:\prime} )
    \right ]=0\:.
    \label{eq:bzbz2}
\end{eqnarray}
As for the four-dimensional commutator of  $ {\cal B}^{Z} ( t, \vec{x}_{\perp}, z  )$,  we would get a lengthy formula containing the generalized invariant $\Delta $-functions  of (\ref{eq:deltafunction}) and (\ref{eq:deltafunction2}).  It is nonetheless still possible to prove the vanishing of the four-dimensional commutator 
    \begin{eqnarray}
    & & 
    \left [  
    {\cal B}^{Z} ( t, \vec{x}_{\perp}, z ), \:\:
     {\cal B}^{Z} ( t^{\:\prime }, \vec{x}_{\perp}^{\:\prime} , z^{\:\prime} )
     \right ]=0\:, 
     \label{eq:bzbz3}
    \end{eqnarray}
by using (\ref{eq:bzbz2}) and the Klein-Gordon type equation of motion   (\ref{eq:afterkleingordon})  of  $ {\cal B}^{Z} ( t, \vec{x}_{\perp}, z  )$. The vanishing property of (\ref{eq:bzbz3}) is due to the cancellation between two contributions that generate positive and negative norm states. This is similar  to the circumstance in  the Abelian Higgs model  \cite{nakanishinoboru1} -  \cite{nishijima1}.  The technique of proving (\ref{eq:bzbz3}) is given in 
Appendix \ref{appa:fourdimcr} in order to avoid digressing from the main stream of the present paper. 

\subsection{The ghost and anti-ghost fields}

The asymptotic ghost and anti-ghost fields in the $R_{\xi}$ gauge with $\xi =1$ satisfy the Klein-Gordon type equations, as can be seen from   the Faddeev-Popov Lagrangian   (\ref{eqappghostlagrangian1}) and   (\ref{eqappghostlagrangian2})
in Appendix \ref{sect:FPghostlagrangian}, expressed in terms of (\ref{eq:anotherghost}) and (\ref{eq:anotherantighost}),  namely, 
\begin{eqnarray}
\left ( \square + M_{W}^{2}  \right ) c^{(\pm)}=0, 
\hskip1.0cm
\left ( \square + M_{Z}^{2}  \right ) c^{Z}=0, 
\hskip1.0cm
 \square \:    c^{A}=0, 
 \label{eq:ghosteqofmotion}
 \\
\left ( \square + M_{W}^{2}  \right ) \overline{c}^{(\pm)}=0, 
\hskip1.0cm
\left ( \square + M_{Z}^{2}  \right ) \overline{c}^{Z}=0, 
\hskip1.0cm
 \square \:    \overline{c}^{A}=0\:.
 \label{eq:antighosteqofmotion}
\end{eqnarray}
The mode expansion of these  fields goes without any change as in Section 
\ref{sebsec:fieldquantization}.  
Since we focus our attention only to the $Z$-boson sector, we write down the mode expansion only 
of $c^{Z} (t, \vec{x}_{\perp},z)$ and $\overline{c}^{Z}(t, \vec{x}_{\perp}, z)$, 
 \begin{eqnarray}
 c^{Z} (t, \vec{x}_{\perp}, z) 
 &=&
  \int d\lambda \sum_{i=1, 2} \int \frac{d^{2}\vec{p}_{\perp}}
 {\sqrt{(2\pi)^{2} \: 2E }}
 \nonumber \\
 & & \times \bigg \{
 \gamma_{i}(\vec{p}_{\perp}, \lambda ) e^{i \vec{p}_{\perp} \cdot \vec{x}_{\perp} -iEt}
  +   \gamma_{i}^{\dag }(\vec{p}_{\perp}, \lambda ) e^{ - i \vec{p}_{\perp} \cdot \vec{x}_{\perp} + iEt}
\bigg \}   \phi_{i}(z; \lambda)   \:,
  \label{eq:ghostmodeexpansion}
 \\
 i \: \overline{c}^{Z} (t, \vec{x}_{\perp}, z) 
 &=&
  \int d\lambda \sum_{i=1, 2} \int \frac{d^{2}\vec{p}_{\perp}}
 {\sqrt{(2\pi)^{2} \: 2E }} 
 \nonumber \\
 & & \times \bigg \{
 \overline{\gamma}_{i}(\vec{p}_{\perp}, \lambda ) e^{i \vec{p}_{\perp} \cdot \vec{x}_{\perp} -iEt} 
+ 
  \overline{\gamma}_{i}^{\: \dag }(\vec{p}_{\perp}, \lambda ) e^{ - i \vec{p}_{\perp} \cdot \vec{x}_{\perp} 
  + iEt}   \bigg \}   \phi_{i}(z; \lambda)    \:.
  \label{eq:antighostmodeexpansion}
 \end{eqnarray}
 Here let us recall the hermiticity of ghost and anti-hermiticity of anti-ghost fields,  (\ref{eq:hermiticity}).   We have therefore put $`` \: i\:  "$ on the left hand side of (\ref{eq:antighostmodeexpansion})\:.
 
 The  anti-commutation relations of the coefficients in    (\ref{eq:ghostmodeexpansion}) and (\ref{eq:antighostmodeexpansion}) that    we postulate here are given by 
 \begin{eqnarray}
 \left \{
 \gamma_{i} (\vec{p}_{\perp}, \lambda ),  \overline{\gamma}_{j}^{\:\dag} (\vec{p}_{\perp}^{\:\prime},  
  \lambda^{\:\prime} )
 \right \} 
 &=&
 i \: \delta^{2}(\vec{p}_{\perp}^{} - \vec{p}_{\perp}^{\:\prime} ) 
 \delta ( \lambda - \lambda ^{\:\prime} )\frac{d \rho_{ij}(\lambda )}{d\lambda }\:,
\nonumber 
\\
\left \{
 \overline{\gamma}_{i} (\vec{p}_{\perp}, \lambda ),  \gamma_{j}^{\:\dag} (\vec{p}_{\perp}^{\:\prime},  
  \lambda^{\:\prime} )
 \right \} 
 &=&
- i \: \delta^{2}(\vec{p}_{\perp}^{} - \vec{p}_{\perp}^{\:\prime} ) 
 \delta ( \lambda - \lambda ^{\:\prime} )\frac{d \rho_{ij}(\lambda )}{d\lambda }\:,
  \hskip0.3cm (i, j=1,2)\:.
  \label{eq:gammagammabar}
 \end{eqnarray}
All other anti-commutators should vanish.  With (\ref{eq:gammagammabar}), we can compute the four-dimensional anti-commutaion relations between ghost and anti-ghost fields as 
\begin{eqnarray}
\left \{  c^{Z}(t, \vec{x}_{\perp}, z), \overline{c}^{Z}(t^{\:\prime}, \vec{x}_{\perp}^{\:\prime}, z^{\:\prime})
\right \}
=
i \Delta (t-t^{\:\prime}, \vec{x}_{\perp} - \vec{x}_{\perp}^{\:\prime} ; M_{Z}(*))\:.
\label{eq:czcbarzcr}
\end{eqnarray}
Thanks to the property  (\ref{eq:formula5}) of our $\Delta$-function, the equal-time anti-commutaion relations 
\begin{eqnarray}
\left \{  \dot{c}^{Z}(t, \vec{x}_{\perp}, z), \overline{c}^{Z}(t^{}, \vec{x}_{\perp}^{\:\prime}, z^{\:\prime})
\right \}
&=&
- i \: \delta ( \vec{x}_{\perp} - \vec{x}_{\perp}^{\:\prime}) \delta (z-z^{\:\prime}) \:,
\\
\left \{  c^{Z}(t, \vec{x}_{\perp}, z), \dot{\overline{c}}^{Z}(t^{}, \vec{x}_{\perp}^{\:\prime}, z^{\:\prime})
\right \}
&=&
i \: \delta ( \vec{x}_{\perp} - \vec{x}_{\perp}^{\:\prime}) \delta (z-z^{\:\prime}) \:,
\end{eqnarray}
follow from (\ref{eq:czcbarzcr}) immediately.

\section{More about the auxiliary field ${\cal B}^{Z}(t, \vec{x}_{\perp}, z)$}
\label{sec:moreabout}

In  Section  \ref{sec:gaugefieldquantization},  we carried out the mode expansion of $Z^{\mu}(t, \vec{x}_{\perp}, z)$ and $\chi^{3} ( t, \vec{x}_{\perp}, z)$ as given in (\ref{eq:zmu012alphacretionannihilation2}) and (\ref{eq:scalaralphacreationannihilation3}). Since the auxiliary field ${\cal B}^{Z}( t, \vec{x}_{\perp}, z)$ is expressed as a linear combination of these fields as in (\ref{eq:auxiliaryfieldbz}), the mode expansion of   ${\cal B}^{Z}( t, \vec{x}_{\perp}, z)$  is also already at our hand. Now by looking at the mode expansion of ${\cal B}^{Z}( t, \vec{x}_{\perp}, z)$ carefully, we will show in  Sections   \ref{subsec:modeexpansionbz}  and   \ref{subsec:connectionbetween}  that the creation operators contained in ${\cal B}^{Z}( t, \vec{x}_{\perp}, z)$,  when applied on the vacuum, give rise to only zero norm states which are unobservable.  We will define another field ${\cal N} (t, \vec{x}_{\perp}, z)$ in Section  \ref{sec:modeexpansionofnfield} and will discuss the BRST quartet mechanism in Section  \ref{sec:slavnovtaylor} by using ${\cal B}^{Z}( t, \vec{x}_{\perp}, z)$ and ${\cal N} (t, \vec{x}_{\perp}, z)$.

\subsection{Mode expansion of  ${\cal B}^{Z}(t, \vec{x}_{\perp}, z)$}
\label{subsec:modeexpansionbz}

Let us write down  the mode expansion of   ${\cal B}^{Z}(t, \vec{x}_{\perp}, z)$,   which is, by combining (\ref{eq:zmu012alphacretionannihilation2}) and (\ref{eq:scalaralphacreationannihilation3}),    cast  into  a  lengthy formula 
\begin{eqnarray}
& & \hskip-1cm 
{\cal B}^{Z}(t, \vec{x}_{\perp}, z)
\nonumber \\
&=&
\partial_{\mu}Z^{\mu} (t, \vec{x}_{\perp}, z)+M_{Z}(z) \chi^{3}   (t, \vec{x}_{\perp}, z)
\nonumber \\
&=& -i \int \sqrt{\lambda} \: d\lambda \sum_{i=1,2} \int \frac{d^{2}\vec{p}_{\perp}}{\sqrt{(2 \pi)^{2} 2E}} 
\nonumber \\
& & \hskip0.5cm  
\times
\bigg \{ \alpha_{i}^{(0)} (\vec{p}_{\perp}, \lambda ) e^{i \vec{p}_{\perp} \cdot \vec{x}_{\perp} -iEt}
- \alpha_{i}^{(0) \dag } (\vec{p}_{\perp}, \lambda ) e^{- i \vec{p}_{\perp} \cdot \vec{x}_{\perp} + iEt}
 \bigg \} \phi_{i}(z; \lambda )
 \nonumber \\
 & &
 + \int  \: d\lambda \sum_{k=1}^{4}  \int \frac{d^{2}\vec{p}_{\perp}}{ \sqrt{ (2 \pi)^{2} 2E}} 
 \left \{
\frac{d}{dz}\Phi_{1}^{(k)} (z; \lambda ) + M_{Z}(z) \Phi_{2}^{(k)} (z; \lambda )
\right \}
 \nonumber \\
 & &   \hskip0.5cm 
 \times 
\bigg \{  \beta_{k} (\vec{p}_{\perp}, \lambda ) e^{i \vec{p}_{\perp} \cdot \vec{x}_{\perp} -iEt}
+ \beta_{k}^{\dag} (\vec{p}_{\perp}, \lambda ) e^{- i \vec{p}_{\perp} \cdot \vec{x}_{\perp}  + iEt}
\bigg \} \:.
\label{eq:modeexpansionfz1}
\end{eqnarray}
There appears  an interesting  and suggestive combination of $\Phi _{1}^{(k)}( z; \lambda )$ and   
$\Phi _{2}^{(k)}( z; \lambda )$    in the last integrand  of (\ref{eq:modeexpansionfz1}), i.e., 
\begin{eqnarray}
\mathcal{F}^{(k)}    (z; \lambda )    \equiv   \frac{d}{dz}\Phi_{1}^{(k)} (z; \lambda ) + M_{Z}(z) \Phi_{2}^{(k)} (z; \lambda )\:,
\hskip1cm (k=1, \cdots , 4)\:.
\label{eq:peculiarcombination}
\end{eqnarray}
By shuffling the equation (\ref{eq:capitalphiequation})  for 
$\Phi _{1}^{(k)}( z; \lambda )$ and  $\Phi _{2}^{(k)}( z; \lambda )$, 
 we find an intriguing formula satisfied by   (\ref{eq:peculiarcombination}), 
\begin{eqnarray}
\left \{  -\frac{d^{2}}{dz^{2}} +M_{Z}(z)^{2} -\lambda   \right \}
\mathcal{F}^{(k)}(z; \lambda )
&=&
M_{Z}(z) \left \{
\frac{v^{\: \prime \prime}(z) }{v(z)} - V^{\:\prime} \left ( \frac{v(z)^{2}}{2}\right )
\right \} \Phi_{2}^{(k)}(z; \lambda )\:,
\nonumber \\
& & \hskip1cm (k=1, \cdots, 4)\:. 
\end{eqnarray}
whose right hand side vanishes, if we impose the absence of tadpole, i.e., 
(\ref{eq:ordinarydiffeq}).   In other words,  under (\ref{eq:ordinarydiffeq}),   the peculiar combination
(\ref{eq:peculiarcombination})   satisfies the same equation as that in 
(\ref{eq:schroedingertypeeq}) and can  be expressed as a linear combination of the basic solutions,  
$\phi_{1}(z; \lambda )$ and $\phi_{2}(z; \lambda )$.
To determine the form of the linear combination, we note the values of (\ref{eq:peculiarcombination})
 and its derivative at $z=0$ by looking at   (\ref{eq:intislconditions1})   -   (\ref{eq:intislconditions4}), , i.e., 
 \begin{eqnarray}
  & & \mathcal{F}^{(k)}    (0; \lambda )  = \mathfrak{f}_{1}^{\: (k)} \:,
\hskip2cm
\frac{d}{dz} \mathcal{F}^{(k)}    (z; \lambda ) \Bigg \vert_{z=0} = \mathfrak{f}_{2}^{\: (k)} \:,
 \end{eqnarray}
 where
\begin{eqnarray}
& &  \mathfrak{f}_{1}^{\: (1)}  = 0\:,
\hskip2cm
 \mathfrak{f}_{2}^{\: (1)}  =  M_{Z}(0)^{2} - \lambda \:,
\\
& &    \mathfrak{f}_{1}^{\: (2)} = 1\:,
\hskip2cm
 \mathfrak{f}_{2}^{\: (2)}  = 0\:,
\\
& &    \mathfrak{f}_{1}^{\: (3)}  = M_{Z}(0)\:,
\hskip1.1cm
 \mathfrak{f}_{2}^{\: (3)}  =  - M_{Z}(0)\frac{v^{\:\prime}(0)}{v(0)} \:,
\\
& &    \mathfrak{f}_{1}^{\: (4)}  = 0\:,
\hskip2cm
 \mathfrak{f}_{2}^{\: (4)}  = M_{Z}(0)\:.
\end{eqnarray}
Here use has been made of the derivative  formula of (\ref{eq:peculiarcombination}), 
\begin{eqnarray}
\frac{d}{dz} \mathcal{F}^{(k)}    (z; \lambda )  
&=&
 \left \{  M_{Z}(z)^{2} -  \lambda \right \}  \Phi_{1}^{(k)}(z; \lambda )
 -M_{Z}(z) \frac{v^{\:\prime}(z)}{v(z)} \Phi_{2}^{(k)} (z; \lambda )
         \nonumber \\
         & & 
 +M_{Z}(z) \frac{d }{dz}  \Phi_{2}^{(k)} (z; \lambda )\:.
  \end{eqnarray}
Comparing  the initial value conditions at $z=0$ in  (\ref{eq:initialconditions1}) and 
  (\ref{eq:initialconditions})  with the above,   we are led to the following equality
  \begin{eqnarray}
  \mathcal{F}^{(k)}    (z; \lambda )   = \sum_{i=1}^{2} \mathfrak{f}_{i}^{(k)} \phi_{i}(z; \lambda)\:.
  \label{eq:identification}
  \end{eqnarray}

Taking     (\ref{eq:identification})  into account,  the mode expansion   (\ref{eq:modeexpansionfz1}) is rendered much simpler, i.e.,  
\begin{eqnarray}
{\cal B}^{Z}(t, \vec{x}_{\perp}, z)
&=&
\int \: d\lambda 
\int \frac{d^{2}\vec{p}_{\perp}}{\sqrt{(2 \pi)^{2} 2E}} 
\nonumber \\
& & \hskip-1cm
\times  \sum_{i=1,2}
\bigg \{ \mathfrak{b}_{i} (\vec{p}_{\perp}, \lambda ) e^{i \vec{p}_{\perp} \cdot \vec{x}_{\perp} -iEt}
+ \mathfrak{b}_{i}^{ \dag } (\vec{p}_{\perp}, \lambda ) e^{- i \vec{p}_{\perp} \cdot \vec{x}_{\perp} + iEt}
 \bigg \} \phi_{i}(z; \lambda )\;,
 \label{eq:modeexpansionoffz2}
\end{eqnarray}
where the coefficients are defined by 
\begin{eqnarray}
 \mathfrak{b}_{i} (\vec{p}_{\perp}, \lambda ) = 
 -i \sqrt{\lambda } \: \alpha_{i}^{(0)}(\vec{p}_{\perp}, \lambda ) +
  \sum_{k=1}^{4} \: \mathfrak{f}_{i}^{(k)} 
 \beta_{k}(\vec{p}_{\perp}    , \lambda) \:, \hskip0.5cm (i=1,2)\:.
 \label{eq:mixtureoftwotypes}
\end{eqnarray}
 The simplification of the mode expansion  (\ref{eq:modeexpansionoffz2}) in comparison with (\ref{eq:modeexpansionfz1}) is, however,   not quite unexpected. We have already noticed before with reference to  (\ref{eq:beforekleingordon1})   and (\ref{eq:afterkleingordon}), that the auxiliary field ${\cal B}^{Z}(t, \vec{x}_{\perp}, z)$ obeys the  Klein-Gordon type equation  under the condition of   (\ref{eq:stability}), and  we have  actually foreseen    there  that the mode expansion of    ${\cal B}^{Z}(t, \vec{x}_{\perp}, z)$  should go along the line in Section  \ref{sec:wavepropagation}.    The expansion in (\ref{eq:modeexpansionoffz2}) is consistent with the previous observation.
 
 \subsection{Connection between $d\rho_{ij}(\lambda )$ and $d\widehat{\rho}_{kl}(\lambda )$}
 \label{subsec:connectionbetween}
 
 Looking at the two terms on the right hand side of (\ref{eq:mixtureoftwotypes}), we notice  immediately that, while   $ \beta_{k}^{\dag }(\vec{p}_{\perp}    , \lambda) $ generates positive norm states when acting  on the vacuum,   the other  $\alpha_{i}^{(0)\:\dag}(\vec{p}_{\perp}, \lambda ) $ introduces  negative norm states    on the contrary.    Such a mixture of qualitatively distinct type of operators is characteristic in the definition of    $ \mathfrak{b}_{i} (\vec{p}_{\perp}, \lambda ) $  and it it extremely interesting to compute  its commutator with   $\mathfrak{b}^{\dag}(\vec{p}_{\perp}^{\:\prime}, \lambda ^{\:\prime}) $.  On applying the formulae (\ref{eq:alphakalphalcommutator}) and   (\ref{eq:betakbetalcommutator}), we get the following provisional result
 \begin{eqnarray}
 & & \hskip-1cm
\left [  \mathfrak{b}_{i} (\vec{p}_{\perp}, \lambda ),  \mathfrak{b}^{\dag}_{j}(\vec{p}_{\perp}^{\:\prime}, 
\lambda ^{\:\prime}) \right ]
\nonumber \\
&=&
\sqrt{\lambda }  \sqrt{\lambda^{\:\prime}  }    \left [ \alpha_{i}^{(0)} (\vec{p}_{\perp}, \lambda ),   \alpha_{j}^{(0)\:\dag } 
(\vec{p}_{\perp}^{\:\prime}, \lambda^{\:\prime} ), 
\right ]
+
\sum_{k,l=1}^{4} \mathfrak{f}_{i}^{(k)} \mathfrak{f}_{j}^{(l)} \left [
\beta_{k}( \vec{p}_{\perp}, \lambda ),    \beta_{l}^{\dag}( \vec{p}_{\perp}^{\:\prime}, \lambda^{\:\prime} )
\right ]
\nonumber \\
&=&
\left \{
- \lambda  \frac{d\rho_{ij}(\lambda )}{d\lambda}
+
 \sum_{k,l=1}^{4} 
  \mathfrak{f}_{i}^{(k)} 
  \frac{d\widehat{\rho}_{kl}(\lambda ) }{d\lambda}
 \mathfrak{f}_{j}^{(l)} 
\right \}
\delta^{2}(\vec{p}_{\perp} - \vec{p}_{\perp}^{\:\prime} ) \delta (\lambda - \lambda^{\:\prime})\:.
\label{eq:anyfurther}
\end{eqnarray}
We would be  unable to go  further from  ($\ref{eq:anyfurther}$)  without the knowledge on the  connection between  $d\rho_{ij}(\lambda )$ and $d\widehat{\rho}_{kl}(\lambda )$. 
 
It is, however, more than likely that there must exist a close connection between  $d\rho_{ij}(\lambda )$ and $d\widehat{\rho}_{kl}(\lambda )$, because the two  differential equations, (\ref{eq:schroedingertypeeq})   and  (\ref{eq:2by2differentialequation}), are both given in terms of the common   function $v(z)$  if we replace $V^{\:\prime}(v^{2}/2)$   in  (\ref{eq:m1m2})   by $v^{\:\prime \prime}(z)/v(z)$   by virtue of   (\ref{eq:ordinarydiffeq}).  The   algorithm of computing    $d\rho_{ij}(\lambda )$ and $d\widehat{\rho}_{kl}(\lambda )$  has  in principle been established by the work of Titchmarsh \cite{titchmarsh}   and of Kodaira   \cite{kodaira1} -  \cite{kodaira2}, and it is most desirable  to look at the connection along the line of their work.  For now, however,  we employ a short-cut method without going into too much details of  mathematical complexities. 
 
 Let us now study the following integration 
 \begin{eqnarray}
 \sum_{k,l=1}^{4}\int d\lambda \: {\cal F}^{(k)}(z; \lambda ) \frac{d\widehat{\rho}_{kl}(\lambda ) }{d\lambda}
 {\cal F}^{(l)}(z^{\:\prime}; \lambda )
 &=&
  \sum_{k,l=1}^{4} \sum_{i, j=1}^{2}
  \int d\lambda \: 
  \mathfrak{f}_{i}^{(k)} \phi_{i}(z; \lambda )
  \frac{d\widehat{\rho}_{kl}(\lambda ) }{d\lambda}
 \mathfrak{f}_{j}^{(l)} \phi_{j}(z^{\:\prime}; \lambda )\:.
 \nonumber \\
    \label{eq:aboveequation0}
  \end{eqnarray}
Putting the definition (\ref{eq:peculiarcombination})  of  ${\cal F}^{(k)}(z; \lambda )$ into the left hand side, and using the integration formula    (\ref{eq:bigformula}), we arrive at a concise formula
 \begin{eqnarray}
 & & 
   \hskip-2cm
 \sum_{k,l=1}^{4}\int d\lambda \: {\cal F}^{(k)}(z; \lambda ) \frac{d\widehat{\rho}_{kl}(\lambda ) }{d\lambda}
 {\cal F}^{(l)}(z^{\:\prime}; \lambda )
     \nonumber \\
    &=&
   \sum_{k,l=1}^{4}    \int d\lambda \: 
  \left \{     \frac{d}{dz} \Phi_{1}^{(k)} ( z; \lambda )+ M_{Z}(z) \Phi_{2}^{(k)} (z; \lambda ) \right \}
  \frac{d\widehat{\rho}_{kl}(\lambda ) }{d\lambda}
  \nonumber \\
  & & \hskip1.5cm    \times 
   \left \{     \frac{d}{dz^{\:\prime}} \Phi_{1}^{(l)} ( z^{\:\prime}; \lambda )+ M_{Z}(z^{\:\prime}) \Phi_{2}^{(l)} (z^{\:\prime}; \lambda ) \right \}
   \nonumber \\
   &=&\left \{     \frac{d }{dz} \frac{d }{d z^{\:\prime}}    + M_{Z}(z)M_{Z}(z^{\:\prime})
   \right \} \delta (z - z^{\:\prime} )
   \nonumber \\
   &=&    \left \{     - \frac{d^{2} }{dz^{2}}    + M_{Z}(z)^{2}   \right \} \delta (z - z^{\:\prime} )\:.
   \label{eq:aboveequation}
  \end{eqnarray}
In the meantime we can express  the right hand side of (\ref{eq:aboveequation}) in the following way,  i.e.,
 \begin{eqnarray}
  \left \{     - \frac{d^{2} }{dz^{2}}    + M_{Z}(z)^{2}   \right \} \delta (z - z^{\:\prime} )
    &=&
  \left \{     - \frac{d^{2} }{dz^{2}}    + M_{Z}(z)^{2}   \right \} 
  \sum_{i,j=1}^{2} \int d\lambda \: \phi_{i}(z; \lambda ) \frac{d\rho_{ij}(\lambda )}{d\lambda}
\phi_{j}(z^{\:\prime}; \lambda ) 
 \nonumber \\
  &=&  
  \sum_{i,j=1}^{2} \int d\lambda \: \phi_{i}(z; \lambda )\: \lambda  \frac{d\rho_{ij}(\lambda )}{d\lambda}
\phi_{j}(z^{\:\prime}; \lambda ) \:.
\label{eq:above2}
 \end{eqnarray}
 Combining (\ref{eq:aboveequation0}),  (\ref{eq:aboveequation})  and (\ref{eq:above2}),    we get  an identity that contains both $d\rho_{ij}(\lambda )$ and   $d\widehat{\rho }_{kl}(\lambda )$, 
 \begin{eqnarray}
  \sum_{k,l=1}^{4} \sum_{i, j=1}^{2}
  \int d\lambda \: 
  \mathfrak{f}_{i}^{(k)} \phi_{i}(z; \lambda )
  \frac{d\widehat{\rho}_{kl}(\lambda ) }{d\lambda}
 \mathfrak{f}_{j}^{(l)} \phi_{j}(z^{\:\prime}; \lambda )
=  \sum_{i,j=1}^{2} \int d\lambda \: \phi_{i}(z; \lambda )\: \lambda  \frac{d\rho_{ij}(\lambda )}{d\lambda}
\phi_{j}(z^{\:\prime}; \lambda ) \:.
\nonumber \\
\label{eq:equality}
 \end{eqnarray}
 For the equality  (\ref{eq:equality}) to be valid for arbitrary values of $z$ and $z^{\:  \prime}$, we deduce  the connection between   $d\rho_{ij}(\lambda )$ and  $d\widehat{\rho }_{kl}(\lambda )$, 
 \begin{eqnarray}
  \sum_{k,l=1}^{4} 
  \mathfrak{f}_{i}^{(k)} 
  \frac{d\widehat{\rho}_{kl}(\lambda ) }{d\lambda}
 \mathfrak{f}_{j}^{(l)} 
&=&
  \: \lambda  \frac{d\rho_{ij}(\lambda )}{d\lambda} \:, \hskip0.5cm (i,j =1,2)\:.
 \end{eqnarray}
This has a deep impact  on  the commutator   (\ref{eq:anyfurther}).  Namely, we get a vanishing commutation relation
\begin{eqnarray}
\left [  \mathfrak{b}_{i} (\vec{p}_{\perp}, \lambda ),  \mathfrak{b}^{\dag}_{j}(\vec{p}_{\perp}^{\:\prime}, 
\lambda ^{\:\prime}) \right ]
&=& 0\:,   \hskip0.5cm (i,j =1,2)\:.
\label{eq:deepimpact}
\end{eqnarray}
This vanishing commutator indicates that  the operator  $\mathfrak{b}^{\dag}_{i}(\vec{p}_{\perp},  \lambda )$,  when applied on the vacuum,  generates only  zero-norm states. It should also be noted that 
(\ref{eq:deepimpact}) is consistent with the four-dimensional commutation relation (\ref{eq:bzbz3}).
 \subsection{Mode expansion of  ${\cal N}(t, \vec{x}_{\perp}, z)$}
 \label{sec:modeexpansionofnfield}

 In the usual BRST analysis in the standard electroweak theory with position-independent masses \cite{aoki1} - \cite{aoki3}, the BRST quartet fields in the $Z$-boson sector  consist of ${\cal B}^{Z}(t, \vec{x}_{\perp}, z)$,  $c^{Z}(t, \vec{x}_{\perp}, z)$, 
$\overline{c}^{Z}(t, \vec{x}_{\perp}, z)$ and the unphysical scalar field   $\chi^{3}(t, \vec{x}_{\perp}, z)$. 
For the decoupling  of the quartet members from the physical S-matrix  to work well, the asymptotic fields of the quartet must satisfy the same equations of  motion with the common mass term. In the presence of the electroweak bubble wall of our case,    ${\cal B}^{Z}(t, \vec{x}_{\perp}, z)$,  $c^{Z}(t, \vec{x}_{\perp}, z)$ and  $\overline{c}^{Z}(t, \vec{x}_{\perp}, z)$ satisfy the same Klein-Gordon type equations of motion with the common mass term $M_{Z}(z)$,  namely, (\ref{eq:afterkleingordon}),   (\ref{eq:ghosteqofmotion}) and (\ref{eq:antighosteqofmotion}),  but $\chi^{3}(t, \vec{x}_{\perp}, z)$ does not. We have already seen that $\chi^{3}(t, \vec{x}_{\perp}, z)$ satisfies the rather complicated  equation  (\ref{eq:2by2differentialequation})   coupled with $Z^{3}(t, \vec{x}_{\perp}, z)$.   We have to seek for something else for the fourth member of the quartet. 
 
There are two candidates for the fourth member, that is,\footnote{We use the summation symbol ($\sum$) for the repeated index $\mu$,  only when the summation is taken over $\mu = 0,1,2$.} 
 \begin{eqnarray}
 \sum_{\mu=0,1,2} \partial_{\mu} Z^{\mu}(t, \vec{x}_{\perp}, z), 
 \hskip1cm
 \frac{\partial }{\partial z} Z^{3}  (  t, \vec{x}_{\perp} , z )  +M_{Z}(z) \chi^{3}  (t, \vec{x}_{\perp} , z ) ,  \end{eqnarray}
 whose asymptotic fields  both satisfy the Klein-Gordon type equations of motion. We can think of various linear combinations of the above two, but   for definiteness we would like  to take   
 \begin{eqnarray}
 \hskip-1cm
 {\cal N}  (  t, \vec{x}_{\perp} , z )  
 & \equiv  &  
 - \sum_{\mu=0,1,2} \partial_{\mu} Z^{\mu}(t, \vec{x}_{\perp}, z) + 
 \frac{\partial }{\partial z} Z^{3}  (  t, \vec{x}_{\perp} , z )  +M_{Z}(z) \chi^{3}  (t, \vec{x}_{\perp} , z ) 
 \label{eq:definitionofn}
 \end{eqnarray}
 as the fourth member of the quartet. Actually in the absence of the tadpole,  the asymptotic field of (\ref{eq:definitionofn}) satisfies the Klein-Gordon type equation of motion with the mass term $M_{Z}(z)$, 
\begin{eqnarray}
\left \{  \square + M_{Z}(z)^{2} \right \} {\cal N}  (  t, \vec{x}_{\perp} , z )  
&=&
M_{Z}(z) \left \{
\frac{v^{\:\prime \prime}(z)}{v(z)}   -   V^{\:\prime} \left ( \frac{v(z)^{2}}{2}  \right ) 
\right \} \chi ^{3} (t, \vec{x}_{\perp}, z )
\nonumber \\
&=&0\:.
\label{eq:nsatisfieskg}
\end{eqnarray}
Note that the definition (\ref{eq:definitionofn}) differs from that of   ${\cal B}^{Z}(t, \vec{x}_{\perp}, z)$ in  (\ref{eq:auxiliaryfieldbz}) only by the minus sign in front of   the first term in  (\ref{eq:definitionofn}) and some of the formulas to be given  below are in parallel with   those of   ${\cal B}^{Z}(t, \vec{x}_{\perp}, z)$ .   The principal reason for choosing (\ref{eq:definitionofn}), however,  is that 
the combination of (\ref{eq:definitionofn})   makes the metric structure of the state vectors very  convenient owing to the commutation relation (\ref{eq:nncommutator}) to be explained later. Other choices would work equally well but the analysis on the BRST quartet mechanism would become a little involved, although the conclusion is unaltered.
 
 Since  ${\cal N}(t, \vec{x}_{\perp}, z)$  satisfies (\ref{eq:nsatisfieskg}), its mode expansion goes in the same way as in Section \ref{sebsec:fieldquantization} with the use of $\phi_{i}(z; \lambda ), (i=1,2)$,  namely, 
  \begin{eqnarray}
 {\cal N}  (  t, \vec{x}_{\perp} , z )
&=&
 \int d\lambda 
\int \frac{d^{2}\vec{p}_{\perp}}
 {\sqrt{(2\pi)^{2} \: 2E }} 
 \nonumber \\
 & & \hskip-1cm 
  \times    \sum_{i=1, 2}   \bigg \{
 \mathfrak{n}_{i}^{} (\vec{p}_{\perp}, \lambda ) 
\:  e^{i \vec{p}_{\perp} \cdot \vec{x}_{\perp} -iEt} 
 +    \mathfrak{n}_{i}^{\: \dag  }(\vec{p}_{\perp}, \lambda ) 
\:   e^{ - i \vec{p}_{\perp} \cdot \vec{x}_{\perp} + iEt}  \bigg \}
    \phi_{i}(z; \lambda)  \:,
\end{eqnarray}
where the coefficients are given by 
\begin{eqnarray}
   \mathfrak{n}_{i}^{  }(\vec{p}_{\perp}, \lambda )&=& 
   i\sqrt{\lambda} \: \alpha_{i}^{(0)} (\vec{p}_{\perp}, \lambda ) + \sum_{k=1}^{4} \mathfrak{f}_{i}^{(k)}
   \beta_{k}^{} (\vec{p}_{\perp}, \lambda )\:.
\label{eq:coefficientni}
  \end{eqnarray}
Comparing (\ref{eq:coefficientni}) with (\ref{eq:mixtureoftwotypes}), we notice that   the difference between    $\mathfrak{n}_{i}^{}(\vec{p}_{\perp}, \lambda )$ and $\mathfrak{b}_{i}^{}(\vec{p}_{\perp}, \lambda )$   is only the sign in front of the first term and therefore    the commutation relations among them are computed  in a parallel way with  the case of (\ref{eq:deepimpact}),   i.e., 
 \begin{eqnarray}
 \left [ \mathfrak{n}_{i}^{  }(\vec{p}_{\perp}, \lambda ),  \mathfrak{n}_{j}^{\: \dag  }(\vec{p}_{\perp}
 ^{\:\prime}, \lambda^{\:\prime} ) \right ]
 &=&0\:,\hskip1.5cm (i,j =1,2)\:, 
 \label{eq:nncommutator}
\\
 \left [ \mathfrak{n}_{i}^{  }(\vec{p}_{\perp}, \lambda ),  \mathfrak{b}_{j}^{\: \dag  }(\vec{p}_{\perp}
 ^{\:\prime}, \lambda^{\:\prime} ) \right ]
 &=&
  \delta^{2}(\vec{p}_{\perp} - \vec{p}_{\perp}^{\:\prime} ) \delta( \lambda - \lambda^{\:\prime})
\times 2     
 \lambda \frac{d\rho_{ij}(\lambda)}{d\lambda}  \:.
\label{eq:extralambda}
 \end{eqnarray}
The vanishing commutator (\ref{eq:nncommutator}) indicates  that the state 
$ \mathfrak{n}_{i}^{\: \dag  }(\vec{p}_{\perp} ^{}, \lambda^{} ) \vert 0 \rangle$ is a zero-norm 
state and is therefore unobservable.   The non-trivial commutator   (\ref{eq:extralambda}) 
implies that two states 
$ \mathfrak{n}_{i}^{\: \dag  }(\vec{p}_{\perp} ^{}, \lambda^{} ) \vert 0 \rangle$ 
and 
$ \mathfrak{b}_{i}^{\: \dag  }(\vec{p}_{\perp} ^{}, \lambda^{} ) \vert 0 \rangle$
have non-vanishing inner product and can communicate with each other.   The following four-dimensional commutation relation comes out of   (\ref{eq:extralambda}), 
\begin{eqnarray}
 & & \hskip-1cm
 \left [     {\cal N}  (  t, \vec{x}_{\perp} , z ), \:   {\cal B}^{Z}  (  t^{\:\prime}, \vec{x}_{\perp}^{\:\prime} , z^{\: \prime} )  \right ]
= 2i \left \{   - \frac{\partial^{2}}{\partial z^{2}} + M_{Z}(z)^{2}   \right \}
 \Delta(t-t^{\:\prime}, \vec{x}_{\perp} - \vec{x}_{\perp}^{\:\prime} z, z^{\:\prime} ; M_{Z}(*))\:.
 \nonumber \\
\label{eq:extradiffoperator}
 \end{eqnarray}
 The differential operator on the right hand side of (\ref{eq:extradiffoperator}) is due to the extra $\lambda$
  in front of the spectral density function in (\ref{eq:extralambda}).
The equal-time commutation relations 
\begin{eqnarray}
 \left [     {\cal N}  (  t, \vec{x}_{\perp} , z ), \:   {\cal B}^{Z}  (  t^{}, \vec{x}_{\perp}^{\:\prime} , z^{\:\prime} )  \right ]
&=&0\;,
\nonumber \\
 \left [     {\cal N}  (  t, \vec{x}_{\perp} , z ), \:  \dot{ {\cal B}}^{Z}  (  t^{}, \vec{x}_{\perp}^{\:\prime} , z^{\:\prime}   )  \right ]
&=&2i \left \{   - \frac{\partial^{2}}{\partial z^{2}} + M_{Z}(z)^{2}   \right \}
\delta^{2}(\vec{x}_{\perp} - \vec{x}_{\perp}^{\:\prime} ) \delta ( z - z^{\:\prime} )\:, 
\label{eq:ndotbdiifop}
 \end{eqnarray}
are attained  immediately from (\ref{eq:extradiffoperator}).   

Incidentally, it is worth remarking that we can  derive (\ref{eq:extradiffoperator}), by starting from the two commutation relations  in (\ref{eq:ndotbdiifop}) and  by employing the method explained in  Appendix \ref{appa:fourdimcr} without referring to (\ref{eq:extralambda}).

  \subsection{The Slavnov-Taylor identity and the BRST quartet mechanism }
  \label{sec:slavnovtaylor}
  
  In previous sections we have worked out all of the non-vanishing commutation    relations between creation and annihilation   operators of wave    propagation modes   associated with  the  asymptotic fields, ${\cal B}^{Z} ( t, \vec{x}_{\perp}, z)$,   ${\cal N}^{Z} ( t, \vec{x}_{\perp}, z)$,    $c^{Z} ( t, \vec{x}_{\perp}, z)$,   and       $\overline{c}^{Z} ( t, \vec{x}_{\perp}, z)$.  Those calculations are suggesting  that the states  obtained by applying four creation operators 
\begin{eqnarray}
\mathfrak{b}_{i}^{\dag}( \vec{p}_{\perp} , \lambda ) \:,
\hskip0.5cm
\mathfrak{n}_{i}^{\dag}( \vec{p}_{\perp} , \lambda ) 
\:, \hskip0.5cm
\gamma_{i}^{\dag}( \vec{p}_{\perp} , \lambda ) 
, \hskip0.5cm
\overline{\gamma}_{i}^{\dag}( \vec{p}_{\perp} , \lambda )
 \:, \hskip0.5cm
\label{eq:brstquartetstates}
\end{eqnarray}
on a physical state   form  a BRST quartet.  We are now in a position to investigate the  decoupling of unphysical states, namely, ghost, anti-ghost, unphysical scalar fields  and unphysical polarization states of gauge fields. The point of the decoupling from the physical S-matrix is that, whenever these states appear in the course of time development,    they always appear   in a particular combination of zero-norm state which    is     unobservable.   The pattern of the combination can be seen by looking at the Slavnov-Taylor type identity    \cite{slavnov} - \cite{jctaylor2}\:.

Let us consider the following identity 
  \begin{eqnarray}
     & & \hskip-1.0cm
\left \{ Q_{B}, {\rm T}\left (   {\cal N} (t, \vec{x}_{\perp}, z ) \overline{c}^{Z} (t^{\:\prime}, \vec{x}_{\perp}^{\:\prime}, z^{\:\prime} )   \right )\right \}  
      \nonumber \\ 
&=&
 {\rm T} \left (  \left [ Q_{B}, \: {\cal N} (t, \vec{x}_{\perp}, z)\right ]
   \overline{c}^{Z} (t^{\:\prime}, \vec{x}_{\perp}^{\:\prime}, z^{\:\prime}) \right )   
   +     {\rm T} \left ( {\cal N} (t, \vec{x}_{\perp}, z) \left \{ Q_{B}, \:   \overline{c}^{Z} (t^{\:\prime}, \vec{x}_{\perp}^{\:\prime}, z^{\:\prime})  \right \} 
   \right ), 
  \end{eqnarray}
 where  $``\: {\rm T}\:  "$ stands for time-ordered product. By sandwiching this identity with physical states,  
 $\vert {\rm phys} \rangle $ and $\langle {\rm phys}^{\:\prime} \vert $,   and by using the subsidiary condition 
    (\ref{eq:subsidiarycondition}) we arrive at 
  \begin{eqnarray}
0  &=&
\langle {\rm phys}^{\:\prime}  \vert {\rm T} \left (  \left [ Q_{B}, \: {\cal N} (t, \vec{x}_{\perp}, z)\right ]
   \overline{c}^{Z} (t^{\:\prime}, \vec{x}_{\perp}^{\:\prime}, z^{\:\prime}) \right )    \vert {\rm phys} \rangle 
      \nonumber \\
         & & \hskip1cm
   +    \langle {\rm phys}^{\:\prime} \vert  {\rm T} \left ( {\cal N} (t, \vec{x}_{\perp}, z) \left \{ Q_{B}, \:   \overline{c}^{Z} (t^{\:\prime}, \vec{x}_{\perp}^{\:\prime}, z^{\:\prime})  \right \} 
   \right ) \vert {\rm phys} \rangle \:.
  \label{eq:slavnovtaylor2}
  \end{eqnarray}
 This identity (\ref{eq:slavnovtaylor2}) indicates that the unwanted  states created by (\ref{eq:brstquartetstates}),  whenever they appear in graphical calculations, must appear  in  a particular combination.   This can be seen by evaluating the commutator and anti-commutator in (\ref{eq:slavnovtaylor2})  explicittly.  Before doing so,   note that the commutation and anti-commutation relations of $Q_{B}$ with various fields as given  in   (\ref{eq:slavnovtaylor2})   and in    (\ref{eq:brstcummmutationrelations}) are those with Heisenberg fields,  while creation and annihilation operators   are defined referring  to the asymptotic fields. We  have to  express  the relations (\ref{eq:brstcummmutationrelations}) by using the  asymptotic fields.

  Let us recall in this connection that Haag \cite{haag1} has once argued that  Heisenberg operators may be expanded   in an infinite series of the asymptotic field products.  Using the technique developed in \cite{lsz2}, Glaser, Lehmann and Zimmermann (GLZ) \cite{glz} refined the Haag's formula further for a wider range of applicability.  (See Appendix C in Ref. \cite{kugoojima5} for more details on the GLZ formula.)   According to the analyses in  Refs. \cite{kugoojima3} and \cite{kugoojima5}, the commutation (or anti-commutation) relations of asymptotic fields with $Q_{B}$     are linear and non-linear interaction terms do not appear. This is plausible since the asymptotic fields are basically free fields.  The present case differs from that discussed in \cite{kugoojima3} and \cite{kugoojima5} in that the mass terms depend on the $z$-coordinate, but still non-linear interaction  terms should not appear and     the relevant commutator and anti-commutator in  (\ref{eq:slavnovtaylor2}) then become 
\begin{eqnarray}
\left \{ Q_{B}, \overline{c}^{Z}(t, \vec{x}_{\perp}, z) \right \} &=&  i\: {\cal B}^{Z}(t, \vec{x}_{\perp}, z) \:,
\label{eq:qbnanticommutator0}
\\
\left [ Q_{B}, {\cal N}(t, \vec{x}_{\perp}, z) \right ] &=& -2i\:\left \{ - \frac{\partial^{2}}{\partial z^{2}} + M_{Z}(z)^{2}  \right \}  c^{Z}(t, \vec{x}_{\perp}, z) \:.
\label{eq:qbnanticommutator}
\end{eqnarray}
We note here that the  differential operator on the right hand side  of (\ref{eq:qbnanticommutator}) is due to the relation (\ref{eq:ndotbdiifop}). 
 This differential operator can be  manipulated  equivalently in the following way. Namely we note that the mode expansion formula  (\ref{eq:ghostmodeexpansion}) gives 
\begin{eqnarray}
& & \hskip-1cm
-2i \left \{ - \frac{\partial^{2}}{\partial z^{2}} + M_{Z}(z)^{2}  \right \}  c^{Z} (t, \vec{x}_{\perp}, z) 
\nonumber \\
 &=&  -2i 
  \int d\lambda \sum_{i=1, 2} \int \frac{d^{2}\vec{p}_{\perp}}
 {\sqrt{(2\pi)^{2} \: 2E }}
 \nonumber \\
 & & \times \bigg \{
\lambda \:  \gamma_{i}(\vec{p}_{\perp}, \lambda ) e^{i \vec{p}_{\perp} \cdot \vec{x}_{\perp} -iEt}
  +   \lambda \: \gamma_{i}^{\dag }(\vec{p}_{\perp}, \lambda ) e^{ - i \vec{p}_{\perp} \cdot \vec{x}_{\perp} + iEt}
\bigg \}   \phi_{i}(z; \lambda)   \:, 
 \end{eqnarray}
and this indicates that annihilation and creation  operators should be simply replaced by 
$-2i \lambda \gamma_{i} ( \vec{p}_{\perp}, \lambda )$ and $-2i \lambda \gamma_{i}^{\dag} ( \vec{p}_{\perp}, \lambda )$, respectively, when we use the relation (\ref{eq:qbnanticommutator}).

\begin{center}
\begin{figure}[bht]
\begin{center}
\unitlength 0.1in
\begin{picture}(36.00,20.50)(13.10,-27.60)
\put(39.5000,-13.1000){\makebox(0,0)[lb]{$\mathfrak{b}_{i}( \vec{p}_{\perp}, \lambda)$}}%
\put(17.0000,-13.1000){\makebox(0,0)[lb]{$\mathfrak{n}_{i}( \vec{p}_{\perp}, \lambda)$}}%
\put(17.0000,-26.4000){\makebox(0,0)[lb]{$\gamma_{i}( \vec{p}_{\perp}, \lambda)$}}%
\put(39.2000,-26.3000){\makebox(0,0)[lb]{$\overline{\gamma}_{i}( \vec{p}_{\perp}, \lambda)$}}%
\put(39.4000,-18.0000){\makebox(0,0)[lb]{$Q_{B}$}}%
%
\special{pn 8}%
\special{pa 3160 2450}%
\special{pa 2650 2450}%
\special{fp}%
\special{sh 1}%
\special{pa 2650 2450}%
\special{pa 2717 2470}%
\special{pa 2703 2450}%
\special{pa 2717 2430}%
\special{pa 2650 2450}%
\special{fp}%
%
\special{pn 13}%
\special{pa 3580 2020}%
\special{pa 4910 2020}%
\special{pa 4910 2760}%
\special{pa 3580 2760}%
\special{pa 3580 2020}%
\special{fp}%
%
\special{pn 8}%
\special{pa 3170 1070}%
\special{pa 2660 1070}%
\special{fp}%
\special{sh 1}%
\special{pa 2660 1070}%
\special{pa 2727 1090}%
\special{pa 2713 1070}%
\special{pa 2727 1050}%
\special{pa 2660 1070}%
\special{fp}%
\put(16.0000,-10.4000){\makebox(0,0)[lb]{${\cal N}(t, \vec{x}_{\perp}, z)$}}%
\put(38.6000,-10.4000){\makebox(0,0)[lb]{${\cal B}^{Z}(t, \vec{x}_{\perp}, z)$}}%
\put(15.9000,-23.6000){\makebox(0,0)[lb]{$c^{Z}(t, \vec{x}_{\perp}, z)$}}%
\put(38.4000,-23.6000){\makebox(0,0)[lb]{$\overline{c}^{Z}(t, \vec{x}_{\perp}, z)$}}%
%
\special{pn 13}%
\special{pa 1310 2020}%
\special{pa 2640 2020}%
\special{pa 2640 2760}%
\special{pa 1310 2760}%
\special{pa 1310 2020}%
\special{fp}%
%
\special{pn 13}%
\special{pa 1320 710}%
\special{pa 2650 710}%
\special{pa 2650 1450}%
\special{pa 1320 1450}%
\special{pa 1320 710}%
\special{fp}%
%
\special{pn 13}%
\special{pa 3580 710}%
\special{pa 4910 710}%
\special{pa 4910 1450}%
\special{pa 3580 1450}%
\special{pa 3580 710}%
\special{fp}%
%
\special{pn 8}%
\special{pa 1990 1450}%
\special{pa 1990 2020}%
\special{fp}%
\special{sh 1}%
\special{pa 1990 2020}%
\special{pa 2010 1953}%
\special{pa 1990 1967}%
\special{pa 1970 1953}%
\special{pa 1990 2020}%
\special{fp}%
%
\special{pn 8}%
\special{pa 4240 2020}%
\special{pa 4240 1450}%
\special{fp}%
\special{sh 1}%
\special{pa 4240 1450}%
\special{pa 4220 1517}%
\special{pa 4240 1503}%
\special{pa 4260 1517}%
\special{pa 4240 1450}%
\special{fp}%
\put(16.9000,-17.9000){\makebox(0,0)[lb]{$Q_{B}$}}%
\put(29.4000,-10.3000){\makebox(0,0)[lb]{(6.27)}}%
\put(29.4000,-24.0000){\makebox(0,0)[lb]{(5.66)}}%
\put(20.7000,-18.1000){\makebox(0,0)[lb]{(6.33)}}%
\put(43.1000,-18.1000){\makebox(0,0)[lb]{(6.32)}}%
\put(29.3000,-12.8000){\makebox(0,0)[lb]{(6.26)}}%
\put(29.5000,-26.6000){\makebox(0,0)[lb]{(5.65)}}%
%
\special{pn 8}%
\special{pa 3080 2450}%
\special{pa 3570 2450}%
\special{fp}%
\special{sh 1}%
\special{pa 3570 2450}%
\special{pa 3503 2430}%
\special{pa 3517 2450}%
\special{pa 3503 2470}%
\special{pa 3570 2450}%
\special{fp}%
%
\special{pn 8}%
\special{pa 2970 1070}%
\special{pa 3580 1070}%
\special{fp}%
\special{sh 1}%
\special{pa 3580 1070}%
\special{pa 3513 1050}%
\special{pa 3527 1070}%
\special{pa 3513 1090}%
\special{pa 3580 1070}%
\special{fp}%
\end{picture}%
\caption{
The relations among the four members of the BRST quartet.  The arrows with $Q_{B}$ means   that the fields and operators are transformed by the BRST transformation along the arrows.   The fields and operators connected by left-right arrows have non-vanishing commutation or anti-commutaion relations.}
\label{fig2:BRSTquartet}
\end{center}
\end{figure}
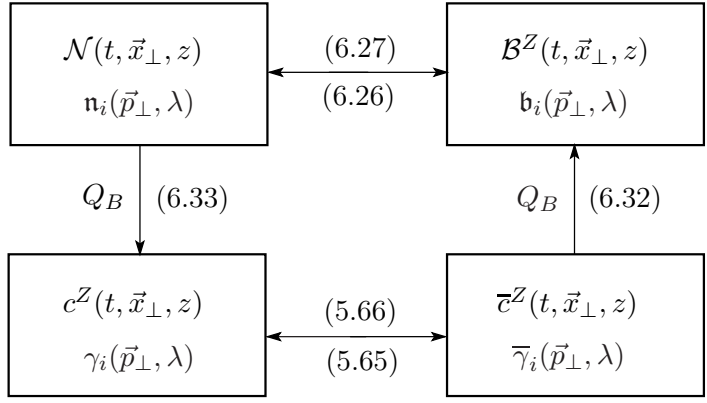
 \end{center}
 
Thus the identity   (\ref{eq:slavnovtaylor2}) amounts to saying that,  whenever unphysical states appear  in the course of time-development,   they are always combined  together  with the following particular relative weight,  
\begin{eqnarray}
\vert (\vec{p}_{\perp}, \lambda , i  ),   (\vec{p}_{\perp}^{\:\prime} , \lambda^{\:\prime}  , i^{\:\prime}   ) \rangle \equiv 
\left \{
\mathfrak{n}_{i}^{\dag} (\vec{p}_{\perp}, \lambda ) \mathfrak{b}_{i^{\:\prime}}^{\dag} (\vec{p}_{\perp}^{\:\prime} , \lambda^{\:\prime} )
+
 2i\lambda \: \gamma_{i}^{\dagger} ( \vec{p}_{\perp} , \lambda ) \overline{\gamma}_{i^{\:\prime} }^{\dag} ( \vec{p}^{\:\prime}_{\perp} , \lambda^{\: \prime} )
\right \} \vert 0 \rangle \:.
\label{eq:slavnovtaylorcombination}
\end{eqnarray}
The important point here is that   (\ref{eq:slavnovtaylorcombination})  is a zero-norm state, i.e., 
 \begin{eqnarray}
 \langle (\vec{q}_{\perp}, \lambda_{2} , j  ),   (\vec{q}_{\perp}^{\:\prime} , \lambda_{2}^{\:\prime}  , j^{\:\prime}   ) \vert (\vec{p}_{\perp}, \lambda_{1} , i  ) ,  (\vec{p}_{\perp}^{\:\prime} , \lambda_{1}^{\:\prime}  , i^{\:\prime}   ) \rangle 
 =0\:,
 \label{eq:zeronormproperty}
 \end{eqnarray}
 for arbitrary labels  of  $(\vec{p}_{\perp}, \lambda_{1} , i )$,   $(\vec{p}_{\perp}^{\:\prime}, \lambda_{1}^{\:\prime} , i^{\prime} )$,   $(\vec{q}_{\perp}, \lambda_{2} , j )$   and $(\vec{q}_{\perp}^{\:\prime}, \lambda_{2}^{\:\prime} , j^{\prime} )$.    This can be confirmed by employing  the commutation and anti-commutation relations given  so far and by computing the following  inner products, 
 \begin{eqnarray}
 & &\hskip-1cm
 \langle 0 \vert 
 \mathfrak{b}_{j^{\:\prime}} (\vec{q}_{\perp}^{\:\prime}, \lambda_{2}^{\:\prime} ) 
 \mathfrak{n}_{j} (\vec{q}_{\perp}^{} , \lambda_{2}^{} ) 
\mathfrak{n}_{i}^{\dag} (\vec{p}_{\perp}, \lambda_{1} ) 
\mathfrak{b}_{i^{\:\prime}}^{\dag} (\vec{p}_{\perp}^{\:\prime} , \lambda_{1}^{\:\prime} ) 
 \vert 0 \rangle 
 \nonumber \\
 &=& \delta^{2}( \vec{p}_{\perp} - \vec{q}_{\perp}^{\:\prime} ) \delta (\lambda_{1} - \lambda_{2}^{\:\prime})
\:  2\lambda_{1} \frac{d\rho_{ij^{\:\prime}} (\lambda_{1})}{d\lambda_{1}}
 \cdot 
 \delta^{2}( \vec{p}^{\:\prime}_{\perp} - \vec{q}_{\perp}^{} ) \delta (\lambda_{1}^{\:\prime} - 
 \lambda_{2}^{})
 \: 2\lambda_{2} \frac{d\rho_{i^{\:\prime}j^{}} (\lambda_{2})}{d\lambda_{2}}\:,
 \nonumber \\
 \label{eq:bnbn}
   \\
 & &\hskip-1cm
 (+ 2i\lambda_{1})  \cdot ( - 2i\lambda_{2}) \cdot 
 \langle 0 \vert 
 \overline{\gamma}_{j^{\:\prime}} (\vec{q}_{\perp}^{\:\prime}, \lambda_{2}^{\:\prime} ) 
 \gamma_{j} (\vec{q}_{\perp}^{} , \lambda_{2}^{} ) 
\gamma _{i}^{\dag} (\vec{p}_{\perp}, \lambda_{1} ) 
\overline{\gamma}_{i^{\:\prime}}^{\dag} (\vec{p}_{\perp}^{\:\prime} , \lambda_{1}^{\:\prime} ) 
 \vert 0 \rangle 
 \nonumber \\
 &=& -4\lambda_{1}\lambda_{2}\cdot 
 \delta^{2}( \vec{p}_{\perp} - \vec{q}_{\perp}^{\:\prime} ) \delta (\lambda_{1} - \lambda_{2}^{\:\prime})
\:   \frac{d\rho_{ij^{\:\prime}} (\lambda_{1})}{d\lambda_{1}}
 \cdot 
 \delta^{2}( \vec{p}^{\:\prime}_{\perp} - \vec{q}_{\perp}^{} ) \delta (\lambda_{1}^{\:\prime} - 
 \lambda_{2}^{})
 \:  \frac{d\rho_{i^{\:\prime}j} (\lambda_{2})}{d\lambda_{2}}\:.
 \nonumber \\
 \label{eq:gammagammagammagamma}
\end{eqnarray}
The cancellation occurring between  (\ref{eq:bnbn}) and  (\ref{eq:gammagammagammagamma})
brings us immediately to the zero-norm property  (\ref{eq:zeronormproperty}).   The state
(\ref{eq:slavnovtaylorcombination}) is thus unobservable. 
It should also be noted that the state (\ref{eq:slavnovtaylorcombination}) is orthogonal to all of the physical states.   The relations among the four members of the BRST quartet are summarized in Figure  \ref{fig2:BRSTquartet}.

 {\color{black}{
\subsection{BRST cohomology classes}

A supplementary remark to be added here is that the zero norm property (\ref{eq:zeronormproperty}) 
 can also be seen by rewriting (\ref{eq:slavnovtaylorcombination})
 in the following way 
\begin{eqnarray}
 \vert (\vec{p}_{\perp}, \lambda , i  ),   (\vec{p}_{\perp}^{\:\prime} , \lambda^{\:\prime}  , i^{\:\prime}   ) \rangle 
 =
 - Q_{B} \: \left (\mathfrak{n}_{i}^{\dag} (\vec{p}_{\perp}, \lambda ) \overline{\gamma}_{i^{\prime}}^{\dag} (\vec{p}_{\perp}^{\: \prime}, \lambda^{\prime} ) \vert 0 \rangle \right )\:.
 \end{eqnarray}
This can be confirmed by noting  $Q_{B} \vert 0 \rangle =0$ and  the following commutation and anti-commutation relations
\begin{eqnarray}
\left [ Q_{B}, \mathfrak{n}_{i} ^{\dag}( \vec{p}_{\perp} , \lambda ) \right ] = -2i \lambda \gamma _{i}^{\dag} ( \vec{p}_{\perp}, \lambda ) , \:\:\:\:
\left \{ 
Q_{B} \:  \overline{\gamma}_{i^{\prime}}^{\dag} (\vec{p}_{\perp}^{\: \prime}, \lambda^{\prime} )\right \}= - \mathfrak{b}_{i^{\prime}}^{\dag} ( \vec{p}_{\perp}^{\prime}, \lambda ^{\prime} ),
\end{eqnarray}
which can be  deduced, respectively,   from  (\ref{eq:qbnanticommutator})
and 
(\ref{eq:qbnanticommutator0}).
A little more generally one can confirm  the identity 
\begin{eqnarray}
& &\hskip-2cm
 \left \{
\mathfrak{n}_{i}^{\dag} (\vec{p}_{\perp}, \lambda ) \mathfrak{b}_{i^{\:\prime}}^{\dag} (\vec{p}_{\perp}^{\:\prime} , \lambda^{\:\prime} )
+
 2i\lambda \: \gamma_{i}^{\dag} ( \vec{p}_{\perp} , \lambda ) \overline{\gamma}_{i^{\:\prime} }^{\dag} ( \vec{p}^{\:\prime}_{\perp} , \lambda^{\: \prime} )
\right \} \vert {\rm phys} \rangle 
\nonumber  \\
&=&
 - Q_{B} \: \left (\mathfrak{n}_{i}^{\dag} (\vec{p}_{\perp}, \lambda ) \overline{\gamma}_{i^{\prime}}^{\dag} (\vec{p}_{\perp}^{\: \prime}, \lambda^{\prime} ) \vert {\rm phys} \rangle \right )
 \label{eq:brstcohomology}
\end{eqnarray}
for all physical states $\vert {\rm phys} \rangle $ because of $Q_{B} \vert {\rm phys} \rangle =0$.  These  are zero norm states and are orthogonal to all of the physical states.
In gauge theories in general,  two physical states,  $\vert {\rm phys } \rangle $ and $\vert {\rm phys }^{\:\prime}  \rangle $,  are said to belong to the same BRST cohomology class if the difference between  these two states are in the form of $Q_{B} \vert \psi \rangle $, i.e.,
\begin{eqnarray}
\vert {\rm phys } \rangle  = \vert {\rm phys }^{\:\prime}  \rangle +   Q_{B} \vert \psi \rangle\:.
\end{eqnarray}
The states (\ref{eq:brstcohomology}) expressed in terms of creation operators all belong to  the set of $Q_{B} \vert \psi \rangle $ states.
}}

 \section{Physical Polarization States}
 \label{sec:physicalpolarization}
 
In the presence of the bubble wall, we are unable to define properly the transverse or longitudinal polarization  of gauge field wave-propagation, because of the lack of  momentum conservation law in the $z$-direction.  However, we are still able to distinguish the  physical polarization states from unphysical ones,   thanks to the subsidiary condition (\ref{eq:subsidiarycondition}) and the decoupling mechanism of quartet fields from the physical S-matrix explained in the previous section.

First of all it is almost obvious that the states
 \begin{eqnarray}
 \alpha_{i}^{(1)\: \dag } (\vec{p}_{\perp}, \lambda ) \vert 0 \rangle \:,
 \hskip1cm 
  \alpha_{i}^{(2)\: \dag } (\vec{p}_{\perp}, \lambda ) \vert 0 \rangle \:,
  \hskip1cm (i=1, 2)\:,
  \label{eq:transeversepolarizationstates}
 \end{eqnarray}
are positive norm states and  satisfy  the physical state condition (\ref{eq:subsidiarycondition}).  (Note the relations $\left [ Q_{B}, \alpha_{i}^{(1)\:\dagger} (\vec{p}_{\perp}, \lambda ) \right ]=0$ and   $\left [ Q_{B}, \alpha_{i}^{(2)\:\dagger} (\vec{p}_{\perp}, \lambda ) \right ]=0$.)   In addition to (\ref{eq:transeversepolarizationstates}),  we have so far discussed the following states 
 \begin{eqnarray}
& &\alpha_{i}^{(0)\:\dag} (\vec{p}_{\perp}^{}, \lambda^{} ) \vert 0 \rangle   
  =
   - \frac{1}{2i \sqrt{\lambda}}\left \{
  \mathfrak{n}_{i}^{\dag}( \vec{p}_{\perp}, \lambda  ) -   \mathfrak{b}_{i}^{\dag}( \vec{p}_{\perp}, \lambda  ) \right \}  \vert 0 \rangle  \:,    \hskip0.8cm (i=1, 2)\:,
\\
& &\sum_{k=1}^{4}   \mathfrak{f}_{i}^{(k)} \beta_{k}^{\dag} (\vec{p}_{\perp}^{}, \lambda^{} ) \vert 0 \rangle   
  =
   \frac{1}{2}\left \{
  \mathfrak{n}_{i}^{\dag}( \vec{p}_{\perp}, \lambda  ) +   \mathfrak{b}_{i}^{\dag}( \vec{p}_{\perp}, \lambda  ) \right \}  \vert 0 \rangle  \:,
   \hskip1cm (i=1, 2)\:.
   \label{eq:unphysicalscalarlikepolarization}
 \end{eqnarray}
We have already seen in Section \ref{sec:slavnovtaylor} that these states appear in the perturbative calculations only in a particular combination of unobservable zero-norm states together with ghost and anti-ghost states. 

Now we would like to argue that there exist other combinations of  $\beta_{k}^{\dag} (\vec{p}_{\perp}^{}, \lambda^{} ) \vert 0 \rangle    $,    i.e.,   
 \begin{eqnarray}
  \sum_{k=1}^{4}   \mathfrak{h}_{i}^{(k)} \beta_{k}^{\dag} (\vec{p}_{\perp}^{}, \lambda^{} ) \vert 0 \rangle \:,\hskip1cm (i=1, 2)\;,
  \label{eq:longitudinallikestate}
 \end{eqnarray}
which are orthogonal  to (\ref{eq:unphysicalscalarlikepolarization}) and have not yet appeared in the foregoing analyses of the present paper.   The coefficients $\mathfrak{h}_{i}^{(k)} (k=1, \cdots , 4)$ will be determined soon.  In order for the states (\ref{eq:longitudinallikestate})   to be physical, we require the condition 
 \begin{eqnarray}
 \left [ Q_{B}, \: 
  \sum_{k=1}^{4}   \mathfrak{h}_{i}^{(k)} \beta_{k}^{\dag} (\vec{p}_{\perp}^{}, \lambda^{} ) 
  \right ] =0, 
  \hskip1cm (i=1, 2)\:.
  \label{eq:longitudinalpolarizationcondition}
 \end{eqnarray}
Looking at the expression  (\ref{eq:brstchargebcterms}) of $Q_{B}$, it is necessary and sufficient to set 
 \begin{eqnarray}
 & & 
   \sum_{k=1}^{4} \left [   \mathfrak{h}_{i}^{(k)} \beta_{k} (\vec{p}_{\perp}, \lambda ),      {\cal B}^{Z}(t, \vec{x}_{\perp}, z)  \right ]=0\;, 
\end{eqnarray}
or equivalently 
\begin{eqnarray}
  \sum_{k=1}^{4} \left [   \mathfrak{h}_{i}^{(k)} \beta_{k} (\vec{p}_{\perp}, \lambda ), 
 \mathfrak{b}_{j}^{\dag} ( \vec{p}_{\perp}^{\:\prime}, \lambda ^{\:\prime} )  \right ]=0
 \label{eq:7777}
\end{eqnarray}
for the condition (\ref{eq:longitudinalpolarizationcondition}) to be satisfied. 
Putting the definition (\ref{eq:mixtureoftwotypes}) of $\mathfrak{b}_{j}^{} ( \vec{p}_{\perp}^{}, \lambda ^{} )$ into (\ref{eq:7777}), we get the physical state condition of the state (\ref{eq:longitudinallikestate})
\begin{eqnarray}
\sum_{k=1}^{4} \sum_{l=1}^{4}\mathfrak{h}_{i}^{(k)} 
\frac{d\widehat{\rho}_{kl}(\lambda)}{d\lambda } \mathfrak{f}_{j}^{(l)}=0\:,
\label{eq:hikortogonaltofjl}
 \end{eqnarray}
which is an algebraic equation to determine the coefficients $\mathfrak{h}^{(k)}_{i}$.   Eq. (\ref{eq:hikortogonaltofjl}) states that the two sets of states, (\ref{eq:unphysicalscalarlikepolarization}) and (\ref{eq:longitudinallikestate}), are orthogonal to each other.   In this way we classify  the four states $\beta^{\dag}_{k} (\vec{p}_{\perp}, \lambda) \vert 0 \rangle , \:\: (k=1, \cdots , 4)$  into a pair of  states given in (\ref{eq:unphysicalscalarlikepolarization}) and another pair of states  (\ref{eq:longitudinallikestate}). The latter pair of states contributes to the physical S-matrix, but the former does not.   To sum up,  physical three pairs of polarization states consist of  (\ref{eq:transeversepolarizationstates})  and  (\ref{eq:longitudinallikestate}).

The polarization vectors corresponding to the two states in (\ref{eq:transeversepolarizationstates})  are of course given by   (\ref{eq:polarizationvectors3})   and   (\ref{eq:polarizationvectors}), respectively.
On the other hand,  the third state   (\ref{eq:longitudinallikestate}) is polarized along  the $z$-direction and therefore the polarization vector is 
\begin{eqnarray}
\varepsilon^{(3)\:\mu  }=\left ( 
\begin{tabular}{c}
$0$
\\
$0$
\\
$0$
\\
$1$
\end{tabular}
\right )\:.
\label{eq:longitudinalpolarizationavector}
\end{eqnarray}
Collecting   (\ref{eq:polarizationvectors3}),    (\ref{eq:polarizationvectors}) and (\ref{eq:longitudinalpolarizationavector}) altogether,  the summation  formula over these three physical polarization vectors turns out to be 
\begin{eqnarray}
\sum_{a=1,2,3} \varepsilon^{(a)\:\mu} \varepsilon^{(a)\:\nu} 
=
-g^{\mu \nu} + \frac{1}{\lambda}\: \widehat{p}^{\: \mu} \widehat{p}^{\: \nu}, 
\hskip1.0cm
\widehat{p}^{\: \mu} = \left ( 
\begin{tabular}{c}
$E$
\\
$p_{x}$
\\
$p_{y}$
\\
$0$
\end{tabular}
\right ) \:.
\label{eq:polarizationsumformula}
\end{eqnarray}
Here let us recall that our notations are $g^{\mu \nu} = {\rm diag} (1, -1,-1,-1)$ and  $  \lambda =E^{2} -  \vert \vec{p}_{\perp}  \vert^{2} $. 
It has been pointed out in   \cite{farrar} that there occurs the rearrangement among  transverse and longitudinal   polarization states during the passage of the wave through the wall, but our summation formula (\ref{eq:polarizationsumformula})   looks rather simple.

\section{Summary}
\label{sec:summary}
 
 In the present paper we have investigated the standard electroweak gauge field theory  in the presence of the bubble wall by taking the $Z^{\mu}$-field sector as the representative  case.  We included the effect of the position dependent  Higgs condensate $v(z)$  into the asymptotic field equations of motion of   $Z^{\mu} ( t, \vec{x}_{\perp}, z) $  and the unphysical scalar field $\chi^{3}  ( t, \vec{x}_{\perp}, z)$.   In the $R_{\xi}$-gauge with $\xi=1$,  the $Z^{\mu}  ( t, \vec{x}_{\perp}, z)$ asymptotic fields with $\mu =0,1$ and $2$ satisfy the Klein-Gordon type equations of motion (\ref{eq:zbosoneqofmotion2}), but  those  of $Z^{3}  ( t, \vec{x}_{\perp}, z)$ and $\chi^{3}  ( t, \vec{x}_{\perp}, z)$  are coupled with each other as shown in   (\ref{eq:z3chi3eqofmotion1}) and   (\ref{eq:z3chi3eqofmotion2}).

 We applied the eigenfunction expansion method developed in Refs. \cite{weyl} - \cite{kodaira2} to the asymptotic field equations and introduced two sets of operators,  $\alpha_{i}^{(a)} (\vec{p}_{\perp}, \lambda ) $ ($i=1, 2$ and $a=0, 1, 2 $) and  $\beta_{k}(\vec{p}_{\perp}, \lambda ) $ ($k=1, \cdots , 4$).  The commutation relations between these operators with their hermitian conjugates are postulated with the help of   the spectral functions,  as in (\ref{eq:alphakalphalcommutator}) and (\ref{eq:betakbetalcommutator}).  The  auxiliary field ${\cal B}^{Z}  ( t, \vec{x}_{\perp}, z)$ that was introduced through  the gauge fixing procedure also satisfies the Klein-Gordon type equation (\ref{eq:afterkleingordon}) in the absence of the tadpole.  The annihilation and creation   operators, $\mathfrak{b}_{i}$ and $\mathfrak{b}^{\dag}_{i}$,  of ${\cal B}^{Z}  ( t, \vec{x}_{\perp}, z)$  are expressed as linear combinations of  $\alpha_{i}^{(0)} (\vec{p}_{\perp}, \lambda )$ and    $\beta_{k}(\vec{p}_{\perp},\lambda )$,  and their hermitian conjugates, as shown in   (\ref{eq:mixtureoftwotypes}). We also defined another field ${\cal N}  ( t, \vec{x}_{\perp}, z)$ by (\ref{eq:definitionofn}) and argued that states created by ghost, anti-ghost, ${\cal N}  ( t, \vec{x}_{\perp}, z)$ and the auxiliary ${\cal B}^{Z}  ( t, \vec{x}_{\perp}, z)$  fields constitute the BRST quartet and that they decouple jointly from the physical S-matrix. 
 
Now that the theoretical framework of the standard model gauge theory in the presence of the bubble wall has been  put into  a good shape, we are now able to  launch with renewed interest into  reanalyses of plasma interactions with the electroweak bubble wall that have been dealt with in 
a large variety  of publications    \cite{boedeker1},  \cite{boedeker2}  - \cite{ai3},   \cite{dine1} - \cite{doomsday}  in the last several decades\:.
{\color{black}{
The microscopic approach 
as given in the present paper is appropriate for   studying momentum transfer in $z$-direction from plasma to the bubble wall in a systemmatic way. 
Note that each wave propagation mode of the gauge boson is characterized by $\vec{p}_{\perp}$ and $\lambda$ and there is no conservation law associated with $\lambda$.  It will be interesting to see how the pressure exerted on the wall by plasma particles  depends on the Lorentz boost factor, when the effects of multi-soft vector boson emission in the transition radiation are summed up effectively  to all orders.   A careful look at the infrared region of  the phase space is required, and the framework given in this paper will be the most reliable,  since the position dependent mass is taken into account without any approximation. 

There exists another type of approach based on hydrodynamical methods under the local thermal equilibrium condition.   This approach  is supposed to be useful for the case of relatively low velocity of the bubble wall.
 Plasma particles in the distribution function  in such approaches  are endowed with definite fixed masses and with definite three-momenta,  and   the momentum non-conservation in the $z$-direction is introduced in a somewhat  ad hoc manner.  In order to improve such a situation,  the field theoretical approach as developed in the present paper
 should be incorporated in the hydrodynamical approach. The quantum version of the Boltzmann equation should be based on the finite temperature Green's functions, which must be expressed in terms of  $\vec{p}_{\perp}$ and $\lambda$  without using such an  ill-defined quantity as the $z$-component of momenta.  Also various techniques, that have been developed in non-equilibrium statistical physics, must be applied extensively to Green's functions.
 We hope to  come to these issues in our future publications. 

}}
\appendix
\section{The ghost and anti-ghost  Lagrangian}
\label{sect:FPghostlagrangian}

For  convenience' sake,  we here rewrite the  Faddeev-Popov Lagrangian  (\ref{eq:symmetryrespecedFPghostLagrangian}) by using the ``physical" ghost and anti-ghost  fields,    i.e.,  (\ref{eq:anotherghost})  and (\ref{eq:anotherantighost}). (See also \cite{aoki2}.)  The gauge fields  are also expressed in terms of the physical combinations  (\ref{eq:wpmza}).  The unphysical scalar field,  $\chi^{1}$ and $\chi^{2}$,  are  combined together in the form of (\ref{eq:chiplusminusdef}).   The Lagrangian  (\ref{eq:symmetryrespecedFPghostLagrangian})  is classified into three terms
\begin{eqnarray}
{\cal L}_{\rm FP}={\cal L}_{\rm FP}^{(2)} + {\cal L}_{\rm FP}^{(V)} + {\cal L}_{\rm FP}^{(S)}\:,
\label{eqappghostlagrangian1}
\end{eqnarray}
where ${\cal L}_{\rm FP}^{(2)}$ is the quadratic part with respect to the ghost and anti-ghost fields
\begin{eqnarray}
{\cal L}_{\rm FP}^{(2)} &=&   
\partial^{\mu} {\overline c}^{(+)} \partial_{\mu}  c^{(-)} -  {\overline c}^{(+)} M_{W}^{\: 2}  c^{(-)}
+  
\partial^{\mu}  {\overline c}^{(-)} \partial_{\mu}   c^{(+)}  -  {\overline c}^{(-)} M_{W}^{\: 2}   c^{(+)}
 \nonumber \\
  & & 
 +  \partial^{\mu} {\overline c}^{Z}  \partial_{\mu}    c^{Z}    -  {\overline c}^{Z}  M_{Z}^{\: 2}   c^{Z}   
 + \partial^{\mu} {\overline c}^{A}   \partial_{\mu}    c^{A}\:.
 \label{eqappghostlagrangian2}
\end{eqnarray}
On the other hand,  the interaction terms   with vector    (${\cal L}_{\rm FP}^{(V)}$) and with scalar fields (${\cal L}_{\rm FP}^{(S)}$) are given,  respectively,     by   
\begin{eqnarray}
 {\cal L}_{\rm FP}^{(V)} &=&
 \frac{ig}{\sqrt{g^{2} + g^{\:\prime \:2}}} \bigg [  \partial^{\mu} {\overline c}^{(+)} \left \{
 -c^{(-)} \left ( g^{\:\prime} A_{\mu} + g Z_{\mu}   \right ) 
 +\left ( g^{\:\prime} c^{A} + gc^{Z}   \right )  W_{\mu}^{-}   \right \}  
 \nonumber \\
 & & + \partial^{\mu}  {\overline c}^{(-)} \left \{
 c^{(+)} \left ( g^{\:\prime} A_{\mu} + g Z_{\mu}    \right )
  -  \left ( g^{\:\prime} c^{A} + gc^{Z}   \right )W_{\mu}^{+}    \right \}
  \nonumber \\
  & & + \left (    g^{\:\prime} \partial^{\mu} {\overline c} ^{A} + g \partial^{\mu} {\overline c}^{Z}    \right )
  \left (    -c^{(+)} W_{\mu}^{-} + c^{(-)} W_{\mu}^{+}    \right )  \bigg ]\:,
 \\ 
  {\cal L}_{\rm FP}^{(S)}&=&
  \frac{i}{4} g^{2} v \chi^{3}  \left (   {\overline c}^{(+)} c^{(-)}  -   {\overline c}^{(-)} c^{(+)}    \right )
  -
    \frac{1}{4} g^{2} v H   \left (   {\overline c}^{(+)} c^{(-)}  +   {\overline c}^{(-)} c^{(+)}    \right )
    \nonumber \\
    & & 
    -\frac{1}{4} \left (  g^{2} + g^{\:\prime \:2} \right )  v H \:  {\overline c}^{Z} c^{Z} 
    \nonumber \\
    & &
    + \frac{igv}{4 \sqrt{g^{2} + g^{\:\prime \:2}}} \left (   -{\overline c} ^{(+)} \chi^{-} 
    + {\overline c}^{(-)} \chi^{+}    \right ) \left \{
    2gg^{\:\prime} c^{A} - \left (  g^{\:\prime \:2} -g^{2} \right )   c^{Z}    \right \}
    \nonumber \\
    & &
    +\frac{igv}{4}  \sqrt{g^{2} + g^{\:\prime \:2}} \: {\overline c}^{Z} \left (
    c^{(+)} \chi^{-} - c^{(-)} \chi^{+}
    \right )\:.
\end{eqnarray}

\section{The BRST transformation}
\label{sec:brsttransformation}

The BRST transformation rules of Heisenberg fields given in Section \ref{sec:43brsttransformation} are  rewritten here by using physical variables,  (\ref{eq:bpmbzba}),     (\ref{eq:chiplusminusdef}),  (\ref{eq:wpmza}), (\ref{eq:anotherghost}) and     (\ref{eq:anotherantighost}).  (The same formulas were previously  listed up  in  \cite{aoki2} by using their  conventions.)   The gauge and scalar field transformation formulas are as follows, 
\begin{eqnarray}
\delta _{B} W_{\mu}^{\pm} &=& \delta \lambda \Big [ \partial_{\mu} c^{(\pm)} \pm \frac{ig}{\sqrt{g^{2} 
+ g^{\:\prime 2}}}  \left \{ 
-(g^{\:\prime} c^{A} + gc^{Z} ) W_{\mu}^{\pm} + (g^{\:\prime} A_{\mu} + g Z_{\mu} ) c^{(\pm)}
\right \} \Big ]\:,
 \\
\delta _{B} Z_{\mu} &=& \delta \lambda \Big [ \partial_{\mu} c^{Z} +  \frac{ig^{2}}{\sqrt{g^{2} 
+ g^{\:\prime 2}}}  \left \{ 
-  c^{(+)} W_{\mu}^{-} + c^{(-)} W_{\mu}^{+} 
\right \} \Big ]\:,
\\
\delta _{B} A_{\mu} &=& \delta \lambda \Big [ \partial_{\mu} c^{A} +  \frac{igg^{\:\prime} }{\sqrt{g^{2} 
+ g^{\:\prime 2}}}  \left \{ 
-  c^{(+)} W_{\mu}^{-} + c^{(-)} W_{\mu}^{+} 
\right \} \Big ]\:,
\\
\delta_{B} \chi^{+}&=& \delta \lambda \Big [  - \frac{i}{2\sqrt{g^{2} + g^{\:\prime 2}}}
\left \{
2gg^{\:\prime} c^{A} -(g^{\:\prime \:2} - g^{2} )c^{Z}
\right \}  \chi^{+}
 + \frac{g}{2} c^{(+)} (v+ H + i\chi^{{\color{black}{3}}}) \Big ]\:,
\nonumber \\
\\
\delta_{B}  \chi^{-} &=& \delta \lambda \Big [  \frac{i}{2\sqrt{g^{2} + g^{\:\prime 2}}}
\left \{
2gg^{\:\prime} c^{A} -(g^{\:\prime \:2} - g^{2} )c^{Z}
\right \}    \chi^{-}
+  \frac{g}{2} c^{(-)} (v+ H - i\chi^{{\color{black}{3}}}) \Big ]\:,
\nonumber \\
\\
\delta_{B}H &=&\delta \lambda \Big [ - \frac{ \: g}{2  } \left (
c^{(-)} \chi^{+}  + c^{(+)} \chi^{-}
\right ) -\frac{1}{2} \sqrt{g^{2} + g^{\:\prime \: 2} }\:  c^{Z} \chi^{{\color{black}{3}}}\Big ] \:,
\\
\delta_{B} \chi^{{\color{black}{3}}}  &=& \delta \lambda \Big [  + \frac{i\:g}{2}  \left (
c^{(-)} \chi^{+}   -   c^{(+)} \chi^{-}
\right ) + \frac{1}{2} \sqrt{g^{2} + g^{\:\prime \: 2} }\:  c^{Z}  (v+ H )   \Big ] \:.
\end{eqnarray}
Here $\delta \lambda$ is an anti-commuting  c-number.
The ghost fields, on the other hand, obey the following transformation rules,
\begin{eqnarray}
\delta _{B} c^{(+)} &=& \delta \lambda \Big [ 
-\frac{ig}{\sqrt{g^{2} + g^{\:\prime \:2}}}
\left ( gc^{Z} + g^{\:\prime} c^{A} \right) c^{(+)} \Big ]\:,
\label{eq:ghostbrs1}
\\
\delta _{B} c^{(-)}&=& \delta \lambda \Big [ 
\frac{ig}{\sqrt{g^{2} + g^{\:\prime \:2}}}
\left ( gc^{Z} + g^{\:\prime} c^{A} \right) c^{(-)} \Big ]\:,
\\
\delta _{B} c^{Z}&=& \delta \lambda \Big [ 
\frac{ig^{2}}{\sqrt{g^{2} + g^{\:\prime \:2}}}
c^{(-)}c^{(+)} \Big ]\:,
\\
\delta _{B} c^{A}&=& \delta \lambda \Big [ 
\frac{igg^{\:\prime} }{\sqrt{g^{2} + g^{\:\prime \:2}}}
c^{(-)}c^{(+)} \Big ]\:,
\label{eq:ghostbrs4}
\end{eqnarray}
while the rules for the anti-ghost fields are simply given by the auxiliary fields, 
\begin{eqnarray}
\delta _{B} {\overline c}^{(\pm)} &=& {\color{black}{-}} \delta \lambda B^{\pm}\:,
\hskip1cm
\delta _{B} {\overline c}^{Z}= {\color{black}{-}} \delta \lambda B^{Z}\:,
 \hskip1cm
\delta _{B} {\overline c}^{A}= {\color{black}{-}} \delta \lambda B^{A}\:,
\end{eqnarray}
and those for the  auxiliary fields are 
\begin{eqnarray}
\delta_{B} B^{\pm}=0 \: ,\hskip1cm 
\delta_{B} B^{Z}=0 \: ,\hskip1cm 
\delta_{B} B^{A}=0\: .
\end{eqnarray}

\section{The four-dimensional commutator of   ${\cal B}^{Z}(t, \vec{x}_{\perp}, z)$ }
\label{appa:fourdimcr}

We now prove the four-dimensional commutation relation (\ref{eq:bzbz3})  by using  Nakanishi's technique employed in \cite{nakanishinoboru1} and \cite{nakanishinoboru2}.   We argued in  Section \ref{subsec:asymptoticfields}  that ${\cal B}^{Z}(t, \vec{x}_{\perp}, z)$ satisfies the Klein-Gordon type equation in the absence of the  tadpole, i.e.,
\begin{eqnarray}
\big \{
\square +M_{Z}(z)^{2} 
\big \} {\cal B}^{Z}(t, \vec{x}_{\perp}, z) =0\:.
\end{eqnarray}
Then we can derive the following  identity 
\begin{eqnarray}
 {\cal B}^{Z}(t, \vec{x}_{\perp}, z)
&=&  \int d^{2}{\vec x}_{\perp}^{\:\prime}   \: dz^{\:\prime}
 \bigg \{ \frac{\partial }{\partial t^{\:\prime} }\Delta (t-t^{\:\prime} , \vec{x}_{\perp} -{\vec x}_{\perp}^{\:\prime}, z,  z^{\:\prime}; M_{Z}(*)) \: 
 {\cal B}^{Z}(t^{\:\prime}, \vec{x}^{\:\prime}_{\perp}, z^{\:\prime} )
 \nonumber \\
 & & 
 -
 \Delta (t-t^{\:\prime} , \vec{x}_{\perp} -{\vec x}_{\perp}^{\:\prime}, z,  z^{\:\prime}; M_{Z}(*) ) \: 
 \frac{\partial }{\partial t^{\:\prime} }  {\cal B}^{Z}(t^{\:\prime}, \vec{x}^{\:\prime}_{\perp}, z^{\:\prime} )
 \bigg \}\:.
 \label{eq:usefulformulae}
\end{eqnarray}
To confirm  (\ref{eq:usefulformulae}), we first note that the right hand side of   (\ref{eq:usefulformulae})
is independent of $t^{\:\prime}$,  as we  see by   the following manipulation
\begin{eqnarray}
& & 
\hskip-1.0cm
\frac{\partial }{\partial t ^{\:\prime}}
 \int d^{2}{\vec x}_{\perp}^{\:\prime}   \: dz^{\:\prime}
 \bigg \{ \frac{\partial }{\partial t^{\:\prime} }\Delta (t-t^{\:\prime} , \vec{x}_{\perp} -{\vec x}_{\perp}^{\:\prime}, z,  z^{\:\prime}; M_{Z}(*)) \: 
    {\cal B}^{Z}(t^{\:\prime}, \vec{x}^{\:\prime}_{\perp}, z^{\:\prime} )
 \nonumber \\
 & & 
 \hskip1.5cm
 -
 \Delta (t-t^{\:\prime} , \vec{x}_{\perp} -{\vec x}_{\perp}^{\:\prime}, z,  z^{\:\prime}; M_{Z}(*) ) \: 
 \frac{\partial }{\partial t^{\:\prime} } {\cal B}^{Z}(t^{\:\prime}, \vec{x}^{\:\prime}_{\perp}, z^{\:\prime} )
 \bigg \}
 \nonumber \\
 &=&
 \int d^{2}{\vec x}_{\perp}^{\:\prime}   \: dz^{\:\prime}
 \bigg \{ 
 \left (
 \nabla ^{\:\prime \: 2}  - M_{Z}(z^{\:\prime} )^{2}
 \right )
 \Delta (t-t^{\:\prime} , \vec{x}_{\perp} -{\vec x}_{\perp}^{\:\prime}, z,  z^{\:\prime}; M_{Z}(*)) \: 
   {\cal B}^{Z}(t^{\:\prime}, \vec{x}^{\:\prime}_{\perp}, z^{\:\prime} )
 \nonumber \\
 & & 
 \hskip1.5cm
-  \Delta (t-t^{\:\prime} , \vec{x}_{\perp} -{\vec x}_{\perp}^{\:\prime}, z,  z^{\:\prime}; M_{Z}(*) ) \: 
 \frac{\partial ^{2}}{\partial t^{\:\prime \: 2} } {\cal B}^{Z}(t^{\:\prime}, \vec{x}^{\:\prime}_{\perp}, z^{\:\prime} )
 \bigg \}
 \nonumber \\
 &=&
 -  \int d^{2}{\vec x}_{\perp}^{\:\prime}   \: dz^{\:\prime}
 \Delta (t-t^{\:\prime} , \vec{x}_{\perp} -{\vec x}_{\perp}^{\:\prime}, z,  z^{\:\prime}; M(*)) \: 
 \left \{
 \square ^{\:\prime } + M_{Z}(z^{\: \prime}) ^{\:2}
 \right \}
   {\cal B}^{Z}(t^{\:\prime}, \vec{x}^{\:\prime}_{\perp}, z^{\:\prime} )
 \nonumber \\
 &=&0\:.
\end{eqnarray}
Here use has been made of  (\ref{eq:formula2}), together with the partial integration twice.  Since (\ref{eq:usefulformulae})  is independent of $t^{\:\prime}$,  we are   allowed to  evaluate the integration in  (\ref{eq:usefulformulae}) by setting $t=t^{\:\prime}$ and arrive at the left hand side of (\ref{eq:usefulformulae})  by using (\ref{eq:formula3}) and  (\ref{eq:formula5}).

Now that (\ref{eq:usefulformulae})  has been established, let us compute the  commutator by using (\ref{eq:usefulformulae}),  
\begin{eqnarray}
& &\hskip-0.7cm
\left [ 
 {\cal B}^{Z}(t, \vec{x}_{\perp}, z),    
\: {\cal B}^{Z}(t^{\: \prime \prime}, \vec{x}_{\perp}^{\: \prime \prime},  z^{\: \prime \prime}  )
 \right ]
 \nonumber \\
&=&  \int d^{2}{\vec x}_{\perp}^{\:\prime}   \: dz^{\:\prime}
 \bigg \{ \frac{\partial }{\partial t^{\:\prime} }\Delta (t-t^{\:\prime} , 
 \vec{x}_{\perp} -{\vec x}_{\perp}^{\:\prime}, z,  z^{\:\prime}; M_{Z}(*)) \: 
 \left [ 
 {\cal B}^{Z}(t^{\:\prime}, \vec{x}^{\:\prime}_{\perp}, z^{\:\prime} ), 
 \: {\cal B}^{Z}(t^{\: \prime \prime}, \vec{x}_{\perp}^{\: \prime \prime},  z^{\: \prime \prime}  )
 \right ]
 \nonumber \\
 & & 
 \hskip0.5cm
 -
 \Delta (t-t^{\:\prime} , \vec{x}_{\perp} -{\vec x}_{\perp}^{\:\prime}, z,  z^{\:\prime}; M_{Z}(*) ) \: 
 \frac{\partial }{\partial t^{\:\prime} } 
 \left [  {\cal B}^{Z}(t^{\:\prime}, \vec{x}^{\:\prime}_{\perp}, z^{\:\prime} ), 
 \: {\cal B}^{Z}(t^{\: \prime \prime}, \vec{x}_{\perp}^{\: \prime \prime},  z^{\: \prime \prime}  )
 \right ]
 \bigg \}\:. 
 \label{eq:A4}
\end{eqnarray}
Since the right hand side of (\ref{eq:A4}) is independent of $t^{\:\prime}$, we can set 
$t^{\:\prime}=t^{\:\prime \prime}$,  use the knowledge of the equal time commutators (\ref{eq:bzbz2}),  and are led to conclude that the four-dimensional commutator
 vanishes, i.e.,
\begin{eqnarray}
\left [ 
 {\cal B}^{Z}(t, \vec{x}_{\perp}, z),    
\: {\cal B}^{Z}(t^{\: \prime \prime}, \vec{x}_{\perp}^{\: \prime \prime},  z^{\: \prime \prime}  )
 \right ]=0\:.
\end{eqnarray}
This completes the proof of  (\ref{eq:bzbz3}).


    \end{document}